\documentclass[a4paper,11pt]{article}
\pdfoutput=1 

\usepackage{jheppub}
\usepackage[T1]{fontenc} 
\usepackage{xcolor}

\usepackage[dvipsnames]{xcolor}

\usepackage{tikz}
\usetikzlibrary{math}
\usetikzlibrary{arrows,shapes,positioning}
\usetikzlibrary{decorations.markings}
\tikzstyle arrowstyle=[scale=1]
\tikzstyle directed=[postaction={decorate,decoration={markings,
    mark=at position .65 with {\arrow[arrowstyle]{stealth}}}}]
\tikzstyle reverse directed=[postaction={decorate,decoration={markings,
    mark=at position .65 with {\arrowreversed[arrowstyle]{stealth};}}}]

\usepackage{multirow}
\usepackage{subcaption}

\usepackage{framed}
\usepackage{amsthm} 
\usetikzlibrary{patterns}
\usetikzlibrary{snakes}

\def\hl{{\text{\tiny HL}}}

\newtheorem{prop}{Property}

\newcommand{\bal}{\begin{equation}\begin{aligned}}
\newcommand{\eal}{\end{aligned}\end{equation}}

\newcommand{\beq}{\begin{equation}}
\newcommand{\eeq}{\end{equation}}
\newcommand{\beqa}{\begin{eqnarray}}
\newcommand{\eeqa}{\end{eqnarray}}

\newcommand{\ov}{\over}
\newcommand{\g}{\gamma}

\def\eps{{\epsilon}}

\definecolor{grey}{rgb}{0.4,0.4,0.5}
\definecolor{darkgreen}{rgb}{0,0.5,0}
\definecolor{darkred}{rgb}{0.6,0.0,0}
\definecolor{lightbrown}{rgb}{1,0.9,0.8}
\definecolor{brown}{rgb}{0.6,0.3,0.3}
\definecolor{darkblue}{rgb}{0,0,0.5}
\definecolor{darkmagenta}{rgb}{0.5,0,0.5}

\newcommand{\ads}{${\rm  AdS}_5\times {\rm S}^5\ $}
\newcommand{\adsst}{$\text{AdS}_3\times \text{S}^3\times\text{T}^4$}

\def\bY{{\dot Y}}

\newcommand{\la}{\label}
\def\a {\alpha}
\def\St{{\widetilde S}}
\def\Sb{{\bar S}}
\def\tSi{\widetilde \Sigma}
\def\Si{ \Sigma}
\def\bes{{\text{\tiny BES}}}
\def\tp{{\widetilde p}}
\def\tE{\widetilde\E}

\def\tK{{\widetilde K}}
\def\hstar{\,\hat{\star}\,}
\def\cstar{\,\check{\star}\,}
\newcommand{\E}{\mathcal E}

\title{Deriving the \adsst \,Quantum Spectral Curve I: Y-system and discontinuity relations}

\author[a]{Andrea Cavagli\`a,}
\author[b]{Davide Polvara,}
\author[a]{Nicol\`o Primi,}
\author[c,d,1]{Alessandro Sfondrini,}
\author[a]{Roberto Tateo}

\affiliation[a]{
Dipartimento di Fisica, Universit\`a di Torino and INFN, Sezione di Torino,
Via Pietro Giuria 1, I-10125 Torino, Italy
}

\affiliation[b]{
Institut f\"ur Theoretische Physik, Universit\"at Hamburg,
Luruper Chaussee 149, 22761 Hamburg, Germany
}

\affiliation[c]{
School of Mathematics, University of Birmingham,
Watson Building, Edgbaston, Birmingham B15 2TT, UK
}

\affiliation[d]{
INFN, Sezione di Padova,
via Marzolo 8, 35131 Padova, Italy
}

\note{On leave from the University of Padova, Italy.}

\emailAdd{andrea.cavaglia@unito.it}
\emailAdd{davide.polvara@gmail.com}
\emailAdd{nicolo.primi@unito.it}
\emailAdd{a.sfondrini@bham.ac.uk}
\emailAdd{roberto.tateo@unito.it}

\abstract{
The mirror Thermodynamic Bethe Ansatz (TBA) describes the spectrum of \adsst\, superstrings supported by a mixture of Ramond-Ramond (RR) and Neveu-Schwarz-Neveu-Schwarz (NSNS) flux. In recent years, a conjecture was put forward regarding the Quantum Spectral Curve (QSC) formulation in the Ramond-Ramond case. In this paper, we initiate the derivation of the Ramond-Ramond QSC from the mirror TBA equations. We describe the Y-system underlying the TBA and its discontinuity relations for pure-RR backgrounds. This provides the foundation for the derivation of the T-system and the Quantum Spectral Curve, which will be presented in a follow-up paper.
}

\begin{document} 
\begin{flushright}\small{ZMP-HH/26-27}\end{flushright}
\maketitle
\section{Introduction}
\label{sec:introduction}

The AdS$_3$/CFT$_2$ correspondence provides one of the first examples of a holographic description~\cite {Maldacena:1997re}. The best-studied instance of this duality is the one involving the \adsst\ background. This preserves sixteen supersymmetries - half of the maximal amount - and therefore presents a richer dynamics than $\text{AdS}_5\times\text{S}^5$ and $\text{AdS}_4\times\text{CP}^3$, which preserve 32 and 24 supersymmetries, respectively. 

\adsst\ can be supported by a mixture of Ramond-Ramond (RR) and Neveu-Schwarz-Neveu-Schwarz (NSNS) fluxes. At the pure NSNS point, the string spectrum can be found by worldsheet-CFT techniques, in terms of a Wess-Zumino-Witten model based on $SL(2,\mathbb{R})$~\cite{Maldacena:2000hw}. It can be expressed in closed form and is highly degenerate. Moving away from this point by turning on RR fluxes, however, the worldsheet-CFT description becomes non-local, and the spectrum more intricate.
While it is in principle possible to study this model by perturbing around the NSNS point (see~\cite{Cho:2018nfn}), integrability is likely to be the best tool for solving the spectral problem, and certainly the only one which can deal with \textit{finite} (as opposed to \textit{infinitesimal}) amounts of RR flux.
Indeed, superstrings propagating on \adsst
supported by mixed RR/NSNS flux are classically integrable \cite{Cagnazzo:2012se}.
\footnote{The study of classical integrability for AdS3/CFT2 was initiated, in the case of RR backgrounds, in~\cite{Babichenko:2009dk}.}

Over the past several years, the integrability machinery has been developed to construct the worldsheet S-matrix in both the pure RR~\cite{Borsato:2013qpa,Borsato:2014exa,Borsato:2014hja} and mixed-flux theories~\cite{Hoare:2013pma,Lloyd:2014bsa}, and dressing factors together with the mirror Thermodynamic Bethe Ansatz (TBA) equations have recently been proposed in both cases (see~\cite{Frolov:2021fmj,Frolov:2021bwp} and~\cite{Frolov:2025uwz,Frolov:2024pkz,Frolov:2025tda,OhlssonSax:2023qrk} respectively).
In particular, the equations for the pure RR case, which is the focus of this paper, have been proposed in~\cite{Frolov:2021bwp}.
These equations govern the spectrum of \adsst\, strings at any string tension. 
They are written in terms of massive left- and right-momentum carrying modes, together with massless modes, the latter being a characteristic feature of AdS$_3$ models and absent in the higher-dimensional AdS$_5$ and AdS$_4$ cases. In addition, there are auxiliary excitations that do not directly contribute to the state's energy.

In the case of $\text{AdS}_5\times\text{S}^5$ and $\text{AdS}_4\times\text{CP}^3$, the mirror TBA equations could be reduced to a more compact system of equations, dubbed the Quantum Spectral Curve (QSC), which proved more efficient for numerical evaluation and more suitable for the analytic continuation of the spectrum e.g.\ in the spin of the operators, see~\cite{Gromov:2017blm} for a review.

In the case of \adsst, the QSC for the pure Ramond-Ramond case was proposed in \cite{Cavaglia:2021eqr, Ekhammar:2021pys} independently from (and indeed, slightly earlier than) the mirror TBA. Instead, the QSC was obtained by gluing two copies of the ``Q--system'' for the supergroup $\text{PSU}(1,1|2)$ via analytic continuation through branch points in the spectral parameter plane, whose position depends on the coupling constant of the theory. The form of these gluing conditions was inspired by intuition coming from the AdS$_5$ and AdS$_4$ cases. Despite this inspiration, a novel feature emerges as compared to previous cases: the non-quadratic nature of the aforementioned branch points, which means novel methods are required to solve the equations. 
 This construction was shown to reproduce the worldsheet S-matrix in the large-volume limit \cite{Ekhammar:2024kzp} and was used to study the spectrum of short operators in \cite{Cavaglia:2022xld} and, more recently,  \cite{Ekhammar:2026ykk}.  The latter paper introduces a method to efficiently handle non-quadratic branch points and solve the equations at finite coupling. 

So far, the exact relation between the TBA and the QSC has remained an open problem: in particular, in the QSC description, the massless excitations are not manifest, whereas in the TBA they play a central role, especially at weak tension~\cite{Brollo:2023pkl,Brollo:2025rgp}.
In this paper and in a subsequent work~\cite{part2}, we will bridge this gap by carefully deriving the Quantum Spectral Curve from the Thermodynamic Bethe Ansatz, as was done in the AdS$_5$ and AdS$_4$ cases \cite{Gromov:2014caa, Bombardelli:2017vhk}.

We will split this program into two steps: 
\begin{enumerate}
    \item Here, we derive the extended Y-system for this model from the TBA equations in \cite{Frolov:2021bwp}. It is composed of two copies of a $\text{PSU}(1,1|2)$ Y-system, i.e. functional relations imposed on a set of Y functions, which are complex functions living in a Riemann surface with infinitely many branch cuts, supplemented by a set of local, state-independent discontinuity relations across these branch cuts. This extended Y-system constitutes a complete local description of the TBA equations, valid for any state, and will serve as the starting point for deriving the QSC. We will also show that the extended Y-system is completely equivalent to the ground-state TBA in \cite{Frolov:2021bwp}, in the sense that the TBA equations can be fully reconstructed from the Y-system and the discontinuity relations.  
    
    {To obtain a consistent extended Y-system, it is necessary to remove a specific contribution from one of the dressing phases that enter the S-matrix, following the proposal of~\cite{Ekhammar:2024kzp,Frolov:2025ozz}.} In fact, this factor would introduce additional branch cuts in the Y functions that are incompatible with the Y-system; see equation \eqref{eq:a_g} and below for details.

Another modification we adopt, compared with the TBA equations presented in \cite{Frolov:2021bwp}, is to use a single species of massless Y functions. An indication that this would be the right choice came already from semiclassical analysis in \cite{Frolov:2023wji}.

    \item In an upcoming work, we will derive the Quantum Spectral Curve from the extended Y-system, following the general strategy of \cite{Gromov:2011cx, Gromov:2014caa}. The main new challenge in the AdS$_3$ case will be to properly incorporate the massless excitations into the QSC framework, which is not straightforward since they are not associated to any particular node of the T-hook diagrams underlying the two $\text{PSU}(1,1|2)$ Y-systems. Instead, they only appear in the discontinuity relations of the other Y functions.
\end{enumerate}
The Y-system is a set of functional finite-difference equations satisfied by the Y functions, which are the solutions to the TBA equations. Y-systems play a central role in the theory of integrable models: they are universal in the sense that the same set of functional equations can describe different integrable models possessing the same global symmetries. To pin down a specific model, one needs to supplement the Y-system with analyticity data on the Y functions; if a TBA description of the model is available, this analyticity data can usually be extracted directly from the TBA equations.

For the AdS$_3\times$S$^3\times$T$^4$ model with pure RR fluxes, the derivation of the Y-system from the mirror TBA equations was partially carried out for the massive Y functions in~\cite{Frolov:2021bwp}.

In this paper, we complete the derivation of the functional Y-system and derive the corresponding discontinuity relations, thereby establishing the extended Y-system in the sense of~\cite{Cavaglia:2010nm}.
This will be done by carefully analysing the analytic structure of all the Y functions and deriving the corresponding discontinuity relations, which describe their behaviour across the branch cuts in the complex rapidity plane and are closely related to the branch-cut structure of the QSC mentioned above.

We relegate the derivation of these discontinuity relations to the appendices, so that the interested reader can focus on the main results without being distracted by the technical details. We will show that the Y-system equations can be placed into two copies of a $\text{PSU}(1,1|2)$ T-hook structure, one for the undotted (left) and one for the dotted (right) Y functions, and that the two halves interact exclusively through the discontinuity relations, a feature that distinguishes this model from its AdS$_5$ counterpart and reflects the richer physical content of the AdS$_3$ theory. We also show that the role of the two copies of the PSU$(1,1|2)$ Y-system is completely interchangeable, in the sense that the TBA equations (and therefore, the discontinuity relations and the Y-system) are invariant under a ``chiral'' symmetry that maps the undotted (left) Y functions to the dotted (right) Y functions and vice versa. This symmetry is fundamental for the derivation of the QSC, as the Q-systems appearing in each ``half'' of the AdS$_3$ QSC are completely interchangeable \cite{Cavaglia:2021eqr}, and it is natural to expect that this symmetry should be visible already at the level of the TBA and Y-system. 

In the second part of the paper, we will show that the TBA equations can be fully reconstructed from the Y-system and the discontinuity relations via a procedure known as ``inversion'' \cite{Cavaglia:2010nm}, which was already carried out in the AdS$_5$ and AdS$_4$ cases. This will establish the complete equivalence between the TBA and the extended Y-system, providing a fundamental consistency check of our derivation of the Y-system and the discontinuity relations. Our derivation will also clarify some features, assumptions, and subtleties for the QSC equations.
As it turns out, the inversion procedure is more involved than in the higher dimensional cases due to the presence of massless excitations and the richer structure of mixed-mass discontinuities, but we will show that it can be successfully carried out by carefully analysing the analytic properties of all Y functions and kernels involved.

The rest of this paper is organised as follows. Section \ref{sec:TBAreview} reviews the ground-state TBA equations of \cite{Frolov:2021bwp} and establishes the notation used throughout. It also proves the chiral symmetry of the TBA equations. Section \ref{sec:Ysystem} introduces the Y-system, states our analyticity assumptions on the Y functions, and lists all the discontinuity relations. Section \ref{sec:inversion_1} performs the inversion for the auxiliary and massless Y functions, respectively. Appendices \ref{app:kinematical_conventions} and \ref{app:Smatrix} collect our kinematical conventions and S-matrix elements, while appendix \ref{app:S_mat_and_kernels_part2} lists the kernels and some useful identities between them. Appendices \ref{appendixYpm}, \ref{appendixBES}, \ref{appendix_discontinuities_Y1}, and \ref{appendixY0disc} contain derivations of all discontinuity relations used in the main text, obtained from the ground-state TBA equations. Finally, appendix~\ref{app:useful_rel} provides identities used to perform the inversions.

\section{Reviewing the TBA equations}
\label{sec:TBAreview}

\subsection{A review of the Y functions}

In this section, we describe the Y functions defined as solutions of the ground-state TBA equations derived in~\cite{Frolov:2021bwp}. In light of the results of~\cite{Frolov:2023wji}, we assume there is a single Y function associated with the massless modes, which we label by $Y_0$. Moreover, differently from~\cite{Frolov:2021bwp}, we introduce the following notation for the massive left and right Y functions:
\begin{equation}
Y_Q \to Y_n \, \qquad \bar{Y}_Q \to \bY_n , \qquad Q=n \,.
\end{equation}
Let us recall the different types of Y functions and their physical regions, namely the domains over which they are integrated in the TBA equations. We have:
\begin{enumerate}

\item The functions $Y_n(u)$ and $\bY_n(u)$ of the massive left and right particles, respectively, 
\begin{equation}
    Y_n(u),\quad n=1,2,\ldots \infty\,,\qquad 
    \bY_n(u)\,,\quad \, n=1,2,\ldots\infty\,,\qquad u\in\mathbb{R}\,.
\end{equation}
The kinematics of these particles is parameterised by the Zhukovsky variables (see appendix~\ref{app:kinematical_conventions} for conventions)
\bal\label{eq:xshifted}
x^{[\pm n]}(u)=x\left(u \pm \frac{i}{h} n\right) \,.
\eal
Here $u$ is a point in the mirror complex plane, and $x(u)$ is the mirror Zhukovsky map:
\begin{equation}\label{eq:xmirror}
x(u)\equiv \frac{1}{2} \left(u - i \sqrt{4-u^2} \right)\,.
\end{equation}
{It is a function naturally defined with two long branch cuts $(-\infty, -2) \cup (+2, +\infty)$, which are called ``mirror cuts'', and satisfies $x(u)^*=1/x(u^*)$, with complex conjugation of $u$ defined on the same section with long cuts. The parameter $h$ appearing in the shifts (\ref{eq:xshifted}) is identified as a coupling constant of the theory. The shifts are performed avoiding the long cuts, which is a convention taken throughout this paper unless specified otherwise.}

\item The function $Y_{0}(u)$ of the massless particles,
\begin{equation}
    Y_{0}(u)\,, \qquad |u|>2\,.
\end{equation}
We assume the physical region of these particles to be the upper edge of the mirror cut $(-\infty, -2) \cup (+2, +\infty)$, where we parameterise their kinematics by $x(u+ i0)$.
 Note that we use only one massless Y function, as found from the analysis of the twisted ground state in \cite{Frolov:2023wji}; this means that in the TBA equations of \cite{Frolov:2021bwp} one should remove the sums over the massless particles, denoted by dotted Greek letters. 
\item The function $Y_-^{(\a)}(u)$ of the auxiliary $y$-particles with negative imaginary part,
\begin{equation}
    \Im(y)<0:\qquad
   Y_-^{(\a)}(u)\,,\qquad -2<u<2\,,
\end{equation}
 {where $\alpha \in \left\{ 1,2 \right\}.$}
The corresponding $y$-coordinate is given by $y=x(u)$. The string and mirror Zhukovsky maps coincide on the lower half of the complex plane.

\item The function $Y_+^{(\a)}(u)$ of the auxiliary $y$-particles with positive imaginary part,
\begin{equation}
    \Im(y)>0:\qquad
    Y_+^{(\a)}(u)\,,\qquad -2<u<2\,,
\end{equation}
 {where $\alpha \in \left\{ 1,2 \right\}.$} 
The corresponding $y$-coordinate is given by $y=1/x(u)$.
\end{enumerate}

\subsection{Ground-state TBA equations}
\label{sec:gstba}

In this section we summarise the TBA equations derived in~\cite{Frolov:2021bwp}. In this paper, we have used a slightly different convention that swaps the auxiliary Y functions, $Y_+ \leftrightarrow Y_-$.  {Moreover, as anticipated in the introduction, we will only have one massless Y function, and we remove from the massless-massless dressing phase the problematic term \eqref{eq:a_g}. }The kernels $K_{ij}$ used in the convolutions over these functions can be found in appendix~\ref{app:aux_kernels}, and follow the conventions of~\cite{Brollo:2023rgp}. 
We use the following notation for the ``star'' convolution 
\bal\la{starp}
(Y_i \star  K_{ij})(u) = \sum_i\int dv\, Y_i(v) K_{ij}(v,u)\,,
\eal
where the integration domain depends on the type of star convolution and summation over the repeated index $i$ is understood. In particular, we define three types of star convolution:
\begin{equation}
    \star \leftrightarrow\int\limits_{-\infty}^{+\infty}\text{d}u\,,\qquad
    \hat{\star} \leftrightarrow\int\limits_{-2}^{+2}\text{d}u\,,\qquad
    \check{\star} \leftrightarrow\Big(\int\limits_{-\infty}^{-2}+\int\limits_{+2}^{+\infty}\Big)\text{d}u\,.
\end{equation}
With these conventions, the TBA equations take the following form. 

\paragraph{Left-particles.}
\bal\label{TbaQ}
\log  Y_n =& -L\, \tE_{n} + \log\left(1+Y_{n'} \right)\star K_{sl}^{n'n} +  \log\left(1+ \bY_{n'} \right)\star \tK_{su}^{n' n} 
+  \log  \left(1+Y_0 \right)\cstar K^{0n} \\
 &+ \log \prod_{\a=1,2} \left(1-{ \frac{e^{i \mu_\a}}{ Y_{+}^{(\a)}} } \right)\hstar K^{yn}_+ +  \log \prod_{\a=1,2} \left(1-\frac{e^{i \mu_\a}} {Y_{-}^{(\a)}} \right)\hstar K^{yn}_-\,.~~~~~~
\eal

\paragraph{Right-particles.}
\bal\label{TbabQ}
\log   \bY_n = &-L\, \tE_{n} + \log\left(1+ \bY_{n'} \right)\star K_{su}^{n'n} +  \log\left(1+Y_{n'} \right)\star \tK_{sl}^{n' n} 
+\log  \left(1+Y_0 \right)\cstar \tK^{0n} \\
 &+ \log \prod_{\a=1,2}\left(1-{e^{i \mu_\a}\ov Y_{+}^{(\a)}} \right)\hstar K^{yn}_- + \log \prod_{\a=1,2} \left(1-{e^{i \mu_\a}\ov Y_{-}^{(\a)}} \right)\hstar K^{yn}_+\,.~~~~~~
\eal

\paragraph{Massless particles.}
\bal\label{Tba0}
\log Y_0 =& -L\, \tE_{0} + \log\left(1+Y_{0}\right)\cstar K^{00} +  \log\left(1+Y_{n} \right)\star K^{n 0} +  \log\left(1+ \bY_{n} \right)\star \tK^{n 0} \\
&+ \log \prod_{\a=1,2} \left(1-\frac{e^{i \mu_\alpha}}{Y^{(\alpha)}_+} \right) \left(1-\frac{e^{i \mu_\alpha}}{Y^{(\alpha)}_-} \right) \hstar K \,.
\eal

\paragraph{y$^{\boldsymbol -}$-particles.}
\bal
\label{Y-particles}
&\log Y_-^{(\a)} =  - \log\left(1+Y_{n} \right)\star K^{ny}_-  + \log\left(1+ \bY_{n} \right)\star K^{n y}_+   -  \log  \left(1+Y_{0}\right)\cstar K \,.
\eal

\paragraph{y$^{\boldsymbol +}$-particles.}
\bal
\label{Y+particles}
&\log Y_+^{(\a)} =  - \log\left(1+Y_{n} \right)\star K^{ny}_+  + \log\left(1+ \bY_{n} \right)\star K^{n y}_-   +   \log  \left(1+Y_{0}\right)\cstar K \,.
\eal
It is also useful to consider TBA equations for combinations of auxiliary Y functions
\bal
\label{Y+mparticles}
&\log (Y_-^{(\a)}/Y_+^{(\a)} ) =   \log\left[\left(1+Y_{n} \right) \left(1+\bY_{n} \right) \right]\star (K^{ny}_+ - K^{ny}_-)  - 2 \log \left(1+Y_{0}\right)\cstar K \,,\\
&\log (Y_-^{(\a)}\; Y_+^{(\a)} ) =   \log\left(\frac{1+\bY_{n} }{1+Y_{n} } \right)\star (K^{ny}_- + K^{ny}_+)  .
\eal
Differently from~\cite{Frolov:2021bwp}, we replaced the massless-auxiliary kernels $K^{0y}_\pm$ and $K^{y0}_\pm$ with the universal kernel $K$ through~\eqref{eq:Kpm_to_universal}.
The ground-state energy is then given by
\begin{equation}
E = -\int^{+\infty}_{-\infty} \frac{d u }{2 \pi} \frac{d p}{du} \log (1+Y_n)(1+\bY_n)-\int_{|u|>2} \frac{d u }{2 \pi} \frac{d p^0}{du} \log (1+Y_0)\,.
\end{equation}
Following~\cite{Fendley:1997ys, Frolov:2021bwp}, we introduced a twist parameter $\mu_\a=(-1)^\a\mu$ in the TBA equations. 
The case $\mu \rightarrow 0$ corresponds to the even-winding number sector with periodic fermions and a supersymmetric vacuum. In this limit, the ground-state energy vanishes, and $ Y_n=\bY_n=Y_0=0$. From the TBA equations of auxiliary particles, this requires $Y^{(\a)}_+ =Y^{(\a)}_- =  1$.

Therefore, while in this paper we leave the twist parameter $\mu$ generic, in the second part~\cite{part2} we will assume that $\mu \to 0$, as this is the case described by the Quantum Spectral Curve of~\cite{Cavaglia:2021eqr,Ekhammar:2021pys}. This relies on the assumption that the $Y$-system and the discontinuity relations established here remain valid throughout the entire spectrum in the $\mu \to 0$ limit, even though the ground state and the protected subsector are represented by singular solutions of the functional equations.

\subsection{Simplified TBA equations}

\label{sec:simplifiedTBA}
The ground-state TBA equations for the massive Y functions~\eqref{TbaQ} and~\eqref{TbabQ} can also be written in the following simplified form, which is completely equivalent once we prescribe the correct asymptotics for the Y functions:  
\beqa
\label{eq:simplified_TBA}
\log Y_n &=& \log \frac{1}{(1+Y_{n-1}^{-1})(1+Y_{n+1}^{-1})}\star s,\quad n\geq 2\\
\log \bY_n &=& \log \frac{1}{(1+\bY_{n-1}^{-1})(1+\bY_{n+1}^{-1})}\star s,\quad n\geq 2\\
\log Y_1 &=&-\log(1+Y_2^{-1})\star s+\log\prod_{\alpha=1,2}\left(1-\frac{e^{-i \mu_\a}}{Y_+^{(\alpha)}}\right)\star s -\Delta_1 \cstar s, \\
\log \bY_1 &=&-\log(1+\bY_2^{-1})\star s+\log\prod_{\alpha=1,2}\left({e^{-i \mu_\a}}-{Y_-^{(\alpha)}}\right)\star s -\bar\Delta_1 \cstar s,
\eeqa
where $s(u)$ is the universal (hyperbolic) kernel introduced in Zamolodchikov's simplified TBA formulation \cite{Zamolodchikov:1991et}:
\begin{equation}
  s(u)=\frac{h}{4 \cosh \frac{h \pi u}{2}}\,.  
\end{equation}
This simplification was already worked out in~\cite{Frolov:2021bwp}, and follows the well-established route of \cite{Zamolodchikov:1991et}. We do not repeat it here. We emphasise that we follow the conventions of~\cite{Brollo:2023rgp} for kernels involving auxiliary roots, ensuring they are positive. We see that the only quantities we need to input to find the TBA equations for all massive Y functions are the functions $\Delta_1$ and $\bar\Delta_1$, which are defined in terms of the discontinuities of $\log Y_1$ and $\log \dot{Y}_1$ across one of the branch cuts closest to the real axis. These discontinuities will be defined and studied in sections \ref{sec:disc_massive_Y} and appendix \ref{appendix_discontinuities_Y1}.

\subsection{Chiral symmetry}
\label{sec:wingexchangesymmetry}
 In this section, we describe a hidden $\mathbb{Z}_2$ symmetry of the ground-state TBA equations, consisting of the exchange of the following sets of Y functions together with the corresponding twists:
\beqa
\label{def:Wing_exch_symm}
Y_{n} \leftrightarrow \dot Y_{n}  \,, \qquad Y_0 \leftrightarrow Y_0 \, , \qquad Y^{(\alpha)}_{\pm} \leftrightarrow \frac{1}{Y^{(\alpha)}_{\mp}}\,, \qquad \mu_\alpha \leftrightarrow -\mu_\alpha \,.
\eeqa
We call this \emph{chiral symmetry}, as in terms of the Y-system that we will introduce in the next section, it is equivalent to exchanging the two $\text{PSU}(1,1|2)$ factors (see figure \ref{fig:thooks}).
This symmetry is expected since the choice of labelling the left- and right-massive Y functions is arbitrary, and we should not expect any change if we switch them. However, this is not manifest
from the form of the TBA equations presented in the previous section and kernels listed in appendix~\ref{app:S_mat_and_kernels_part2}. Therefore, we now prove that the ground-state TBA equations are invariant under \eqref{def:Wing_exch_symm}.

Since chiral symmetry holds at the level of the TBA equations, it also holds for the Y-system and for the discontinuities of the Y functions presented in the next section, as these can all be derived from the TBA equations. 

\paragraph{Auxiliary TBA equations.}
The chiral symmetry for these equations is immediately evident, since the LHS of~\eqref{Y-particles} is equal to minus the LHS of~\eqref{Y+particles} and the map~\eqref{def:Wing_exch_symm} sends $\log {Y_{\pm }^{(\alpha)}}$ to $-\log { Y_{\mp}^{(\alpha)}}$.
\paragraph{Massive TBA equations.}
We set out to prove that the TBA equation for left ${n}$-particles~\eqref{TbaQ} reduces to~\eqref {TbabQ} upon applying the map~\eqref{def:Wing_exch_symm}. First, breaking 
\beq
\label{def:Y_splitting}\log\prod_{\alpha=1,2}\left(1 - \frac{e^{i \mu_{\alpha}}}{Y_{\pm}^{(\alpha)} }\right)  = - \sum_{\alpha=1,2}\log{Y_{\pm}^{(\alpha)}}  
+ \log\prod_{\alpha=1,2}\left(1 -  {e^{-i \mu_{\alpha}}}{Y_{\pm}^{(\alpha)} }\right),\eeq
where we used $\mu_{\alpha} = (-1)^{\alpha} \mu$, we rewrite~\eqref{TbaQ} as:
\bal
\label{eq:TBALR1}
\log  Y_{n} =& -L\, \tE_{n} + \log\left(1+Y_{n'} \right)\star K_{sl}^{n'n} +  \log\left(1+ \bY_{n'} \right)\star \tK_{su}^{n' n} 
\\
&+  \log \left(1+Y_0\right)\cstar K^{0n}- \sum_{\alpha} \log  Y^{(\alpha)}_{-}\hstar K^{yn}_--\sum_{\alpha}\log Y_{+}^{(\alpha)}\hstar K^{yn}_+\\
 &+ \sum_{\alpha}\log \left(1-e^{-i \mu_{\alpha}}{ Y_{+}^{(\alpha)}} \right)\hstar K^{yn}_+ 
 +  \sum_{\alpha}\log \left(1- e^{-i \mu_{\alpha}}{ Y_{-}^{(\alpha)}} \right)\hstar K^{yn}_-\,.~~~~~~
\eal
Using the TBA equations for $Y_{-}^{(\alpha)}$ \eqref{Y-particles} and $Y_{+}^{(\alpha)}$ \eqref{Y+particles}, we then find the following expression:
\beqa
\log  Y_{n} =&& -L\, \tE_{n} +\log\left(1+Y_{n'} \right)\star (K_{sl}^{n'n}+2K_+^{n'y}\hstar K_+^{yn} +2K_-^{n'y}\hstar K_-^{yn})\\&&
 +\log\left(1+ \bY_{n'} \right)\star (\tK_{su}^{n' n}-2K_-^{n'y}\hstar K_+^{yn} -2K_+^{n'y}\hstar K_-^{yn})\\&&
 + \log  \left(1+Y_0 \right)\cstar (K^{0n}-2K_+^{0y}\hstar K_-^{yn}+2K_+^{0y}\hstar K_+^{yn})\\
 &&+ \sum_{\alpha}\log \left(1-e^{-i \mu_{\alpha}}{ Y_{+}^{(\alpha)}} \right)\hstar K^{yn}_+ 
 +  \sum_{\alpha}\log \left(1-e^{-i \mu_{\alpha}}{ Y_{-}^{(\alpha)}} \right)\hstar K^{yn}_-
\eeqa
where we have used the relation between the universal kernel $K$ and the kernels $K_{\pm}^{0y}$ through~\eqref{eq:Kpm_to_universal}. 

Then we can use the identities in appendix~\ref{sec:convolutions_shortcuts} to prove that:
\beqa
2(K_+^{n'y}\hstar K_+^{yn} +K_-^{n'y}\hstar K_-^{yn})&=&K_{su}^{n'n}-K_{sl}^{n'n}\,,\\
-2(K_-^{n'y}\hstar K_+^{yn} +K_+^{n'y}\hstar K_-^{yn})&=&\tK_{sl}^{n'n}-\tK_{su}^{n' n}\,,\\
2K_+^{0y}\hstar (K_+^{yn}-K_-^{yn})&=&\tK^{0n}-K^{0n}\,.
\eeqa
Hence~\eqref{eq:TBALR1} becomes:
\bal
\label{eq:TBALR2}
\log  Y_{n} =& -L\, \tE_{n} + \log\left(1+Y_{n'} \right)\star K_{su}^{n'n} +  \log\left(1+ \bY_{n'} \right)\star \tK_{sl}^{n' n} +  \log  \left(1+Y_0\right)\cstar \tK^{0n}
\\
&+ \sum_{\alpha}\log \left(1-e^{-i \mu_{\alpha}}{ Y_{+}^{(\alpha)}} \right)\hstar K^{yn}_+ 
 +  \sum_{\alpha}\log \left(1-e^{-i \mu_{\alpha}}{ Y_{-}^{(\alpha)}} \right)\hstar K^{yn}_-\,.
\eal
Applying the map~\eqref{def:Wing_exch_symm} to this equation then yields exactly the TBA equation for the right $n$-particles~\eqref{TbabQ}.

A similar calculation can be performed starting from the TBA equation for the right $n$-particles~\eqref{TbabQ}, obtaining the ones for left $n$-particles \eqref{TbaQ}.
\paragraph{Massless TBA equations.}
In this case, the TBA equation should be invariant under the map~\eqref{def:Wing_exch_symm}. This is easy to prove by using a similar splitting as done before for the fermionic Y functions $Y_{\pm}^{(\alpha)}$, and thus rewriting~\eqref{Tba0} as:
\beqa \log Y_0 =&& -L\, \tE_{0} + \log\left(1+Y_{0}\right)\cstar K^{00} +  \log\left(1+Y_{n} \right)\star K^{n 0} +  \log\left(1+ \bY_{n} \right)\star \tK^{n 0} \nonumber\\
&&-\sum_{\alpha}\log{(Y_{-}^{(\alpha)}Y_{+}^{(\alpha)})}\hstar K + \sum_{\alpha}\log  \left[\left(1-e^{-i \mu_{\alpha}}{Y_{-}^{(\alpha)}} \right)\left(1-e^{-i \mu_{\alpha}}{Y_{+}^{(\alpha)}} \right)\right] \hstar K \,.\nonumber
\eeqa
Using the TBA equation~\eqref{Y+mparticles}, we arrive at:
\begin{align}
\label{eq:TBALRmassless}
\log Y_0 =
& -L\, \tE_{0} + \log\left(1+Y_{0}\right)\cstar K^{00} \nonumber
+  \log\left(1+Y_{n} \right)\star (K^{n 0}+2 K_n \hstar K) \\&+  \log\left(1+ \bY_{n} \right)\star (\tK^{n 0}-2 K_n \hstar K) \nonumber
\\&
+ \sum_{\alpha}\log  \left[\left(1-e^{-i \mu_{\alpha}}{ Y_{-}^{(\alpha)}} \right)\left(1-e^{-i \mu_{\alpha}}{ Y_{+}^{(\alpha)}} \right)\right] \hstar K \,.
\end{align}
Since $K^{n 0}+2 K_n \hstar K=\tK^{n0}$, it is immediate to see that~\eqref{eq:TBALRmassless} is equivalent to the expression we have started from~\eqref{Tba0} after we use the map~\eqref{def:Wing_exch_symm}.

\subsubsection{TBA equations in explicitly symmetric form}

In this section, we rewrite the TBA equations in a way that makes the chiral symmetry explicit. 
Although the two formulations are equivalent, the derivation from the extended Y-system, as written in this paper, naturally yields the TBA equations in the form of section~\ref{sec:gstba}.

\paragraph{Left-particles.}
\bal
\log  Y_n =& -L\, \tE_{n} + \log\left(1+Y_{n'} \right)\star K_{sl}^{n'n} +  \log\left(1+ \bY_{n'} \right)\star \tK_{su}^{n' n} 
+  \log  \left(1+Y_0 \right)\cstar K^{0n} \\
 &+ \log \prod_{\a=1,2} \left(1-{ \frac{e^{i \mu_\a}}{ Y_{+}^{(\a)}} } \right)\hstar K^{yn}_+ +  \log \prod_{\a=1,2} \left(1-\frac{e^{i \mu_\a}} {Y_{-}^{(\a)}} \right)\hstar K^{yn}_-\,.~~~~~~
\eal

\paragraph{Right-particles.}
\bal
\log  \dot{Y}_n =& -L\, \tE_{n} + \log\left(1+\bY_{n'} \right)\star K_{sl}^{n'n} +  \log\left(1+ Y_{n'} \right)\star \tK_{su}^{n' n} 
+  \log  \left(1+Y_0 \right)\cstar K^{0n} \\
 &+ \log \prod_{\a=1,2} \left(e^{i \mu_\a}- Y_{-}^{(\a)} \right)\hstar K^{yn}_+ +  \log \prod_{\a=1,2} \left(e^{i \mu_\a}- Y_{+}^{(\a)} \right)\hstar K^{yn}_-\,.~~~~~~
\eal
\paragraph{Massless particles.}
\beqa
\log Y_0 =&& -L\, \tE_{0} + \log\left(1+Y_{0}\right)\cstar K^{00} \\&&+  \frac{1}{2}\log\left(1+Y_{n} \right)\star( K^{n 0} +\tK^{n0})+  \frac{1}{2}\log\left(1+ \bY_{n} \right)\star (K^{n0}+\tK^{n 0})\\
&&+ \dfrac{1}{2}\log \prod_{\a=1,2} \left(1 -e^{-i \mu_\alpha}Y^{(\alpha)}_+ \right) \left(1 -e^{-i \mu_\alpha} Y^{(\alpha)}_-\right)\left(1-\frac{e^{i \mu_\alpha}}{Y^{(\alpha)}_-} \right)\left(1-\frac{e^{i \mu_\alpha}}{Y^{(\alpha)}_+} \right)  \hstar K \,.\nonumber
\eeqa
\paragraph{y$^{\boldsymbol -}$-particles.}
\bal\label{eq:TBAaux-}
&\log Y_-^{(\a)} =  - \log\left(1+Y_{n} \right)\star K^{ny}_-  + \log\left(1+ \bY_{n} \right)\star K^{n y}_+   -  \log  \left(1+Y_{0}\right)\cstar K \,.
\eal

\paragraph{y$^{\boldsymbol +}$-particles.}
\bal\label{eq:TBAaux+}
&\log Y_+^{(\a)} =  - \log\left(1+Y_{n} \right)\star K^{ny}_+  + \log\left(1+ \bY_{n} \right)\star K^{n y}_-   +   \log  \left(1+Y_{0}\right)\cstar K \,.
\eal

\section{Y-system and discontinuity relations}
\label{sec:Ysystem}

In this section, we introduce the AdS$_3\times$S$^3\times$T$^4$ extended Y-system, i.e. a closed set of state-independent functional equations and discontinuity relations satisfied by the Y functions introduced above.

The extended Y-system can be derived from the TBA equations. The functional equations connecting the different Y functions on the first Riemann sheet are dictated by the symmetry of the underlying theory, which in this case is $\mathrm{PSU}(1,1|2)^{\otimes 2}$, and were partially derived in~\cite{Frolov:2021bwp} for the massive Y functions using the fusion properties of the kernels. The remaining ingredients are the analytic structure of the Y functions, encoded in their branch-cut structure, and the functional equations connecting Y functions on \emph{different} Riemann sheets, which are the discontinuity relations.

The functional equations on the first Riemann sheet alone are not sufficient to reconstruct the \adsst\ ground-state TBA. To derive the latter, these equations must be supplemented by the analytic structure of Y functions and the discontinuity equations. 

We establish these properties in section~\ref{sec:Ysystem}. In section~\ref{sec:inversion_1}, we show how the extended Y-system can be used to reconstruct the \adsst\ ground-state TBA of section~\ref{sec:gstba} via a procedure known as ``inversion''~\cite{Cavaglia:2010nm}.

Although the discontinuity relations are derived from the ground-state TBA equations, we expect them to be universally satisfied by all excited states of the theory as well. This is indeed the case in higher-dimensional AdS/CFT models~\cite{Cavaglia:2010nm,Cavaglia:2013lgg}.

\subsection{The  
\texorpdfstring{$\text{PSU}(1,1|2)^{\otimes 2}$}{PSU112}
Y-system}
\label{sec:Yfun_relations}

\begin{figure}[t]
    \centering
    \begin{subfigure}{0.48\textwidth}
        \centering
        \includegraphics[width=\linewidth]{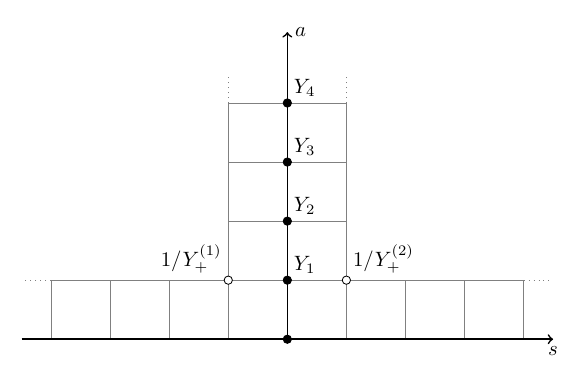}
        \label{fig:thookl}
    \end{subfigure}
    \hfill
    \begin{subfigure}{0.48\textwidth}
        \centering
        \includegraphics[width=\linewidth]{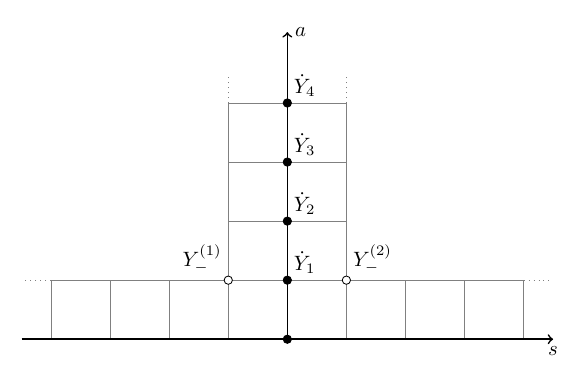}
        \label{fig:thookr}
    \end{subfigure}
    \caption{The Y functions of the \adsst\, model drawn on two $\text{PSU}(1,1|2)$ T-hooks. The chiral symmetry of section \eqref{sec:wingexchangesymmetry} exchanges the Y functions between the two T-hooks.}
    \label{fig:thooks}
\end{figure}
We start by describing what is traditionally known as the Y-system, the part of the construction that is largely dictated by the underlying symmetry algebra, and that connects the Y functions on their first Riemann sheet.
The $\text{PSU}(1,1|2)^{\otimes 2}$ Y-system can be read off from the two T-hooks in figure~\ref{fig:thooks}, with the left (right) Y functions on the left (right) T-hook. It is given by the following functional equations:
\bal
\label{eq:YQ_relations}
&\frac{Y^+_n Y^-_n}{Y_{n-1}Y_{n+1}} = \frac{1}{(1+Y_{n-1}) (1+Y_{n+1})} \,, \qquad  n \ge 2\,, \\
&\frac{\bY^+_n \bY^-_n}{\bY_{n-1} \bY_{n+1}} = \frac{1}{(1+\bY_{n-1}) (1+\bY_{n+1})} \,, \qquad n \ge 2 \,,
\eal
\bal
\label{eq:Y12_relations}
&\frac{Y^+_1 Y^-_1}{Y_2} = \frac{\left(1-{e^{-i \mu}\ov Y_{+}^{(1)}} \right) \left(1-{e^{+i \mu}\ov Y_{+}^{(2)}} \right)}{1+Y_{2}} \,,\\
&\frac{\bY^+_1 \bY^-_1}{\bY_2} = \frac{\left(e^{-i \mu}- Y_{-}^{(1)} \right) \left(e^{+i \mu}-Y_{-}^{(2)} \right)}{1+\bY_{2}} \,,
\eal
where we use the convention $Y^\pm = Y(u\pm \frac{i}{h})$.
This Y-system can be derived from the TBA equations reviewed in section~\ref{sec:gstba}, as proven in~\cite{Frolov:2021bwp}. The only difference compared to~\cite{Frolov:2021bwp}, is that we are swapping the notation between $Y^{(\a)}_-$ and $Y^{(\a)}_+$.

Notice that in this Y-system, the massless Y function $Y_0$ does not appear; however, it will play a fundamental role in the discontinuities of the other Y functions. Furthermore, the chiral and antichiral halves are independent and will interact only through the discontinuities.\footnote{This is expected, since the same Y-system can describe other models with the same $\text{PSU}(1,1|2)^{\otimes 2}$ symmetry but very different features. One such example is a pair of independent copies of the Hubbard model~\cite{Cavaglia:2015nta, Ekhammar:2021pys}, which consist of two non-interacting copies and do not contain massless modes.}

We note that, since the Y functions have branch points, we need to be careful when defining their shifts that appear in the Y-system equations. The Y-system is valid on the ``mirror'' Riemann section, where all branch points are connected to $\pm \infty$ via horizontal branch cuts, leaving the strip $|\Re(u)|<2$ free of singularities.

\subsection{Analyticity assumptions on Y functions}
\label{sec:analytic_Yfunc}

In this section, we list the analytic structure of the Y functions appearing in the Y-system, which can be read off from careful analysis of the ground-state TBA equations and which we assume is valid for excited states as well. All the Y functions possess infinite series of branch points. We do not specify the precise nature of the corresponding branch points at $\pm 2 + i \mathbb{Z}/h$, except that they connect to the same sheet in pairs. We expect all these branch points to be \emph{logarithmic}, i.e. of infinite order.  The position of the branch points of all Y functions on the Riemann sheet where the TBA integration contours are taken can be read off in table \ref{tab:tablebp}. Our default choice of branch cuts is the mirror section described above, where the Y-system holds. 
\begin{table}[htbp]
    \centering
    \renewcommand{\arraystretch}{1.5} 
    \begin{tabular}{|c|l|}
        \hline
         Function &  Position of branch points \\ 
        \hline\hline
        \large $Y_n(u)$ & \multirow{2}{*}{\large \quad $u = \pm 2 + i\frac{J}{h}, \hspace{2.5cm} J = \pm n, \pm (n+2), \pm (n+4), \dots$}  \\ 
        \cline{1-1}
        \large $\bY_{n}(u)$ & \\ 
        \cline{1-1}
        \hline
        \large $Y_0(u)$ & \large \quad $u = \pm 2 + i\frac{2J}{h}, \hspace{2.5cm} J = 0, \pm 1, \pm 2, \dots$ \\ 
        \hline
        \large $Y_{\pm}^{(\alpha)}(u)$ & \large \quad $u = \pm 2 + i\frac{2J}{h}, \hspace{2.5cm} J = 0, \pm 1, \pm 2, \dots$ \\ 
        \hline
        \large $\frac{Y_{+}^{(\alpha)}}{Y_{-}^{(\alpha)}}(u)$ & \large \quad $u = \pm 2 + i\frac{2J}{h}, \hspace{2.5cm} J = 0, \pm 1, \pm 2, \dots$
        \\ 
        \hline
        \large ${Y_{+}^{(\alpha)}}{Y_{-}^{(\alpha)}}(u)$ & \large \quad $u = \pm 2 + i\frac{2J}{h}, \hspace{2.5cm} J =  \pm 1, \pm 2, \pm 3\dots$ \\
        \hline
        \end{tabular}
        \caption{Branch-point singularities of the Y functions in the spectral-parameter domain. The default choice of branch cuts connects the branch points to $\pm \infty$ via long horizontal cuts, leaving the strip $|\Re(u)| < 2$ analytic. 
        Branch points with the same imaginary part on the same sheet lead to the same hidden sheets.}  \label{tab:tablebp}
\end{table}
Moreover, since massless particles have a non-analytic dispersion relation at zero momentum, the function $\log{Y_0}$ has an infinite vertical branch cut on the imaginary axis, corresponding to the splitting between chiral and anti-chiral massless excitations on the worldsheet. This cut is discussed in more detail later and is absent from the function $Y_0$ as a consequence of the quantization of the volume $L$, similar to a phenomenon observed for AdS$_5$ in \cite{Cavaglia:2010nm}.

\subsection{Conventions}

\begin{figure}[htbp]
    \centering
    \begin{subfigure}{0.45\textwidth}
        \centering
        \includegraphics[width=\linewidth]{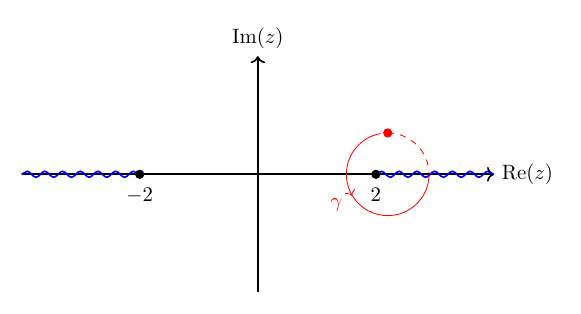}
        \label{fig:gamma}
    \end{subfigure}
    \hfill
    \begin{subfigure}{0.45\textwidth}
        \centering
        \includegraphics[width=\linewidth]{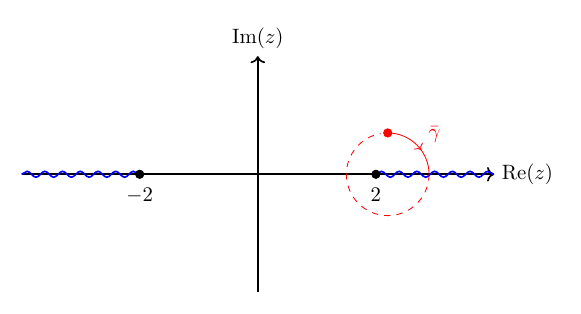}
        \label{fig:gammabar}
    \end{subfigure}
    \caption{The analytic continuation paths $\gamma$ and $\bar\gamma$ for functions with `long' branch cuts. The path $\gamma$ corresponds to crossing the long cuts from below; this is associated with the counterclockwise convention for the right cuts and the clockwise convention for the left cuts. The path $\bar\gamma$ follows the opposite convention. Since the branch points are logarithmic, $\gamma$ and $\bar{\gamma}$ define different continuations.}
    \label{fig:gamma_path}
\end{figure}

Before giving a list of the different discontinuity relations for the Y functions, we provide the definitions we will use in the paper. We define the discontinuity of $f$ on its long branch cut at $\Im(u)=N/h$ as follows:
\bal
\la{eq:def_discontin_f_N}
\bigl[ f \bigr]_N(u) \equiv f(u+\frac{i}{h}N+i0)-f(u+\frac{i}{h}N-i0) \, ,\\
\eal
where the RHS is a definition valid for $u  \in (-\infty, -2)$ or $u  \in  (+2, +\infty)$, $N \in \mathbb{Z}$. 
Focusing on the cuts located on the RHS of the complex plane, this definition is equivalent to
\bal
\bigl[ f \bigr]_N(u) \equiv f(u+\frac{i}{h}N+i0)-f^\gamma(u+\frac{i}{h}N+i0)\,,
\eal
where the continuation path $\gamma$ follows the counterclockwise direction and is shown in figure~\ref{fig:gamma_path}. This relation can then be analytically continued to any value of $u \in \mathbb{C}$. Similarly, we define the symmetric discontinuity of $f$ on its long branch cut at $\Im(u)=N/h$ as
\bal
\{f\}_N & \equiv f(u+\frac{i}{h}N+i0)+f(u+\frac{i}{h}N-i0)\\
&=f(u+\frac{i}{h}N+i0)+f^\gamma(u+\frac{i}{h}N+i0)\,,
\eal
where the latter can be analytically continued to define a function in the full complex plane $\{f\}_N(u)$.

We will also use the following shift convention
\begin{equation}
f^{[\pm m]}=f\left(u \pm \frac{i}{h}m\right)\,,
\end{equation}
where, importantly, in this paper all shifts will be taken on Riemann sheets with long cuts.
Finally, we will denote by $\bar{\gamma}$ the path obtained by going around the cuts in the half-plane $\Re(u)>0$ in the clockwise direction, as depicted in figure \ref{fig:gamma_path}. 
Circling a branch point with real part $-2$ clockwise is equivalent to circling its twin with real part $+2$ anticlockwise, and vice versa.

\subsection{Discontinuity equations}

In this section, we list the discontinuities of the Y functions that are necessary to retrieve the TBA from the Y-system. For the auxiliary and massless Y functions, there are discontinuities along all branch cuts on the first sheet.

For the massive Y functions, it is sufficient to consider the functions $\Delta_1\equiv [\log Y^+_{1}]_0$ and $\bar\Delta_1\equiv [\log \bY^+_{1}]_0$. These same quantities will play a fundamental role in the T-system we introduce in \cite{part2}. 

\subsubsection{Discontinuity equations of auxiliary functions.}

To describe the discontinuities of the auxiliary Y functions, it is convenient to consider their product and ratios separately. 
Their product has the peculiarity that it does not have branch points on the real axis, i.e. its monodromies are trivial
\bal
\label{eq:Yaux_product}
&\left(Y_-^{(\a)}\right)^{\bar{\gamma}}  \left(Y_+^{(\a)}\right)^{\bar{\gamma}}  = \left(Y_-^{(\a)}\right)^\gamma \left(Y_+^{(\a)}\right)^\gamma = Y_-^{(\a)}Y_+^{(\a)}\, ,
\eal
while the monodromies of the ratio are described by
\bal
\label{eq:Yaux_ratio}
\frac{\left(Y_+^{(\a)}\right)^{\bar{\gamma}} }{\left(Y_-^{(\a)}\right)^{\bar{\gamma}} } = \frac{Y_-^{(\a)} }{Y_+^{(\a)}} \left(1 + Y_0^{\bar{\gamma}} \right)^2 \, , \qquad \frac{\left(Y_+^{(\a)}\right)^{{\gamma}} }{\left(Y_-^{(\a)}\right)^{{\gamma}} } = \frac{Y_-^{(\a)} }{Y_+^{(\a)}} \; \left(1 + Y_0 \right)^2 ,
\eal
where we notice the appearance of the massless Y function on the RHS. 
The discontinuities on the long cuts at $\left(-\infty, -2 \pm \frac{2i}{h} M\right) \cup \left(+2 \pm \frac{2i}{h} M, +\infty\right)$, with $M\in\mathbb{N}^+$, are instead
\bal
\label{eq:disc_Ymp_2M}
\left[ \log \frac{Y_-^{(\a)}}{Y_+^{(\a)}} \right]_{\pm 2 M}&=+\sum^{M}_{n=1} \left[\log\left(1+\bY_{n}\right) \right]_{\pm(2M-n)}+\sum^{M}_{n=1} \left[\log\left(1+Y_{n}\right) \right]_{\pm(2M-n)} \,,\\
\left[ \log Y_-^{(\a)}Y_+^{(\a)}\right]_{\pm 2 M}&=+\sum^{M}_{n=1} \left[\log\left(1+\bY_{n}\right) \right]_{\pm(2M-n)}-\sum^{M}_{n=1} \left[\log\left(1+Y_{n}\right) \right]_{\pm(2M-n)} \,.
\eal
We derive these discontinuity equations in appendix \ref{appendixYpm}. They are all valid for $\alpha = 1,2$.

\subsubsection{Discontinuity equations of $Y_0$.}
\label{sec:disc_massless}

Similarly to the auxiliary functions $Y_\pm^{(\alpha)}$, the massless function $Y_0$ has discontinuities on the cut $(-\infty, -2) \cup (+2, +\infty)$ and on the long cuts at $\pm 2N$. We provide details on the derivation of these discontinuities in appendix~\ref{appendixY0disc}.  On the real axis we can express in a universal form its \emph{symmetrised discontinuity}
\bal
\label{Y0symmdisc}
\{\log{Y_0}\}_0=&\sum_{\alpha=1,2} \log{Y^{(\a)}_-}+2\log (1+Y_0) + \sum_{\a=1,2}\log  \left(1-\frac{e^{i \mu_\alpha}}{Y^{(\alpha)}_+}\right)^{\gamma}\left(1-\frac{e^{i \mu_\alpha}}{Y^{(\alpha)}_-} \right)^{\gamma} \, .
\eal
This equation can be put in a manifestly symmetric form between the two halves, as we did for the TBA equations in section \ref{sec:wingexchangesymmetry}. This is done by using relations \eqref{eq:Yaux_product} and \eqref{eq:Yaux_ratio} in \eqref{Y0symmdisc}, obtaining:
\bal
\{\log{Y_0}\}_0=&\sum_{\alpha=1,2} \log  \left(1-e^{-i \mu_\alpha} Y^{(\alpha)}_+ \right)^\gamma \left(1-\frac{e^{i \mu_\alpha}}{Y^{(\alpha)}_-} \right)^\gamma\,,
\eal
For the discontinuities of $Y_0$ on the further cuts with imaginary part $\pm 2 N/h$, $N\in \mathbb{N}^+$,
\bal
\label{eq:lnY0_disc_p2N2}
[\log{Y_0}]_{+2N}= &- \left[ \log \frac{Y_-^{(\a)}}{Y_+^{(\a)}} \right]_{+ 2 N}+ \sum_{\alpha=1,2} \log \frac{Y^{(\a)}_-}{Y^{(\a)}_+}+ 2 \log \left(1+Y_{0} \right)\,,\\
[\log{Y_0}]_{-2N}= &+\left[ \log Y_-^{(\a)}Y_+^{(\a)}\right]_{- 2 N}- \sum_{\alpha=1,2} \log \frac{Y^{(\a)}_-}{Y^{(\a)}_+}- 2 \log \left(1+Y_{0} \right)\\
&+\sum_{\a=1,2}\left[ \log  \left(1-\frac{e^{i \mu_\a}}{Y^{(\a)}_+} \right) \left( 1-\frac{e^{i \mu_\a}}{Y^{(\a)}_-} \right) \right]_{-2N}\,.
\eal
While the first equation is manifestly symmetric under the chiral symmetry, the second is not. Nevertheless, it can be brought to a manifestly invariant form: 
\bal
[\log{Y_0}]_{-2N}= &+ \sum^{N}_{n=1} \left[ \log(1+Y_n) \right]_{-2N+n} + \sum^{N}_{n=1} \left[ \log(1+\dot{Y}_n) \right]_{-2N+n}\\
&- \sum_{\a=1,2} \log \frac{Y^{(\a)}_-}{Y^{(\a)}_+}- 2 \log \left(1+Y_{0} \right)\\
&+\left[ \log \prod_{\a=1,2} \left(1-e^{-i \mu_\a}Y^{(\a)}_+ \right) \left( 1-\frac{e^{i \mu_\a}}{Y^{(\a)}_-} \right) \right]_{-2N}\,.
\eal
Finally, there is a discontinuity on the line $(-i \infty, + i \infty)$ due to the energy term $\tE_0$ in the TBA equation of $Y_0$ \eqref{Tba0}, given by (see equation~\eqref{eq:tE0_disc_on_im_axis})
\bal
\label{eq:Y0_disc_on_im_axis}
\log Y_{0}(it+ \eps)-\log Y_{0}(it- \eps)=-2 i \pi L \,, \qquad t \in \mathbb{R}\,,
\eal
where $L$ is the volume. 
This discontinuity is a consequence of the dispersion relation of massless modes not being analytic at $\tp = 0$, as discussed in appendix~\ref{app:en_and_mom}. Massive Y functions do not have this discontinuity. {We notice in any case that this additional discontinuity along the imaginary axis is emerging purely in $\log{ Y_0}$, but that it is absent in $Y_0$, as follows from the fact that $L$ is a (positive) integer. One can in fact see the regularity of $Y_0$ as a quantisation condition for $L$, similar to what was already observed in \cite{Cavaglia:2010nm}.\footnote{In that case, the discontinuity appeared in other quantities, analogous to $\Delta_1$ and $\bar{\Delta}_1$ we define here. Also, in our case, these quantities have the same type of logarithmic branch cut, inherited from the term $\log Y_0$ in \eqref{eq:Delta1_local}. }
For the same reason, the term $\log(1+Y_0)$ that appears in the convolution integrals of the TBA equations also does not exhibit this discontinuity.}

\subsubsection{Discontinuities of massive Y functions.}
\label{sec:disc_massive_Y}

The fundamental objects needed to reconstruct the massive TBA equations from the Y-system are the discontinuities of $Y_1$ and $\bY_1$ on their cuts at $\Im(u)=i/h$, which we label
\bal
\Delta_1(u) &\equiv [\log  Y_1]_1(u)\,, \qquad \bar{\Delta}_1(u) \equiv [\log  \bY_1]_1(u)\,.
\eal
The derivation of these discontinuities from TBA is reported in appendix~\ref{appendix_discontinuities_Y1},
where
we obtain the following remarkably simple local expressions for them:
\bal
\label{eq:Delta1_local}
\Delta_1\equiv[\log  Y_1]_1(u)= &- \log Y_0  + \log  (1+Y_0) + \log  \prod_{\a=1,2}\left(1-{e^{i \mu_\a}\ov Y_{+}^{(\a)}} \right) -c \,,\\
\bar{\Delta}_1\equiv [\log  \bY_1]_1(u)=&- \log Y_0  + \log (1+Y_0) +  \log  \prod_{\a=1,2}\left(1-{e^{-i \mu_\a} Y_{-}^{(\a)}} \right)+c\,,
\eal
where $c$ is a constant which, according to our derivation from the twisted ground-state TBA equations, is given by the following convolution:
\bal
c \equiv \frac{1}{4} \log \frac{1+Y_{m}(v)}{1+\bY_{m}(v)} \star (K_{m}(2-v)+K_{m}(2+v))\,.
\eal
We can express this constant in a different form as follows. Since $K_+^{ny}(u,v)+K_-^{ny}(u,v)=K_n(u-v)$ (see~\eqref{Krationalrels}), the second TBA equation in \eqref{Y+mparticles} becomes, when evaluated at $\pm2$:
\beq
\log{Y^{(\alpha)}_- Y^{(\alpha)}_+}(\pm 2)=-\log{\frac{1+Y_n(z)}{1+\bY_n(z)}}\star K_n(z\mp2)\,.
\eeq
Therefore, we obtain the following  expression for $c$:\footnote{We inserted a sum over $\alpha$ since we expect that this is the correct generalization for excited states with $Y_{\pm}^{(\alpha=1)}\neq Y_{\pm}^{(\alpha=2)}$.}
\beq
\label{def:constant_c_final}
c=-\frac{1}{2}\sum_{\alpha=1,2}\left(\log{Y^{(\alpha)}_- Y^{(\alpha)}_+}( 2)+\log{Y^{(\alpha)}_- Y^{(\alpha)}_+}(-2)\right)\,.
\eeq
Note that the RHS of this expression is perfectly well defined, since the product $Y^{(\a)}_+ Y_-^{(\a)}$ is regular on the real axis. We now promote equation \eqref{def:constant_c_final} to a universal, state-independent definition of $c$. While the value of $c$ itself may depend on the TBA solution, we expect the relation expressing it in terms of the $Y$ functions to hold universally. Equations \eqref{eq:Delta1_local}, together with this definition of $c$, therefore form part of our fundamental set of monodromy equations extending the $Y$-system.

Although all the information is already encoded in the equations above, it is useful to state several consequences of \eqref{eq:Delta1_local} explicitly, as they will play a central role in the follow-up paper~\cite{part2}. These concern the discontinuities of $\Delta_1$ and $\bar\Delta_1$ across the cuts away from the real axis at $\Im(u)=\pm2N/h$, $N=1,2,\dots$, together with their symmetrised discontinuity across the cuts on the real axis.

These discontinuities are immediate to obtain from \eqref{eq:Delta1_local}, and read:
\begin{equation}
\label{eq:Disc_Delta_1_0}
\{ \Delta_1 \}_0 = - \{\log Y_0\}_0  + \{ \log  (1+Y_0) \}_0 + \{ \log  \prod_{\a=1,2}\left(1-{e^{i \mu_\a}\ov Y_{+}^{(\a)}} \right) \}_0 -2c \,,
\end{equation}
\begin{equation}
\label{eq:Disc2_Delta_1_0}
[ \Delta_1 ]_0 = - [\log Y_0]_0  + [ \log  (1+Y_0) ]_0 + [ \log  \prod_{\a=1,2}\left(1-{e^{i \mu_\a}\ov Y_{+}^{(\a)}} \right) ]_0  \,,
\end{equation}
\begin{equation}
\label{eq:Disc_Delta_1_2N}
[\Delta_1 ]_{\pm 2N} = - [\log Y_0 ]_{\pm 2N}  + [ \log  (1+Y_0) ]_{\pm 2N} + \left[ \log  \prod_{\a=1,2}\left(1-{e^{i \mu_\a}\ov Y_{+}^{(\a)}} \right) \right]_{\pm2N} \,,
\end{equation}
\begin{equation}
\label{eq:Disc_Delta_1bar_0}
\{ \bar\Delta_1 \}_0 = - \{\log Y_0\}_0  + \{ \log  (1+Y_0) \}_0 + \{ \log  \prod_{\a=1,2}\left(1-{e^{-i \mu_\a} Y_{-}^{(\a)}} \right) \}_0 +2c \,,
\end{equation}
\begin{equation}
\label{eq:Disc2_Delta_1bar_0}
[ \bar\Delta_1 ]_0 = - [\log Y_0]_0  + [ \log  (1+Y_0) ]_0 + [ \log  \prod_{\a=1,2}\left(1-{e^{-i \mu_\a} Y_{-}^{(\a)}} \right) ]_0  \,,
\end{equation}
\begin{equation}
\label{eq:Disc_Delta_1bar_2N}
[\bar\Delta_1 ]_{\pm 2N} = - [\log Y_0 ]_{\pm 2N}  + [ \log  (1+Y_0) ]_{\pm 2N} + \left[ \log  \prod_{\a=1,2}\left(1-{e^{-i \mu_\a} Y_{-}^{(\a)}} \right) \right]_{\pm2N} \,,
\end{equation}
where $c$ is the constant defined in \eqref{def:constant_c_final}.

In \cite{part2}, we will analyse in more depth the role of the constant $c$ and its connection to the regularity properties of the Q functions contained in the QSC.

\section{Reconstructing TBA equations via inversion}
\label{sec:inversion_1}

In this section, we explain how to reconstruct the simplified TBA equations of section \ref{sec:TBAreview} using only the extended Y-system, i.e., the functional equations \eqref{eq:YQ_relations} and \eqref{eq:Y12_relations}, together with the discontinuities \eqref{eq:Yaux_product}-\eqref{eq:lnY0_disc_p2N2}, \eqref{eq:Delta1_local}, \eqref{def:constant_c_final} 
from the previous section. In this way, we prove that our extended Y-system is the local reformulation of the \adsst\, TBA.

\subsection{Auxiliary TBA equations from discontinuities of \texorpdfstring{$Y_\pm^{(\a)}$}{Ypm}}
\la{sec:TBAinversionYpm}

In this section, we describe the inversion procedure for the auxiliary particles. In particular, we use the the discontinuity relations~\eqref{eq:Yaux_product}, \eqref{eq:Yaux_ratio} and \eqref{eq:disc_Ymp_2M} to retrieve the TBA equations for the auxiliary functions $Y_-^{(\a)}$ and $Y_+^{(\a)}$ \eqref{Y-particles}-\eqref{Y+particles}. Our strategy relies on obtaining the TBA equations for the two independent combinations $\frac{Y_{-}^{(\a)}}{Y_+^{(\a)}}$ and ${Y_{-}^{(\a)}}{Y_+^{(\a)}}$ \eqref{Y+mparticles}, from which \eqref{Y-particles}-\eqref{Y+particles} follow trivially.

\subsubsection{Inversion of 
\texorpdfstring{${Y_{-}^{(\a)}/Y_+^{(\a)}}$}{Yratio}}

\begin{figure}
\begin{center}
\includegraphics*[width=0.7\textwidth]{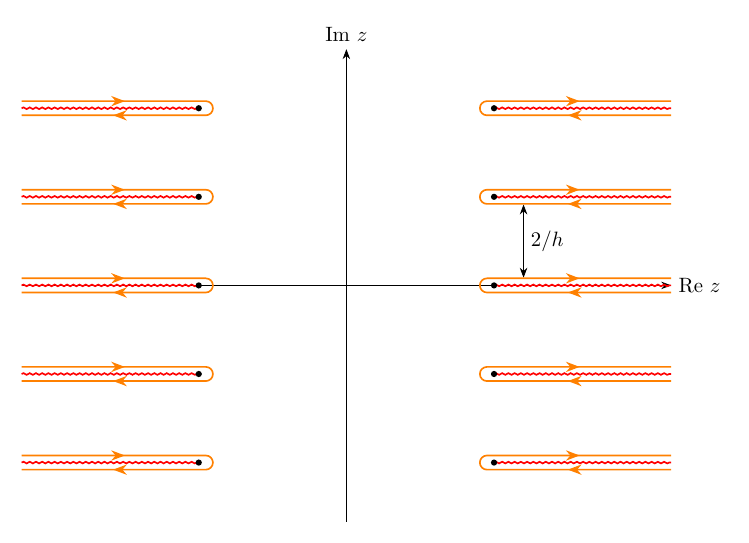} 
\end{center}
\caption{The $\Gamma_\infty$ integration contour, used in eq.~\eqref{eq:integrandgammainf0}, is shown in orange.
} 
\label{fig:ginf_int_contour}
\end{figure}

We start by considering the function 
$$
\frac{1}{\sqrt{4-u^2}} \log{\frac{Y^{(\a)}_-(u)}{Y^{(\a)}_+(u)}}
$$
in the strip $0<\Im(u)<1/h$, away from the branch cuts.
Using Cauchy's theorem on a small counterclockwise circle $\Gamma$ around $u$ we can write the following identity
\beq
\frac{1}{\sqrt{4-u^2}}\, \log{\frac{Y^{(\a)}_-(u)}{Y^{(\a)}_+(u)}}=\oint_{\Gamma} \, \dfrac{dz}{2\pi i} \, \log{\frac{Y^{(\a)}_-(z)}{Y^{(\a)}_+(z)}} \, \frac{1}{\sqrt{4-z^2}(z-u)} \,.
\eeq
We expand the contour to infinity, enclosing all branch cuts of the integrand.
Assuming that $Y_-^{(\a)}/Y_+^{(\a)}$ have asymptotics such that the integrand on the r.h.s. decays uniformly as $|z|\rightarrow \infty$, we can use Jordan's lemma to suppress the contribution of the circle of infinite radius. Concretely, we end up with
\beq
\la{eq:integrandgammainf0}
\log \frac{Y^{(\a)}_-(u)}{Y^{(\a)}_+(u)} =\oint_{\Gamma_{\infty}} dz \, \log \frac{Y^{(\a)}_-(z)}{Y^{(\a)}_+(z)} \, K(z, u)\,,
\eeq
where the universal kernel $K(z,u)$ is defined in~\eqref{def:Kuniversal} and $\Gamma_{\infty}$ is the orange integration contour in figure~\ref{fig:ginf_int_contour}.
Since $K(z,u)$ has a branch cut along the real axis, with $K(z-i0,u)=-K(z+i0,u)$,
the integral above can be written as
\bal
\la{eq:integrandgammainf1}
\log \frac{Y^{(\a)}_-(u)}{Y^{(\a)}_+(u)} = &+\sum_{M=1}^{+\infty}\left(\int_{-\infty}^{-2} + \int_{+2}^{+\infty} \right) dz \, \left[ \log \frac{Y^{(\a)}_-}{Y^{(\a)}_+}\right]_{+2M}(z+i0) \, K(z+\frac{2i}{h}M, u)\\
&+\sum_{M=1}^{+\infty}\left(\int_{-\infty}^{-2} + \int_{+2}^{+\infty} \right) dz \, \left[ \log \frac{Y^{(\a)}_-}{Y^{(\a)}_+}\right]_{-2M}(z+i0) \, K(z-\frac{2i}{h}M, u)\\
&+\left(\int_{-\infty}^{-2} + \int_{+2}^{+\infty} \right) dz \, \left\{ \log \frac{Y^{(\a)}_-}{Y^{(\a)}_+}\right\}_{0}(z+i0) \, K(z+i0, u)\,.
\eal
The first two lines of the expression above are analogous to what appears in the \ads case. However, for \ads the last line is zero,
while here we must deal with a nontrivial contribution from the massless modes (see equation~\eqref{eq:Yaux_ratio}).

Substituting the discontinuity relations \eqref{eq:Yaux_ratio} and~\eqref{eq:disc_Ymp_2M} into the expression above, and using the properties \ref{disc_prop1} and~\ref{disc_prop2} of appendix~\ref{app:useful_rel}, we obtain
\bal
\la{eq:integrandgammainf2}
\log \frac{Y^{(\a)}_-(u)}{Y^{(\a)}_+(u)} = &+\sum_{n=1}^{+\infty} \int_{-\infty}^{+\infty}  dz \, \log(1+Y_n(z))(1+\bY_n (z)) \left( K(z-\frac{i}{h}n, u) - K(z+\frac{i}{h}n, u) \right)\\
&-\left(\int_{-\infty}^{-2} + \int_{+2}^{+\infty} \right) dz \, 
\log  \left(1+Y_0(z+ i0) \right)^2\, K(z+i0, u)\,.
\eal
Finally, using identity~\eqref{Krationalrels}, we recover exactly the TBA equation in the first line of~\eqref{Y+mparticles}:
\bal
\la{eq:integrandgammainf3}
\log \frac{Y^{(\a)}_-(u)}{Y^{(\a)}_+(u)} = &+\log(1+Y_n(z))(1+\bY_n(z))  \star \left( K^{n y}_{+}(z, u) - K^{n y}_{-}(z, u) \right)\\
&-2 \log  \left(1+Y_0(z+i0) \right) \cstar K(z+i0, u)\,.
\eal

\subsubsection{Inversion of \texorpdfstring{$\log{Y_-Y_+}$}{Yprod}}

The inversion for the function $\log{({Y_-}{Y_+})}$ is identical to the one appearing in the AdS$_5$ case of~\cite{Cavaglia:2010nm} since the TBA equation is very similar (except for the presence of both $Y_n$ and $\bY_n$ terms) and especially this function is regular on the real axis. One then applies Cauchy's theorem to $\log{({Y_-}{Y_+})}$ using the standard Cauchy kernel. Repeating the steps of \cite{Cavaglia:2010nm}, one then blows up the integration contour as before, and integrating the discontinuity relations in the second line of~\eqref{eq:disc_Ymp_2M} reconstructs the TBA equation in the second line of~\eqref{Y+mparticles}.

\subsection{Massless TBA equations from discontinuities of 
\texorpdfstring{$\log Y_0$}{lnY0inv}}
\label{sec:inversionY0}

In this section, we obtain the TBA equation for the massless particles \eqref{Tba0} starting from the discontinuities of $\log Y_0$. Since this is the most technical inversion we will deal with, it is convenient to rewrite both the TBA equation \eqref{Tba0} and the discontinuities of $\log Y_0$ \eqref{Y0symmdisc} and \eqref{eq:lnY0_disc_p2N2} in a slightly different form, as we describe below.

\subsubsection{Rewriting the TBA equation of \texorpdfstring{$\log Y_0$}{lnY0inv}}

We will rewrite the massless TBA equation \eqref{Tba0} by
splitting the kernels appearing in it into a BES and non-BES part:
\bal
K^{00}(v, u)&=-2 \, K_\bes^{00}(v, u) + \hat{K}^{00}(v, u)\,,\\
K^{n0}(v, u)&=-2 \, K_\bes^{n0}(v, u) + \hat{K}^{n0}(v, u)\,,\\
\tK^{n0}(v, u)&=-2 \, K_\bes^{n0}(v, u) + \hat{\tK}^{n0}(v, u)\,.
\eal
The non-BES part of the kernels can be read from~\eqref{eq:leftmassive_massless_in_Y0_from_TBA}, \eqref{eq:rightmassive_massless_in_Y0_from_TBA} and~\eqref{eq:massless_massless_in_Y0_from_TBA}, and are given by the following expressions, in terms of quantities defined in appendix \ref{app:S_mat_and_kernels_part2}:
\bal
\label{eq:app_discY0_hat_Ks}
\hat{K}^{n0}(v, u)= &-\frac{1}{2}  K_{n}(v-u) \\
&+\frac{1}{i \pi}  \left( \g^{-n \circ}_{v u} - \frac{i \pi}{2}\right) K(v- \frac{i}{h}n,u)-\frac{1}{i \pi}  \left( \g^{+n \circ}_{v u} - \frac{i \pi}{2}\right) K(v+ \frac{i}{h}n,u) \,,\\
\hat{\tK}^{n0}(v, u)=&+\frac{1}{2} K_{n}(v-u)\\
&+\frac{1}{i \pi} \left( \g^{-n \circ}_{v u} + \frac{i \pi}{2} \right) K(v- \frac{i}{h}n,u)-\frac{1}{i \pi} \left( \g^{+n \circ}_{v u} - \frac{3}{2} i \pi\right) K(v+ \frac{i}{h}n,u) \,,\\
\hat{K}^{00}(v, u)=&+\frac{2 i}{\pi} K(v, u) \g^{\circ \circ}_{vu} \,.
\eal
The BES kernels are instead given by~\eqref{eq:BES_Kernel_as_double_conv_Q02} and~\eqref{eq:BES_Kernel_as_double_conv_002}. With these redefinitions, the TBA equation becomes
\bal
\label{eq:TBAY0_bes_nonbes}
\log Y_0 =& -L\, \tE_{0} + \underbrace{ \log \prod_{\alpha=1,2} \left(1-\frac{e^{i \mu_\alpha}}{Y^{(\alpha)}_+} \right) \left(1-\frac{e^{i \mu_\alpha}}{Y^{(\alpha)}_-} \right) \hstar K }_{\text{(1)}} \\
&+  \underbrace{\log\left(1+Y_{n} \right)\star \hat{K}^{n 0} }_{\text{(2)}} +\underbrace{  \log\left(1+ \bY_{n} \right)\star \hat{\tK}^{n 0} }_{\text{(3)}}+ \underbrace{\log  \left(1+Y_{0}\right)\cstar \hat{K}^{00} }_{\text{(4)}}\\
&\underbrace{-2\log  \left(1+Y_{0}\right)\cstar K_\bes^{00} -2  \log\left(1+Y_{n} \right)\left(1+ \bY_{n} \right)\star K_\bes^{n 0}}_{\text{(5)}}\,.
\eal
This expression is what we will reconstruct from the discontinuities of $\log Y_0$ in the next section.
We indicate the different contributions above with different numbers. We will use the same numbers for the terms generated by the inversion. In this manner, it will be easier to follow the computation.

\subsubsection{Rewriting the discontinuity equations of \texorpdfstring{$\log Y_0$}{lnY0inv}}

In this section,
we will rewrite the discontinuity equations for $\log Y_0$ \eqref{Y0symmdisc} and \eqref{eq:lnY0_disc_p2N2} in a form better suited to inversion (even if it is apparently more involved).
The new symmetrised discontinuity on the real axis is obtained by substituting the TBA equation for $\log Y_-^{(\a)}$ (see~\eqref{Y-particles}) into the first row of~\eqref{Y0symmdisc}:
\bal
\label{Y0symmdisc_for_inv}
\{\log{Y_0}\}_0=&-2 \log (1+Y_0)\cstar K+2\log (1+Y_0)\\
&-2 \log{(1+Y_n)}\star K_-^{ny}+2 \log{(1+\bY_n)}\star K_+^{ny}\\
&+ \log \prod_{\alpha=1,2} \left(1-\frac{e^{i \mu_\alpha}}{Y^{(\alpha)}_+}\right)^\gamma\left(1-\frac{e^{i \mu_\alpha}}{Y^{(\alpha)}_-} \right)^\gamma\,.
\eal
Notice that there is no circularity here: we are merely substituting into the discontinuity relation a TBA equation that has already been derived. Thus, this step remains fully consistent with the reconstruction of the complete TBA from the Y-system and the discontinuity relations.

We will also massage the expression for the  discontinuities on the cuts at $\Im(u)=\pm \frac{2N}{h}$: by inserting in their expression \eqref{eq:lnY0_disc_p2N2}
the TBA equation for the ratio $
Y_-^{(\a)}/Y_+^{(\a)}$
(which we have already reobtained) and equation~\eqref{eq:disc_Ymp_2M} we get
\bal
\label{eq:lnY0_disc_p2N_for_inv}
[\log{Y_0}]_{+2N}= &- \sum^{N}_{n=1} \left[ \log(1+Y_n) \right]_{2N-n} - \sum^{N}_{n=1} \left[ \log(1+\bY_n) \right]_{2N-n}+A(u)\,,
\eal
\bal
\label{eq:lnY0_disc_m2N_for_inv}
[\log{Y_0}]_{-2N}= &- \sum^{N}_{n=1} \left[ \log(1+Y_n) \right]_{-2N+n} + \sum^{N}_{n=1} \left[ \log(1+\bY_n) \right]_{-2N+n}-A(u)\\
&+\left[ \log \prod_{\a=1,2} \left(1-\frac{e^{i \mu_\a}}{Y^{(\a)}_+} \right) \left( 1-\frac{e^{i \mu_\a}}{Y^{(\a)}_-} \right) \right]_{-2N}\,,
\eal
where we define a quantity $A(u)$, which is simply the ground-state TBA expression for $-\sum_{\a}\log Y_-^{(\a)}/Y_+^{(\a)}$ (obtained from \eqref{eq:TBAaux-}-\eqref{eq:TBAaux+}):
\bal
\label{eq:Aterm_def}
A \equiv & - 4 \log \left(1+Y_{0}(v)\right) \cstar K(v, u)   + 2 \log \left(1+Y_{0}(u)\right) \\
&- 2\log (1+Y_{n}) (1+\bY_{n}) \star \left(K^{ny}_-(v, u) - K^{ny}_+(v, u)\right)\,.
\eal
We will also make use of the already mentioned  vertical cut on the imaginary axis
\bal
\label{eq:en_disc_Y0}
\log{Y_0}(i u+ \eps) - \log{Y_0}(i u- \eps)= -2 i \pi L \,.
\eal

\subsubsection{The inversion of \texorpdfstring{$\log Y_0$}{lnY0inv}}

\begin{figure}
\centering
\includegraphics[width=0.49\textwidth]{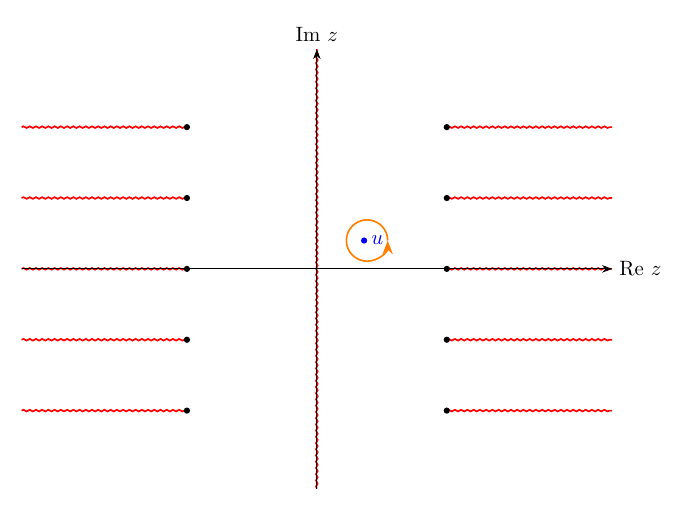}
\hfill
\includegraphics[width=0.49\textwidth]{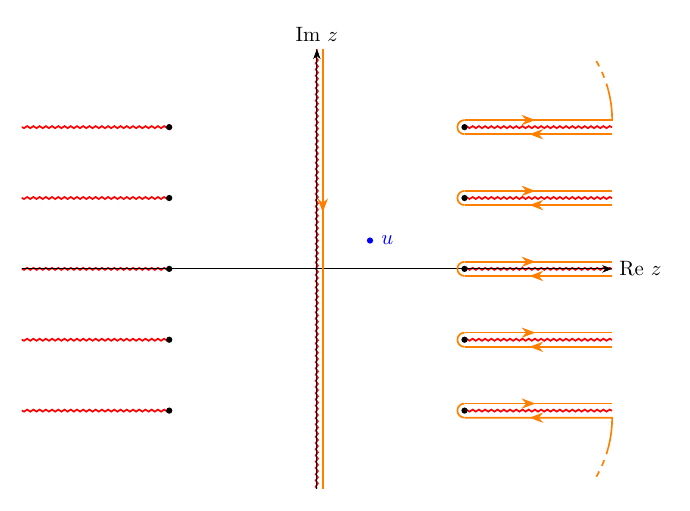}

\vspace{3mm}

\includegraphics[width=0.55\textwidth]{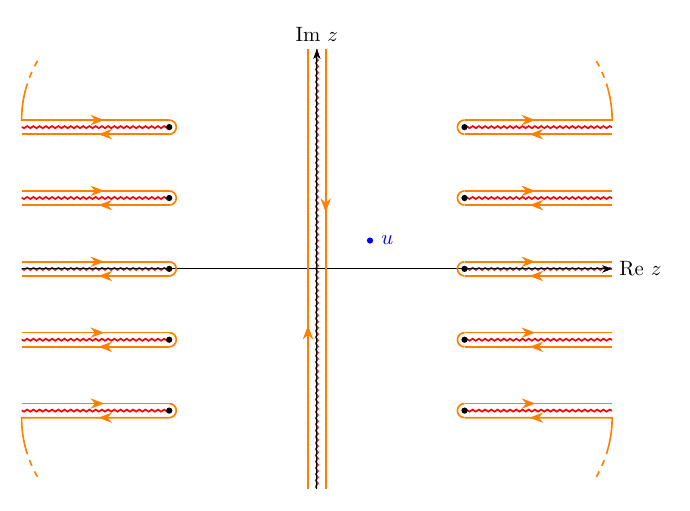}
\caption{Different steps of contour deformation to invert $\log Y_0$. Starting from the leftmost panel, we first blow up the orange integration contour, and then (in the last panel) we add a  (blown-up) contour in the $\Re(z)<0$ half-plane which does not encompass any singularity.}
\label{im:invY0disc}
\end{figure}

Starting from the discontinuities of $\log Y_0 (u)$ \eqref{Y0symmdisc_for_inv}, \eqref{eq:lnY0_disc_p2N_for_inv}, \eqref{eq:lnY0_disc_m2N_for_inv} and~\eqref{eq:en_disc_Y0}, let us now retrieve  the TBA equation \eqref{eq:TBAY0_bes_nonbes}. We start by using  Cauchy's theorem on the function $\frac{\log Y_0(u)}{\sqrt{4-u^2}}$, assuming that $0<\Im(u)<1/h$, and integrating on the small circle contour $\gamma$ depicted on the top left of figure~\ref{im:invY0disc}  we obtain:
\beq
\la{eq:integrandgammaY0}
\log Y_0(u)  =\oint_{\gamma} dz \, \log Y_0(z) \, K(z, u)\,,
\eeq
with the kernel defined in \eqref{def:Kuniversal}.  
Then we blow up the contour; note that we can do this only in the half-complex plane where $\Re(z)>0$, since on the imaginary axis $\log Y_0$ has the energy discontinuity~\eqref{eq:en_disc_Y0}.  After this deformation, we obtain the contour in the top right of figure~\ref{im:invY0disc}. 

At this point, we can add a reflected closed contour on the left side of the complex plane; this is possible because no poles are included on the left contour, and therefore the additional integral vanishes. We end up with the contour on the bottom of figure~\ref{im:invY0disc}, which corresponds to the following contributions, including all the horizontal cuts, as well as the log cut along the imaginary axis:
\bal\label{eq:firstCauchyY0}
\log Y_0 (u)=& \sum_{\substack{N=-\infty\\N\neq 0}}^{\infty}\left(\int_{-\infty}^{-2} + \int_{+2}^{+\infty} \right) dz \,  K(z+\frac{2i N}{h},u) [\log Y_0]_{2N}(z+ i \epsilon)\\
&+\left(\int_{-\infty}^{-2} + \int_{+2}^{+\infty} \right) dz \, K(z+i \epsilon ,u) \{\log Y_0 \}_{0}(z+ i \epsilon)\\
&+\int^{+i\infty}_{-i\infty} \, dz \left(\log Y_0(z- \eps) - \log Y_0(z+ \eps) \right) K(z,u) \,.
\eal
The last line of this equation can be easily computed using~\eqref{eq:en_disc_Y0} and reproduces the energy contribution to $\log Y_0$:
\bal
\label{eq:en_contribution_to_lnY0_inv}
&&\int^{+i\infty}_{-i\infty} \, dz \left(\log Y_0(z- \eps) - \log Y_0(z+ \eps) \right) K(z,u)=2 i \pi L \int^{+i\infty}_{-i\infty} \, dz   K(z,u)\\
&&=L \ln x^2(u)=-L \tE_0(u)\,.
\eal
The remaining terms are more complicated. Using \eqref{eq:lnY0_disc_p2N_for_inv} and~\eqref{eq:lnY0_disc_m2N_for_inv}, the first line of~\eqref{eq:firstCauchyY0} becomes 
\bal
\label{eq:colored_lnY0_inv_disc}
&\sum_{\substack{N=-\infty\\N\neq 0}}^{\infty}\left(\int_{-\infty}^{-2} + \int_{+2}^{+\infty} \right) dz \,  K(z+\frac{2i N}{h},u) [\log Y_0]_{2N}(z+ i \epsilon)=\\
&\underbrace{\scriptstyle{\sum_{N=1}^{+\infty}\left(\int_{-\infty}^{-2} + \int_{+2}^{+\infty} \right) dz \,  K(z-\frac{2i N}{h},u) \left[ \log \prod_{\a=1,2} \left(1-\frac{e^{i \mu_\a}}{Y^{(\a)}_+} \right) \left( 1-\frac{e^{i \mu_\a}}{Y^{(\a)}_-} \right) \right]_{-2N}(z+i \epsilon)}}_{\text{(1)}}\\
&\underbrace{\scriptstyle{-\sum_{N=1}^{+\infty} \sum_{n=1}^{N}\left(\int_{-\infty}^{-2} + \int_{+2}^{+\infty} \right) dz \,  K(z+\frac{2i N}{h},u) \left[ \log(1+Y_n)  \right]_{2N-n}(z+i \epsilon)}}_{\text{(2)}}\\
&\underbrace{\scriptstyle{-\sum_{N=1}^{+\infty}\sum_{n=1}^{N} \left(\int_{-\infty}^{-2} + \int_{+2}^{+\infty} \right) dz \,  K(z-\frac{2i N}{h},u) \left[ \log (1+Y_n)  \right]_{-2N+n}(z+i \epsilon)}}_{\text{(2)}}\\
&\underbrace{\scriptstyle{-\sum_{N=1}^{+\infty} \sum_{n=1}^{N}\left(\int_{-\infty}^{-2} + \int_{+2}^{+\infty} \right) dz \,  K(z+\frac{2i N}{h},u) \left[ \log (1+\bY_n)  \right]_{2N-n}(z+i \epsilon)}}_{\text{(3)}}\\
&\underbrace{\scriptstyle{+\sum_{N=1}^{+\infty}\sum_{n=1}^{N} \left(\int_{-\infty}^{-2} + \int_{+2}^{+\infty} \right) dz \,  K(z-\frac{2i N}{h},u) \left[ \log (1+\bY_n)  \right]_{-2N+n}(z+i \epsilon)}}_{\text{(3)}}\\
&\underbrace{\scriptstyle{-\sum_{N=-\infty}^{-1}\left(\int_{-\infty}^{-2} + \int_{+2}^{+\infty} \right) dz \,  K(z+\frac{2i N}{h},u) A(z+ i \epsilon)}}_{\text{(5)}} +  \underbrace{\scriptstyle{\sum_{N=1}^{+\infty}\left(\int_{-\infty}^{-2} + \int_{+2}^{+\infty} \right) dz \,  K(z+\frac{2i N}{h},u) A(z+ i \epsilon)}}_{\text{(5)}}.
\eal
Using~\eqref{Y0symmdisc_for_inv}, the second line of~\eqref{eq:firstCauchyY0} becomes
\bal
\label{eq:colored_lnY0_inv_symdisc}
&\left(\int_{-\infty}^{-2} + \int_{+2}^{+\infty} \right) dz \, K(z+i \epsilon ,u) \{\log Y_0 \}_{0}(z+ i \epsilon)=\\
&\underbrace{\left(\int_{-\infty}^{-2} + \int_{+2}^{+\infty} \right) dz \, K(z+i \epsilon ,u) \log \prod_{\alpha=1,2} \left(1-\frac{e^{i \mu_\alpha}}{Y^{(\alpha)}_+}\right)\left(1-\frac{e^{i \mu_\alpha}}{Y^{(\alpha)}_-} \right)(z^\gamma+i \epsilon)}_{\text{(1)}}\\
&\underbrace{-2\left(\int_{-\infty}^{-2} + \int_{+2}^{+\infty} \right) dz \, K(z+i \epsilon ,u) \left(\log{(1+Y_n )}\star K_-^{ny}\right)(z+ i \epsilon)}_{\text{(2)}}\\
&\underbrace{+2\left(\int_{-\infty}^{-2} + \int_{+2}^{+\infty} \right) dz \, K(z+i \epsilon ,u) \left(\log{(1+\bY_n )}\star K_+^{ny}\right)(z+ i \epsilon)}_{\text{(3)}}\\
&\underbrace{-2\left(\int_{-\infty}^{-2} + \int_{+2}^{+\infty} \right) dz \, K(z+i \epsilon ,u)  \left(\log (1+Y_0 )\cstar K\right)( z+ i \epsilon)}_{\text{(4)}}\\
&\underbrace{+ 2\left(\int_{-\infty}^{-2} + \int_{+2}^{+\infty} \right) dz \, K(z+i \epsilon ,u)  \log (1+Y_0(z+ i \epsilon)) }_{\text{(4)}}\,.
\eal
We label the contributions in the equations above with distinct numbers, then combine and compute them separately. These numbers match those in~\eqref{eq:TBAY0_bes_nonbes}, highlighting the distinct terms arising from the inversion.

Plugging~\eqref{eq:en_contribution_to_lnY0_inv}, \eqref{eq:colored_lnY0_inv_disc} and~\eqref{eq:colored_lnY0_inv_symdisc} into~\eqref{eq:firstCauchyY0}, we obtain
\bal
\log Y_0 (u)=&-L \tE_0(u) + F_{(1)}+ F_{(2)} + F_{(3)} +   F_{(4)}  +   F_{(5)}\,,
\eal
where we combined together contributions carrying the same number, so that
\bal
\label{eq:app_Y0_aux_inversion1}
& F_{(1)}=\\
&\sum_{N=1}^{+\infty}\left(\int_{-\infty}^{-2} + \int_{+2}^{+\infty} \right) dz \,  K(z-\frac{2i N}{h},u) \left[ \log \prod_{\a=1,2} \left(1-\frac{e^{i \mu_\a}}{Y^{(\a)}_+} \right) \left( 1-\frac{e^{i \mu_\a}}{Y^{(\a)}_-} \right) \right]_{-2N}(z+i \epsilon)\\
&+\left(\int_{-\infty}^{-2} + \int_{+2}^{+\infty} \right) dz \, K(z+i \epsilon ,u) \log \prod_{\alpha=1,2} \left(1-\frac{e^{i \mu_\alpha}}{Y^{(\alpha)}_+}\right)\left(1-\frac{e^{i \mu_\alpha}}{Y^{(\alpha)}_-} \right)(z^\gamma+i \epsilon)\,.
\eal
\bal
\label{eq:app_Y0_Q_inversion1}
& F_{(2)}=\\
&-\sum_{N=1}^{+\infty} \sum_{n=1}^{N}\left(\int_{-\infty}^{-2} + \int_{+2}^{+\infty} \right) dz \,  K(z+\frac{2i N}{h},u) \left[ \log(1+Y_n)  \right]_{2N-n}(z+i \epsilon)\\
&-\sum_{N=1}^{+\infty}\sum_{n=1}^{N} \left(\int_{-\infty}^{-2} + \int_{+2}^{+\infty} \right) dz \,  K(z-\frac{2i N}{h},u) \left[ \log (1+Y_n)  \right]_{-2N+n}(z+i \epsilon)\\
&-2\left(\int_{-\infty}^{-2} + \int_{+2}^{+\infty} \right) dz \, K(z+i \epsilon ,u) \log{(1+Y_n(v))}\star K_-^{ny}(v,z+ i \epsilon)\,.
\eal
\bal
\label{eq:app_Y0_bQ_inversion1}
& F_{(3)}=\\
&-\sum_{N=1}^{+\infty} \sum_{n=1}^{N}\left(\int_{-\infty}^{-2} + \int_{+2}^{+\infty} \right) dz \,  K(z+\frac{2i N}{h},u) \left[ \log (1+\bY_n)  \right]_{2N-n}(z+i \epsilon)\\
&+\sum_{N=1}^{+\infty}\sum_{n=1}^{N} \left(\int_{-\infty}^{-2} + \int_{+2}^{+\infty} \right) dz \,  K(z-\frac{2i N}{h},u) \left[ \log (1+\bY_n)  \right]_{-2N+n}(z+i \epsilon)\\
&+2\left(\int_{-\infty}^{-2} + \int_{+2}^{+\infty} \right) dz \, K(z+i \epsilon ,u) \log{(1+\bY_n(v))}\star K_+^{ny}(v,z+ i \epsilon) \,,
\eal
\bal
\label{eq:app_Y00_inversion1}
 F_{(4)}=& -2\left(\int_{-\infty}^{-2} + \int_{+2}^{+\infty} \right) dz \, K(z+i \epsilon ,u)  \log (1+Y_0(v))\cstar K(v,z+ i \epsilon)\\
&+ 2\left(\int_{-\infty}^{-2} + \int_{+2}^{+\infty} \right) dz \, K(z+i \epsilon ,u)  \log (1+Y_0(z+ i \epsilon)) \,.
\eal
Finally, using~\eqref{eq:Aterm_def}, the fifth contribution (associated to the last line of~\eqref{eq:colored_lnY0_inv_disc}) is given by
\bal
\label{eq:app_Y0_BES_inversion1}
& F_{(5)}=\\
&+2\log\left(1+Y_{n}(v) \right) \left(1+\bY_{n}(v) \right)\star\sum_{N=1}^{\infty} \, \left(K^{ny}_+(v, z+ i \epsilon) - K^{ny}_-(v, z+ i \epsilon)\right) \check{\star}  K(z+\frac{2i N}{h},u) \\
&-2\log\left(1+Y_{n}(v) \right) \left(1+\bY_{n}(v) \right)\star\sum_{N=1}^{\infty} \, \left(K^{ny}_+(v, z+ i \epsilon) - K^{ny}_-(v, z+ i \epsilon)\right) \check{\star}  K(z-\frac{2i N}{h},u) \\
&-4 \log \left(1+Y_{0}(v)\right) \cstar K(v, z+ i \epsilon)  \check{\star} \sum^{+ \infty}_{N=1} \left( K(z+\frac{2i N}{h},u) - K(z-\frac{2i N}{h},u) \right) \\
&+2 \log \left(1+Y_{0}(z+ i\epsilon)\right)  \check{\star} \sum^{+ \infty}_{N=1} \left( K(z+\frac{2i N}{h},u) - K(z-\frac{2i N}{h},u) \right) \,.
\eal
Let us compute each of the terms listed above separately and show that their sum equals the TBA equation~\eqref{eq:TBAY0_bes_nonbes} for $\log Y_0$.

\paragraph{First contribution.}
We start considering the quantity in~\eqref{eq:app_Y0_aux_inversion1}. The integral in the second line, performed above the cuts on the second Riemann sheet, can be written as an integral on the first Riemann sheet below the cuts, and we have
\bal
\label{eq:app_Y0_aux_inversion2}
&F_{(1)}=\\
&\sum_{N=1}^{+\infty}\left(\int_{-\infty}^{-2} + \int_{+2}^{+\infty} \right) dz \,  K(z-\frac{2i N}{h},u) \left[ \log \prod_{\a=1,2} \left(1-\frac{e^{i \mu_\a}}{Y^{(\a)}_+} \right) \left( 1-\frac{e^{i \mu_\a}}{Y^{(\a)}_-} \right) \right]_{-2N}(z+i \epsilon)\\
&-\left(\int_{-\infty}^{-2} + \int_{+2}^{+\infty} \right) dz \, K(z-i \epsilon ,u) \log \prod_{\alpha=1,2} \left(1-\frac{e^{i \mu_\alpha}}{Y^{(\alpha)}_+}\right)\left(1-\frac{e^{i \mu_\alpha}}{Y^{(\alpha)}_-} \right)(z-i \epsilon)\,.
\eal
After adding and subtracting an integral between $-2$ and $+2$ on the real line, this expression can be written as an integral on the path $\Gamma$ depicted in figure~\ref{im:path_gamma_for_iniv_Y0aux} and a piece which only involves the integration between $-2$ and $+2$:
\begin{figure}
\begin{center}
\includegraphics*[width=0.66\textwidth]{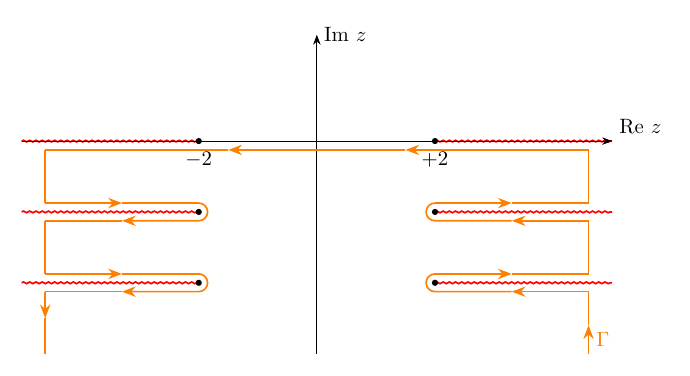} 
\end{center}
\caption{Integration contour for the inversion of the auxiliary contribution to $\log Y_0$.} 
\label{im:path_gamma_for_iniv_Y0aux}
\end{figure}
\bal
\label{eq:app_Y0_aux_inversion3}
F_{(1)}&=\int_\Gamma dz \,  K(z,u)  \log \prod_{\a=1,2} \left(1-\frac{e^{i \mu_\a}}{Y^{(\a)}_+(z)} \right) \left( 1-\frac{e^{i \mu_\a}}{Y^{(\a)}_-(z)} \right) \\&+\int_{-2}^{+2} dz \,  K(z,u)  \log \prod_{\a=1,2} \left(1-\frac{e^{i \mu_\a}}{Y^{(\a)}_+(z)} \right) \left( 1-\frac{e^{i \mu_\a}}{Y^{(\a)}_-(z)} \right) \,.
\eal
Since the only pole of $K(z,u)$ is at $z=u$ and we assume $\Im(u)>0$, $\Gamma$ encloses no poles. Hence, the equation reduces to
\bal
\label{eq:app_Y0_aux_inversion4}
F_{(1)}&=\int_{-2}^{+2} \,  K(z,u)  \log \prod_{\a=1,2} \left(1-\frac{e^{i \mu_\a}}{Y^{(\a)}_+(z)} \right) \left( 1-\frac{e^{i \mu_\a}}{Y^{(\a)}_-(z)} \right)\\
&=   \log \prod_{\a=1,2} \left(1-\frac{e^{i \mu_\a}}{Y^{(\a)}_+} \right) \left( 1-\frac{e^{i \mu_\a}}{Y^{(\a)}_-} \right) \hat{\star}  K \,.
\eal
In the expression above, we have exactly the term $(1)$ of~\eqref{eq:TBAY0_bes_nonbes}.

\paragraph{Second contribution.}

We analyse~\eqref{eq:app_Y0_Q_inversion1}. The last line of this expression can be simplified by using the identity \eqref{eq:final_result_KmQyconvK2}, giving
\bal
\label{eq:invY0_first_Q_contribution}
&-2 \log{(1+Y_n(v))}\star \left(\int_{-\infty}^{-2} + \int_{+2}^{+\infty} \right) dz K_-^{ny}(v,z+ i \epsilon)  \, K(z+i \epsilon ,u) \\
&=\log{(1+Y_n(v))}\star \biggl(-\frac{1}{2} K_{n}(v-u)\\
&+\frac{1}{i \pi}  \left( \g^{-n \circ}_{v u} + \frac{i \pi}{2}\right) K(v- \frac{i}{h}n,u)-\frac{1}{i \pi}  \left( \g^{+n \circ}_{v u} + \frac{i \pi}{2}\right) K(v+ \frac{i}{h}n, u) \biggr) \,.
\eal
For the other terms in~\eqref{eq:app_Y0_Q_inversion1}, we use the two properties in section \ref{app:formula_disc_inversion} to simplify the infinite sums, obtaining
\bal
\label{eq:invY0_second_Q_contribution}
\sum_{n=1}^{+\infty} \log(1+Y_n(v)) K(v+\frac{i n}{h},u)-\sum_{n=1}^{+\infty}   \log (1+Y_n(v)) * K(v-\frac{i n}{h},u) \,.
\eal
Summing~\eqref{eq:invY0_first_Q_contribution} and~\eqref{eq:invY0_second_Q_contribution} we obtain
\bal
\nonumber
&F_{(2)}=\log{(1+Y_n(v))} \\
& \star \biggl( -\frac{1}{2}  K_{n}(v-u) +\frac{1}{i \pi}  \left( \g^{-n \circ}_{v u} - \frac{i \pi}{2}\right) K(v- \frac{i}{h}n,u)-\frac{1}{i \pi}  \left( \g^{+n \circ}_{v u} - \frac{i \pi}{2}\right) K(v+ \frac{i}{h}n,u) \biggr) \,.
\eal
In the expression above, we recognise the kernel $\hat{K}^{n0}$ in~\eqref{eq:app_discY0_hat_Ks}, and therefore we reproduced the term $(2)$ of \eqref{eq:TBAY0_bes_nonbes}.

\paragraph{Third contribution.}

The treatment of \eqref{eq:app_Y0_bQ_inversion1} is similar. The last line of this expression can be simplified by using the identity \eqref{eq:final_result_KpQyconvK2}, giving
\bal
\label{eq:invY0_first_bQ_contribution}
&+2\log{(1+\bY_n(v))}\star \left(\int_{-\infty}^{-2} + \int_{+2}^{+\infty} \right) dz \,K_+^{ny}(v,z+ i \epsilon)K(z+i \epsilon ,u)  \\
&=\log{(1+\bY_n(v))}\star \biggl( \frac{1}{2} K_{n}(v-u)\\
&+\frac{1}{i \pi} \left( \g^{-n \circ}_{v u} - \frac{i \pi}{2}\right) K(v- \frac{i}{h}n,u)-\frac{1}{i \pi} \left( \g^{+n \circ}_{v u} - \frac{i \pi}{2}\right) K(v+ \frac{i}{h}n,u) \biggr) \,.
\eal
For the other terms in~\eqref{eq:app_Y0_bQ_inversion1}, we use the properties in appendix~\ref{app:formula_disc_inversion} to simplify the infinite sums, obtaining
\bal
\label{eq:invY0_second_bQ_contribution}
\sum_{n=1}^{+\infty} \log(1+\bY_n(v)) K(v+\frac{i n}{h},u)+\sum_{n=1}^{+\infty}   \log (1+\bY_n(v)) * K(v-\frac{i n}{h},u) \,.
\eal
Summing~\eqref{eq:invY0_first_bQ_contribution} and~\eqref{eq:invY0_second_bQ_contribution} we obtain
\bal
&F_{(3)}=\log{(1+\bY_n(v))}\star \biggl( \frac{1}{2} K_{n}(v-u)\\
&+\frac{1}{i \pi} \left( \g^{-n \circ}_{v u} + \frac{i \pi}{2} \right) K(v- \frac{i}{h}n,u)-\frac{1}{i \pi} \left( \g^{+n \circ}_{v u} - \frac{3}{2} i \pi\right) K(v+ \frac{i}{h}n,u) \biggr) \,.
\eal
In the expression above, we recognise the kernel $\hat{\tK}^{n0}$ in~\eqref{eq:app_discY0_hat_Ks} and therefore we reproduced the term $(3)$ in \eqref{eq:TBAY0_bes_nonbes}.

\paragraph{Fourth contribution.}

We analyse~\eqref{eq:app_Y00_inversion1}. The last line of this expression can be simplified by using the identity \eqref{eq:app_massless_convolutions}, giving
\bal
&-2 \log (1+Y_0(v))\cstar \left(\int_{-\infty}^{-2} + \int_{+2}^{+\infty} \right) dz \, K(v,z+ i \epsilon)\, K(z+i \epsilon ,u)  \\
&=2 \log (1+Y_0(v))\cstar K(v+ i 0, u) \left(-1 +\frac{i}{\pi} \g^{\circ \circ}_{vu} \right) \,.
\eal
The integral was computed by keeping $v$ just above the mirror cuts and $ z + i \eps$ at a small distance $\eps$ from the cuts (above the $v$-integration contour), ensuring that there are no poles on the integration contour.
Plugging this relation into~\eqref{eq:app_Y00_inversion1} we obtain
\bal
\label{eq:app_Y00_inversion2}
F_{(4)}&= \frac{2 i}{\pi} \log (1+Y_0(v))\cstar K(v+ i 0, u)  \g^{\circ \circ}_{vu} \,,
\eal
and we recognise the kernel $\hat{K}^{00}$ in~\eqref{eq:app_discY0_hat_Ks}. Therefore, we reproduced the term $(4)$ in \eqref{eq:TBAY0_bes_nonbes}.

\paragraph{Fifth contribution.}

We finally consider the contribution~\eqref{eq:app_Y0_BES_inversion1} coming from the BES discontinuities.
The first two rows of~\eqref{eq:app_Y0_BES_inversion1} can be written as
\begin{multline}
\label{eq:invY0mixBESrep}
    2\log\left(1+Y_{n}(v) \right) \left(1+\bY_{n}(v) \right) \star\sum_{N=1}^{\infty} \int_{\rm cuts} dz K^{ny}_+(v, z)
    \left( K(z-\frac{2i N}{h},u) - K(z+\frac{2i N}{h},u) \right)\\
    = -2\log\left(1+Y_{n}(v) \right) \left(1+\bY_{n}(v) \right)\star K^{n0}_\bes(v,u)
\end{multline}
where the integrals around the mirror cuts are performed by integrating to the right along the lower edge of the cuts (the opposite of the prescription in figure~\ref{im:path_gamma_for_iniv_Y0aux}), and we used the relation~\eqref{eq:BES_Kernel_as_double_conv_Q02} for the mixed-mass BES kernel.
Similarly, in the last two rows of~\eqref{eq:app_Y0_BES_inversion1} we identify the massless-massless BES kernel~\eqref{eq:BES_Kernel_as_double_conv_002}. Then they can be written as
\bal
\label{eq:invY0masslessBESrep}
-2 \log  \left(1+Y_0 \right) \cstar K_\bes^{00} \,.
\eal
Replacing the first two rows on the r.h.s.  of~\eqref{eq:app_Y0_BES_inversion1} with~\eqref{eq:invY0mixBESrep}, and the last two rows with~\eqref{eq:invY0masslessBESrep}, we obtain
\bal
\label{eq:app_Y0_BES_inversion_completed}
F_{(5)}=&-2\log\left(1+Y_{n}(v) \right) \left(1+\bY_{n}(v) \right)\star K^{n0}_\bes(v,u) \\
&-2 \log \left(1+Y_{0}(v+ i \epsilon)\right) \cstar  K^{00}_\bes(v+ i \epsilon,u) \,.
\eal
Thus we reproduced the last line of~\eqref{eq:TBAY0_bes_nonbes}.

Summing all the terms we have found so far, we can reconstruct exactly the equation~\eqref{eq:TBAY0_bes_nonbes}, thereby completing the inversion of $\log Y_0$.

\subsection{Reconstructing the TBA equations for massive particles}

Recovering the TBA equations for massive particles can be achieved with well-established methods. 
A crucial step of this inversion process for the massive Y functions (see e.g. \cite{Cavaglia:2010nm}) is to reconstruct the discontinuities $\Delta_1$ and $\bar{\Delta}_1$, but in the present model, remarkably, these are already known in the universal form (\ref{eq:Delta1_local}), thanks to the very nontrivial simplification proved in appendix \ref{appendix_discontinuities_Y1}. 

With these ingredients in place, let us summarise the main steps needed to recover the remaining TBA equations for the massive Y functions, without going into the full details.

First, one can use the standard arguments of~\cite{Zamolodchikov:1991et}, based on the analyticity strips of the massive $Y$ functions, to derive the simplified TBA equations~\eqref{eq:simplified_TBA} from the Y-system. These equations contain convolution terms involving the discontinuities of $\log Y_1$ and $\log \dot{Y}_1$, since these functions have branch cuts at the same distance from the real axis as the shift appearing in the Y-system. These discontinuities are precisely $\Delta_1$ and $\bar{\Delta}_1$, for which the local expression~\eqref{eq:Delta1_local} can be used. Next, using only the TBA equations derived so far, in particular the one for $\log Y_0$, one can invert the derivation of Appendix~\ref{appendix_discontinuities_Y1} and replace the local expressions for $\Delta_1$ and $\bar{\Delta}_1$ with the representation~\eqref{eq:discfromTBA},\eqref{eq:residues}, which is equivalent to~\eqref{eq:Delta1_local}.

The TBA equations in Zamolodchikov’s simplified form are equivalent to the canonical ones, provided that the leading asymptotics of the Y functions are specified. In this sense, we could already regard the derivation as complete. To recover the original form of the TBA equations presented in Section~\ref{sec:gstba}, one should apply the method of~\cite{Zamolodchikov:1991et}, exploiting the relations between the canonical kernels and the universal kernel, which can be found for the present model in~\cite{Frolov:2021bwp}. 

\section{Conclusion}
\label{sec:conclusion}
In this paper we have taken a significant step towards bridging the gap between the
Thermodynamic Bethe Ansatz and the Quantum Spectral Curve for strings on
\adsst \, supported by pure Ramond-Ramond flux. Starting from the ground-state TBA equations of~\cite{Frolov:2021bwp}, deformed by a (purely imaginary) chemical potential to render the solutions nontrivial, we have derived the extended Y-system for this model, comprising a set of functional relations supplemented by local discontinuity relations.

In particular, we have shown that the Y-system takes the form of
two independent $\text{PSU}(1,1|2)$ Y-systems, which can be drawn on two different T-hooks, one for the left (undotted) and one
for the right (dotted) Y functions, interacting only through the discontinuity
relations. We have proved that the massless modes, a distinctive feature of the AdS$_3$ models, do not reside on either T-hook but play a fundamental role in connecting the massive and auxiliary Y functions through their discontinuities. 

Furthermore,
we have identified and proved a $\mathbb{Z}_2$-symmetry of the ground-state TBA
equations under the exchange of left and right Y functions,
\begin{equation}
  Y_n \leftrightarrow \dot{Y}_n\,, \quad Y_0 \leftrightarrow Y_0\,, \quad
  Y_{\pm}^{(\alpha)} \leftrightarrow \bigl(Y_{\mp}^{(\alpha)}\bigr)^{-1}\,, \quad
  \mu_\alpha \leftrightarrow -\mu_\alpha\,.
\end{equation}
This \emph{chiral symmetry} is a fundamental consistency requirement, reflecting
the arbitrariness of the labelling of left and right particles, and is also found in the known Quantum Spectral Curve, since it is composed of two interchangeable PSU$(1,1|2)$ Q-systems.

Finally, we have shown that the extended Y-system is fully equivalent to the \adsst\, ground-state TBA upon removing the problematic factor \eqref{eq:a_g} from the dressing phases. Concretely, we reconstructed all TBA equations via an inversion
procedure, starting from the Y-system functional equations and the discontinuities of the
Y functions. This inversion is considerably more involved than in the AdS$_5 \times S^5$
case~\cite{Cavaglia:2010nm}, due to the presence of massless excitations and the more
intricate structure of the discontinuities. Nevertheless, we have shown that
the procedure can be carried out successfully for all particles, providing a fundamental consistency check of our derivation of the discontinuity
relations.

 The next obvious step is to use this extended Y-system to derive the Quantum Spectral Curve. This will be done in an upcoming companion paper~\cite{part2}.
 
More generally, it would also be interesting to investigate how this derivation applies to \adsst\, with mixed RR and NS-NS fluxes. In that case, the mirror TBA is known~\cite{Frolov:2025tda}, but the QSC is not.
The rather involved cut structure of mixed-flux models~\cite{Frolov:2023lwd} is likely to make such a derivation more difficult.
Finally, it is interesting to notice that the (excited-state) mirror TBA of this model simplifies drastically in the tensionless limit $h\rightarrow 0$, see~\cite{Brollo:2023rgp, Brollo:2023pkl, Brollo:2025rgp}. At leading order $\mathcal{O}(h)$, the TBA is of difference form, and it is immediate to derive a local Y-system involving only massless and auxiliary Y functions. This suggests that the $h\to0$ limit and our derivation of the Y-system do not commute. It would be interesting to investigate this limit further. We hope to return to some of these questions in the near future.

\section*{Acknowledgements}

We thank S.~Ekhammar, S.~Frolov, N.~Gromov, B.~Stefa\'nski Jr, for many helpful discussions.
NP, DP and AS thank the organisers and participants of the workshop ``Higher-dimensional Integrability and Holography'', held in Buggerru, Italy, and of the ``Workshop on Higher-d Integrability'' held in Favignana, Italy, for stimulating discussions that contributed to this work.
AC, NP and RT participate in the project HORIZON-MSCA-2023-SE-01-101182937-HeI.
AS gratefully acknowledges support from the CARIPARO Foundation under grant No. 68079. This work was also partially supported by the INFN initiatives GAST and SFT.
DP acknowledges support from the Deutsche Forschungsgemeinschaft (DFG, German Research Foundation) through SFB 1624 (Project No.~506632645).

\appendix

\section{Kinematical conventions}
\label{app:kinematical_conventions}

\subsection{Zhukovsky and \texorpdfstring{$\g$}{g} variables}
To obtain the mirror $u$-plane parameterisation, we need to solve
\begin{equation}
x+\frac{1}{x} = u \quad, \quad x(u)^*=\frac{1}{x(u^*)}\,.
\end{equation}
The solution on the mirror $u$ plane is given by
\begin{equation}
\label{eq:inv_mirror_zhuk_map}
x(u)=\frac{1}{2} \left(u - i \sqrt{4-u^2} \right)\,,
\end{equation}
and has long cuts $(-\infty, -2)$ and $(+2, +\infty)$.
Moving to the antimirror $u$-plane corresponds to crossing one of the two long cuts; crossing these cuts and coming back to the starting point $u$, we obtain
\begin{equation}
x(u) \to \frac{1}{x(u)}\,.
\end{equation}
We also define the string solution $x_s(u)$ as
\begin{equation}
\label{eq:xstring_sol}
x_s(u)=\begin{cases}
&x(u) \quad \text{if} \ \Im(u)<0\,,\\
&\frac{1}{x(u)} \quad \text{if} \ \Im(u)>0\,.
\end{cases}
\end{equation}
We have that on the first sheet $|x_s(u)|>1$. This solution has a short cut $(-2, +2)$. The string solution is important to parameterise the auxiliary roots.

In order to solve the crossing equations for the mirror theory, it is useful to define the following $\g$ function
\begin{equation}
\label{eq:gamma_u}
\g(u)=\frac{1}{2} \log \left( \frac{2-u}{2+u} \right)\,,
\end{equation}
having long cuts $(-\infty, -2)$ and $(+2,+\infty)$.  We introduced a parameterisation $\g(u)$ which is common to all the particles and differs from the one used in equation (A.9) of~\cite{Frolov:2021bwp} by shifts of $\pm \frac{i \pi}{2}$.
Inverting the expression above, we obtain
\bal
\label{eq:x_funct_of_g}
x(u)= \frac{i+e^{\g(u)}}{i-e^{\g(u)}}\,,
\eal
where we chose the branch where $\Im(x(u))<0$. Under continuation across the long cuts from below $\gamma(u)$ is continued as follows
\bal
\gamma(u^\gamma) = \gamma(u)+ i \pi \,.
\eal

In the following, we define
\begin{equation}
x_i^\pm=x(u_i\pm \frac{i}{h})\,, \qquad x_i=x(u_i+ i 0)\,,
\end{equation}
and
\begin{equation}
\g_i^\pm=\g(u_i\pm\frac{i}{h}) \,, \qquad \g_i^\circ=\g(u_i+ i 0)\,.
\end{equation}
These definitions extend to bound states as follows
\begin{equation}
x_i^{\pm n}=x(u_i\pm \frac{i}{h} n) \,, \qquad \g_i^{\pm n}=\g(u_i\pm \frac{i}{h} n)\,.
\end{equation}

\subsection{Energy and momentum}
\label{app:en_and_mom}
In the mirror region, the particle energies and momenta are defined by
\begin{equation}
\label{eq:energy_Q_particle}
\tE_{n}(u)=\log x^{-n}(u) - \log x^{+n}(u)
\end{equation}
and
\begin{equation}
\label{eq:momentum_Q_particle}
\tp_{n}(u)=\frac{h}{2} \left( x^{-n}(u) - \frac{1}{x^{-n}(u)} -x^{+n}(u) + \frac{1}{x^{+n}(u)} \right) \,,
\end{equation}
respectively. They are both real for $u\in \mathbb{R}$. 
The energy and momentum of a massless particle correspond to the limit $n \to 0^+$ of the expressions above. In this limit they are nonzero only if $u$ takes values on the mirror cut $(-\infty, -2)$ or $(+2, +\infty)$. 
For $u \in (-\infty, -2)$ then in the limit $\tE_{0}(u)>0$ and $\tp_{0}(u)<0$, and the massless particle is on the antichiral branch. For $u \in (+2, +\infty)$ then in the limit $\tE_{0}(u)>0$ and $\tp_{0}(u)>0$, and the massless particle is on the chiral branch. 
Using that on the cuts $x(u+i0)=\frac{1}{x(u-i0)}$ we get the following relations
\bal
\label{eq:mom_en_0_particle}
&\tp_{0}(u)= h \left(\frac{1}{x(u+ i0)}- x(u+ i0) \right) \,, \qquad \tE_{0}(u)=- \log x^2(u+ i0)\,,
\eal
which are valid both on the branch $u\in (+2, +\infty)$ and on the branch $u\in (-\infty,-2)$. As a function of $x$, in the mirror $x$ plane (corresponding to $\Im(x)<0$) the function $\tE_{0}$ has a cut $(-i\infty, 0)$. This cut is mapped to the cut $(-i \infty, + i \infty)$ in the $u$ plane, giving the following discontinuity
\bal
\label{eq:tE0_disc_on_im_axis}
\tE_{0}(it+ \eps)-\tE_{0}(it- \eps)=2 i \pi \,, \qquad t \in \mathbb{R}\,.
\eal
Since the kinematical branches $u<-2$ and $u>2$ are separated by this infinite cut, we cannot analytically continue one branch to the other, as happens for relativistic massless particles. This infinite vertical cut is a feature of $\log Y_0$, which contains the energy of the massless particles. In contrast, we do not have such a cut in $\log Y_n$ and $\log \dot{Y}_n$.

\section{S-matrix elements}
\label{app:Smatrix}

\subsection{Building blocks for the dressing factors}

\paragraph{Improved BES phase as an integral over mirror cuts.}
The improved BES phase can be written in terms of the following function
\begin{equation}
\label{eq:tildePhi_definition}
\begin{aligned}
\tilde\Phi(x_1,x_2)&=-i\int_{\text{cuts}}\frac{ \text{d} v_1}{2\pi i}\int_{\text{cuts}} \frac{\text{d} v_2}{2\pi i}\frac{x'(v_1) \, x'(v_2)}{(x(v_1)-x_1)(x(v_2)-x_2)} \log \frac{\Gamma \bigl[1+\frac{ih}{
2}\big(v_1-v_2\bigl)\bigl]}{\Gamma \bigl[1-\frac{ih}{
2}\big(v_1-v_2\bigl)\bigl]}\,.
\end{aligned}
\end{equation}
By `cuts' we mean that both integrals are performed around the mirror theory cuts $(-\infty, -2)$ and $(+2, +\infty)$; on the lower edge of the cuts we integrate to the right and on the upper edge we integrate to the left.
In the expression above, $x(v)$ is the mirror Zhukovsky variable defined in~\eqref{eq:inv_mirror_zhuk_map}.
Then we define the improved BES phase $\Sigma^{n n'}_\bes$ to be
\begin{equation}
\la{eq:improved_BES_Q_Qp}
{1\ov i}\log\Sigma^{n n'}_\bes (u_1,u_2) = \tilde\Phi(x^{+n}_1,x^{+n'}_2)+\tilde\Phi(x^{-n}_1,x^{-n'}_2)-\tilde\Phi(x^{+n}_1,x^{-n'}_2)-\tilde\Phi(x^{+n}_1,x^{-n'}_2)\,,
\end{equation}
where fundamental massive particles correspond to the case $n=1$ and \textit{massless particles correspond to the limit $n \to 0^+$}.
It is possible to show that the expression above is equivalent to equation (C.4) of~\cite{Frolov:2021bwp}, as proven, for example, in appendix B of~\cite{Frolov:2025uwz}, and that it applies to massless particles as well. In the massive case, this fact was already known from~\cite{Gromov:2009bc}.
A massless particle has
\bal
x_i^{\pm}=x(u_i \pm i \epsilon) \quad i=1, \, 2\,,
\eal
where $u_i+ i \epsilon$ is just above the integration contour and $u_i- i \epsilon$ just below the integration contour.
The expression in~\eqref{eq:improved_BES_Q_Qp} is valid when all Zhukovsky variables are in the mirror region (which is $\Im (x^{\pm n}_1)<0$ and $\Im (x^{\pm n}_2)<0$). Outside this region, we need to continue the phase appropriately by selecting the residues that arise from crossing the integration contour.

\paragraph{Barnes Gamma functions.}

We introduce the following functions
\begin{equation}
\label{eq:R_function}
R (\g)\equiv {G(1- \frac{\g}{2\pi i})\ov G(1+ \frac{\g}{2\pi i}) } =
   \left({e\ov 2\pi}\right)^{+\frac{\gamma}{2\pi i}}\prod_{\ell=1}^\infty \frac{\Gamma(\ell+\frac{\gamma}{2\pi i})}{\Gamma(\ell-\frac{\gamma}{2\pi i})}\,e^{-\frac{\gamma}{\pi i}\,\psi(\ell) }\,,\quad R (-\g) ={1\ov R (\g)}\,,
\end{equation}
used in~\cite{Frolov:2023lwd,Frolov:2024pkz,Frolov:2025uwz, Frolov:2025tda} to solve the crossing equations.
In the expression above, $G$ is the Barnes Gamma function (`BarnesG' in Mathematica).
The $R$ function defined above satisfies 
\bal
\la{eq:properties_R_functions}
R (\g-2\pi i) =i\,  {\pi \ov \sinh{\g\ov2}}R(\g)\,,\quad R (\g+2\pi i) = i\, { \sinh{\g\ov2}\ov \pi }R(\g)\,,\quad R (\g+\pi i) =  { \cosh{\g\ov2}\ov \pi }R(\g-\pi i)\,.
\eal
and has a nice kernel, namely
\begin{equation}
\label{KernelR}
\frac{d}{dz} \log(R(z))= -\frac{i}{4 \pi} \left(z \coth\frac{z}{2} - \log(4 \pi^2) \right)\,.
\end{equation}
Moreover, $R(z)$ is a meromorphic function in the complex plane having poles located at 
\begin{equation}
\label{eq:R_pole_location}
z=-2\pi i n\,, \qquad n=1,\,2,\,\dots
\end{equation}
and zeros located at
\begin{equation}
z=+2\pi i n\,, \qquad n=1,\,2,\,\dots\,.
\end{equation}

\subsection{Dressing factors}
All the dressing factors can be expressed in terms of the previously defined functions. We split the dressing factors into: massive, mixed-mass and massless.

\paragraph{Massive dressing factors.}
Dressing factors for massive particles can be written in the following form\footnote{These phases are equivalent to the ones given in equation (C.6) of~\cite{Frolov:2021bwp}  after having shifted 
\begin{equation}
\gamma^+ \to \gamma^+ - \frac{i \pi}{2} \quad, \gamma^- \to \gamma^- + \frac{i \pi}{2}\,,
\end{equation}
according to the conventions used in that paper.
}
\begin{equation}
\label{eq:massive_dressing_factors}
\begin{split}
\left(\Sigma^{n n'}(u_1, u_2)\right)^{-2}&=\frac{R(\g_{12}^{++}+ i\pi) R(\g_{12}^{++}-i \pi) R(\g_{12}^{--}+ i\pi) R(\g_{12}^{--}-i \pi)}{R(\g_{12}^{+-}+ i\pi) R(\g_{12}^{+-}-i \pi) R(\g_{12}^{-+}+ i\pi) R(\g_{12}^{-+}-i \pi)} \left(\Sigma^{n n'}_\bes(u_1, u_2) \right)^{-2}\,,\\
\left(\widetilde{\Sigma}^{n n'}(u_1, u_2) \right)^{-2}&=\frac{R^2(\g_{12}^{+-}) R^2(\g_{12}^{-+})}{R^2(\g_{12}^{++})R^2(\g_{12}^{--})} \left(\Sigma^{n n'}_\bes(u_1, u_2) \right)^{-2}\,,
\end{split}
\end{equation}
where we defined
\begin{equation}
\g^{\mu \nu}_{12}\equiv \g^{\mu n}_1-\g^{\nu n'}_2 \, \quad \mu, \, \nu=\pm\,.
\end{equation}
When we cross a mirror cut, the only effect is that $\g$ is shifted by $\pm i\pi$; the sign depends on whether we cross the cut from above or below, according to the definition in~\eqref{eq:gamma_u}.

 \paragraph{Mixed-mass and massless dressing phases.}
 
 In~\cite{Frolov:2021bwp} the mixed-mass and massless phases have been expressed in terms of the Sine-Gordon dressing factor
 \begin{equation}
\varPhi (\gamma)=\frac{R(\gamma+ i \pi) R(\gamma- i \pi)}{R^2(\gamma)}\,.
\end{equation}
Writing these dressing factors explicitly in terms of Barnes functions, we obtain
\begin{equation}
\label{eq:massive_massless_full_dressing}
 \left(\Sigma^{n 0}(u_1, u_2) \right)^{-2}=-i \tanh \left(\frac{\g^{+ \circ}_{12}}{2} \right) \frac{R^2(\gamma_{12}^{-\circ})}{R^2(\gamma_{12}^{+\circ})}  \  \frac{R(\gamma_{12}^{+\circ}- i \pi) R(\gamma_{12}^{+\circ}+ i\pi)}{R(\gamma_{12}^{-\circ}- i \pi) R(\gamma_{12}^{-\circ}+ i \pi)} \big(\Sigma_\bes^{n 0}(u_1, u_2) \big)^{-2}\,,
\end{equation}
\begin{equation}
\left(\Sigma^{00}(u_1,u_2) \right)^{-2}= \frac{R^2( \g^{\circ \circ}_{12} - i \pi) R^2( \g^{\circ \circ}_{12} + i \pi)}{R^4(\g^{\circ \circ}_{12})}    \ \big(\Sigma_\bes^{00}(x_1,x_2) \big)^{-2}\,.
\end{equation}
Here, we removed the factor
\bal
\label{eq:a_g}
a(\g^{\circ \circ}_{12})=-i \tanh \left( \frac{\gamma^{\circ \circ}_{12}}{2} - \frac{i \pi}{4} \right) = \frac{(x_1 - x_2) -i (x_1 x_2 -1)}{(x_1 - x_2) + i (x_1 x_2 -1)} \,.
\eal
in the massless-massless S-matrix, in comparison with~\cite{Frolov:2021bwp}; such a factor was introduced in~\cite{Frolov:2021fmj} to remove a sign in the crossing equations. However, the function $a(\g^{\circ \circ}_{12})$ is not compatible with the QSC proposed in~\cite{Ekhammar:2024kzp} and cannot be generalised to the background in the presence of mixed RR and NSNS fluxes~\cite{Frolov:2025tda}. In the study of the Y system performed in this paper, this function would generate complicated cuts starting at non-standard positions, which are incompatible with the standard analytic Y-systems found in any  AdS/CFT model; this fact confirms the absence of the $a(\g^{\circ \circ}_{12})$ function. We should also mention that the sign ambiguity in the crossing equations detected in~\cite{Frolov:2021fmj} can be justified by assuming nontrivial exchange relations in the free theory limit~\cite{Frolov:2025ozz}, allowing for a solution to crossing without the function $a(\g^{\circ \circ}_{12})$.

\subsection{S matrices}
\label{app:S_mat_and_kernels}

Let us start by introducing a standard S matrix for bound states, namely
\bal
S^{nn'}(u-u')&= \left(\frac{u-u' -{i\ov h}(n+n')}{u-u' +{i\ov h}(n+n')} \right) \left(\frac{u-u' -{i\ov h}(n'-n)}{u-u' +{i\ov h}(n'-n)}\right)\\
 &\qquad\qquad\times\prod_{j=1}^{n-1}\left(\frac{u-u' -{i\ov h}(n'-n+2j)}{u-u' +{i\ov h}(n'-n+2j)}\right)^2  \, ,
 \eal
which is of difference form in terms of $u$. In particular
 \bal
S^{1n'}(u-u')&= \left( \frac{u-u' -{i\ov h}(1+n')}{u-u' +{i\ov h}(1+n')} \right) \left( \frac{u-u' -{i\ov h}(n'-1)}{u-u' +{i\ov h}(n'-1)} \right) \, . 
 \eal
Using the above expression and the dressing factors, we can define the following S matrices that appear in the fused Bethe-Yang equations.

\paragraph{Left-physical scattering.}
These are the scattering matrices where the first particle is of type ``left'':
\bal
S_{sl}^{n n'}( u_1, u_2)&= S^{n n'}(u_1-u_2)^{-1}\big(\Sigma^{n n'}(u_1 ,u_2) \big)^{-2}\,, 
\eal
\bal
{\St}_{sl}^{n n'}(  u_1, u_2) &=\frac{x_1^{+}}{x_1^{-}} \frac{1-\frac{1}{x^+_1x^+_2}}{1-\frac{1}{x^-_1x^-_2}}
    \frac{1-\frac{1}{x^+_1x^-_2}}{1-\frac{1}{x^-_1x^+_2}}  \big(\tSi^{nn'}(u_1, u_2)\big)^{-2}\,,
\eal
\bal
\label{eq:Sn0_SMat_app}
{S}^{n 0}(  u_1, u_2)  =\,i\, \sqrt{\frac{x_1^{-}}{x_1^{+}}}   \frac{x^+_1 x_2-1}{x^-_1-x_2} \big(\Sigma^{n 0}(u_1,u_2) \big)^{-2} \,.
\eal

\paragraph{Right-physical scattering.}
These are the scattering matrices where the first particle is of type ``right'':
\bal
S_{su}^{n n'}( u_1, u_2)
&=\frac{x_1^{+}}{x_1^{-}}\frac{x_2^{-}}{x_2^{+}}\left(  {x_1^+-x_{2}^-\ov  x_1^--x_2^+ }\right)^{-2} S^{n n'}(u_1-u_2)^{-1}\left( \Si^{n n'}(u_1, u_2)\right)^{-2} ,
\eal
\bal
{\St}_{su}^{n n'}(  u_1, u_2)    &=\frac{x_2^{-}}{x_2^{+}}
  \frac{1-\frac{1}{x^-_1 x^-_2}}{1-\frac{1}{x^+_1 x^+_2}}
    \frac{1-\frac{1}{x^+_1 x^-_2}}{1-\frac{1}{x^-_1 x^+_2}} \big(\tSi^{nn'}(u_1,u_2)\big)^{-2},
\eal     
 \bal
 {\bar S}^{n 0}(  u_1, u_2)  &=i\,\sqrt{\frac{x_1^{+}}{x_1^{-}}} \,  \frac{x^-_1-x_2}{x^+_1 x_2-1} \big(\Sigma^{n 0}(u_1, u_2) \big)^{-2} .
\eal

\paragraph{Massless-physical scattering.}
These are the scattering matrices where the first particle is massless:
\bal
S^{00}(u_1,u_2)=\left(\Sigma^{00}(u_1,u_2) \right)^{-2} ,
\eal
\bal
{S}^{0n}(  u_1, u_2)  {S}^{n 0}( u_2, x_1)=1\,,\quad {\Sb}^{0 n}(  u_1, u_2)  {\Sb}^{n 0}( u_2, u_1)=1 \,.
\eal

\section{Kernels}
\label{app:S_mat_and_kernels_part2}

The kernel $K_{ij}$ is defined through the corresponding S-matrix by 
\bal
K_{ij}(u_1,u_2)={1\ov 2\pi i}{d\ov du_1}\log S_{ij}(u_1,u_2)\,.
\eal
We split the kernels into physical and auxiliary, depending on whether they involve scattering only of physical particles or scattering with auxiliary excitations.

\subsection{Auxiliary kernels}
\label{app:aux_kernels}

Below, we list the kernels for the auxiliary particles
\bal
   \la{eq:SQy_def}
 {K}_+^{n y}( u_1, u_2) &= \frac{1}{2 \pi i } \frac{d}{d u_1} \ \log \left(\sqrt{\frac{x^{+n}(u_1)}{x^{-n}(u_1)}} \ \frac{x(u_1-\frac{i}{h}n) - x(u_2)}{x(u_1+\frac{i}{h}n) - x(u_2)} \right)\,,\\
  {K}_-^{n y}( u_1, u_2) &= \frac{1}{2 \pi i } \frac{d}{d u_1} \ \log \left(\sqrt{\frac{x^{+n}(u_1)}{x^{-n}(u_1)}} \ \frac{x(u_1-\frac{i}{h}n) - \frac{1}{x(u_2)}}{x(u_1+\frac{i}{h}n) - \frac{1}{x(u_2)}} \right)\,,
\eal
\bal
   \la{eq:SyQ_def}
 {K}_+^{y n}( u_1, u_2) &= \frac{1}{2 \pi i } \frac{d}{d u_1} \ \log \left(\sqrt{\frac{x^{-n}(u_2)}{x^{+n}(u_2)}} \ \frac{x(u_1) - x(u_2+\frac{i}{h}n)}{x(u_1) - x(u_2-\frac{i}{h}n)} \right)\,,\\
 {K}_-^{y n}( u_1, u_2) &= \frac{1}{2 \pi i } \frac{d}{d u_1} \ \log \left(\sqrt{\frac{x^{+n}(u_2)}{x^{-n}(u_2)}} \ \frac{\frac{1}{x(u_1)} - x(u_2-\frac{i}{h}n)}{\frac{1}{x(u_1)} - x(u_2+\frac{i}{h}n)} \right)\,.
\eal
The notation is the one in~\cite{Brollo:2023rgp} (eqs. (A.50)-(A.52)).
If we define the weighted Cauchy kernel  
\bal
\label{def:Kuniversal}
K(u, v) \equiv \frac{1}{2 \pi i} \frac{\sqrt{4-v^2}}{\sqrt{4 - u^2}} \, \frac{1}{u-v}
\eal
and
\bal
\label{def:KQ}
K_n(u) \equiv \frac{1}{\pi} \frac{h n}{n^2+h^2 u^2}\,, \qquad n=1, \, 2, \, \dots
\eal
then the kernels above satisfy the following relations (which are valid for arbitrary points $u_1$ and $u_2$ in the complex plane)
\bal
\label{Krationalrels}
&{K}_+^{n y}( u_1, u_2)+{K}_-^{n y}( u_1, u_2)=K_n(u_1-u_2)\,,\\
&{K}_+^{n y}( u_1, u_2)-{K}_-^{n y}( u_1, u_2)=K(u_1-\frac{i}{h}n,u_2)-K(u_1+\frac{i}{h}n,u_2)
\eal
and
\bal
\label{eq:disc_relations_Kpm_KQ_K_massive}
&{K}_+^{yn}( u_1, u_2)-{K}_-^{yn}( u_1, u_2)=K_n(u_1-u_2)\,,\\
&{K}_+^{yn}( u_1, u_2)+{K}_-^{yn}( u_1, u_2)=K(u_1,u_2+\frac{i}{h}n)-K(u_1,u_2-\frac{i}{h}n)\,.
\eal

The kernels of the massless particles correspond to the limit $n \to 0^+$ of the kernels above in which the $u$ associated to the physical particles takes values on the mirror cuts $(-\infty, -2) \cup (+2, +\infty)$. If this were not the case, then the kernels would be zero in this limit.
For $u_1 \in (-\infty, -2) \cup (+2, +\infty)$ the kernel is given by
\bal
 {K}_+^{0 y}( u_1, u_2) &= \frac{1}{2 \pi i } \frac{d}{d u_1} \ \log \left(\sqrt{\frac{x(u_1+i0)}{x(u_1-i0)}} \ \frac{x(u_1-i0) - x(u_2)}{x(u_1+i0) - x(u_2)} \right)\,,\\
  {K}_-^{0 y}( u_1, u_2) &= \frac{1}{2 \pi i } \frac{d}{d u_1} \ \log \left(\sqrt{\frac{x(u_1+i0)}{x(u_1-i0)}} \ \frac{x(u_1-i0) - \frac{1}{x(u_2)}}{x(u_1+i0) - \frac{1}{x(u_2)}} \right)\,.
\eal
Similarly for $u_2 \in (-\infty, -2) \cup (+2, +\infty)$ we have
\bal
 {K}_+^{y 0}( u_1, u_2) &= \frac{1}{2 \pi i } \frac{d}{d u_1} \ \log \left(\sqrt{\frac{x(u_2-i0)}{x(u_2+i0)}} \ \frac{x(u_1) - x(u_2+i0)}{x(u_1) - x(u_2-i0)} \right)\,,\\
 {K}_-^{y 0}( u_1, u_2) &= \frac{1}{2 \pi i } \frac{d}{d u_1} \ \log \left(\sqrt{\frac{x(u_2+i0)}{x(u_2-i0)}} \ \frac{\frac{1}{x(u_1)} - x(u_2-i0)}{\frac{1}{x(u_1)} - x(u_2+i0)} \right)\,.
\eal
Once these auxiliary-massless kernels are evaluated for values of $u$ of the massless particles on the mirror cuts, then we can continue them away from the cut. Continuing the kernels above the cut, we get
\bal
\label{eq:Kpm_to_universal}
 {K}_+^{0 y}( u_1, u_2) &= \frac{1}{2 \pi i } \frac{d}{d u_1} \ \log \left( \frac{1 - x(u_1) x(u_2)}{x(u_1) - x(u_2)} \right)=-K(u_1, u_2)\,,\\
  {K}_-^{0 y}( u_1, u_2) &= \frac{1}{2 \pi i } \frac{d}{d u_1} \ \log \left( \ \frac{x(u_1) - x(u_2)}{1-x(u_1) x(u_2)} \right)=+K(u_1, u_2)\,,\\
   {K}_+^{y 0}( u_1, u_2) &= \frac{1}{2 \pi i } \frac{d}{d u_1} \ \log \left(\frac{x(u_1) - x(u_2)}{1 - x(u_1) x(u_2)} \right)=+K(u_1, u_2)\,,\\
 {K}_-^{y 0}( u_1, u_2) &= \frac{1}{2 \pi i } \frac{d}{d u_1} \ \log \left( \frac{x(u_1) - x(u_2)}{1-x(u_1)x(u_2)} \right)=+K(u_1, u_2)\,.
\eal

The auxiliary kernels for massive particles satisfy
\bal
\label{eq:app_disc_auxmas_kernels}
K_+^{y n}(v, u \pm \frac{i}{h}n+ i0) - K_+^{y n}(v, u \pm \frac{i}{h}n- i0)= \mp K(v,u) \,,\\
K_-^{y n}(v, u \pm \frac{i}{h}n+ i0) - K_-^{y n}(v, u \pm \frac{i}{h}n- i0)= \mp K(v,u) \,.
\eal

\subsection{Massive kernels}
\label{subsec:massive_kernels}

By an explicit computation of the S-matrices previously listed, we can write the kernels for the scattering of physical particles in terms of the universal kernels in~\eqref{def:Kuniversal} and~\eqref{def:KQ}, and of the BES kernel obtained from~\eqref{eq:improved_BES_Q_Qp}
\bal
K^{n n'}_{\bes}(v,u) &\equiv \frac{1}{2 \pi i} \frac{d}{d v} \log \Sigma^{n n'}_{\bes}(v, u)\,.
\eal

We start by evaluating the dressing factors of massive excitations. We define the kernel 
\bal
K_{\rm mas}(v,u) \equiv \frac{1}{2 \pi i} \, \frac{4 - v u}{4 - v^2} \, \frac{1}{v-u}=\frac{1}{2 \pi i} \frac{d}{dv} \log \sinh \left( \g(v)-\g(u) \right)
\eal
appearing in the massive-massive interactions. Then it holds that
\bal
K_{R}(v,u)&\equiv \frac{1}{2 \pi i} \frac{d}{d v} \log R(\g(v)- \g(u))\\
&=-\frac{i}{4 \pi} \g_{vu} \left( K(v,u)+K_{\rm mas}(v,u) \right)  - \frac{\log 2 \pi}{ 2\pi^2} \frac{1}{4-v^2} \,,
\eal
where we define $\g_{vu} \equiv \g(v) - \g(u)$ and $R$ is given in~\eqref{eq:R_function}.
We also define 
\bal
K^+_{R}(v,u)&\equiv \frac{1}{2 \pi i} \frac{d}{d v} \log R(\g(v)- \g(u)+ i \pi)\\
&=\frac{i}{4 \pi} (\g_{vu} + i \pi) \left( K(v,u)-K_{\rm mas}(v,u) \right)  - \frac{\log 2 \pi}{ 2 \pi^2} \frac{1}{4-v^2}\,,
\eal
and
\bal
K^-_{R}(v,u)&\equiv \frac{1}{2 \pi i} \frac{d}{d v} \log R(\g(v)- \g(u)- i \pi)\\
&=\frac{i}{4 \pi} (\g_{vu} - i \pi) \left( K(v,u)-K_{\rm mas}(v,u) \right)  - \frac{\log 2 \pi}{ 2 \pi^2} \frac{1}{4-v^2}\,.
\eal
Using the notation $v^{\pm n}=v \pm \frac{i}{h}n$ then the kernels of the massive dressing factors are
\bal
\label{eq:app_K_dressing1}
&\frac{1}{2 \pi i} \frac{d}{d v} \log \left(\Sigma^{n n'}(v, u) \right)^{-2}=-2 \, K_\bes^{n n'}(v, u)+K^{n n'}_{\rm Barnes}(v,u)\,,
\eal
\bal
\label{eq:app_K_dressing2}
&\frac{1}{2 \pi i} \frac{d}{d v} \log \left(\widetilde{\Sigma}^{n n'}(v, u) \right)^{-2}=-2 \, K_\bes^{n n'}(v, u) + \widetilde{K}^{n n'}_{\rm Barnes}(v,u)\,,
\eal
where
\bal
\label{eq:app_K_Barnes1}
K^{n n'}_{\rm Barnes}(v,u)=&+ K^+_R (v^{+n}, u^{+n'})+K^+_R (v^{-n}, u^{-n'})-K^+_R (v^{+n}, u^{-n'})-K^+_R (v^{-n}, u^{+n'})\\
&+ K^-_R (v^{+n}, u^{+n'})+K^-_R (v^{-n}, u^{-n'})-K^-_R (v^{+n}, u^{-n'})-K^-_R (v^{-n}, u^{+n'})
\eal
and
\bal
\label{eq:app_K_Barnes2}
\widetilde{K}^{nn'}_{\rm Barnes}(v,u)=& 2 K_R (v^{+n}, u^{-n'})+2 K_R (v^{-n}, u^{+n'})-2 K_R (v^{+n}, u^{+n'})- 2 K_R (v^{-n}, u^{-n'}) \,.
\eal

The improved HL phase can be written as
\bal
\left( \Sigma^{n n'}_\hl(v,u) \right)^2=\frac{R(\g_{12}^{++}+ i\pi) R(\g_{12}^{++}-i \pi) R(\g_{12}^{--}+ i\pi) R(\g_{12}^{--}-i \pi)}{R(\g_{12}^{+-}+ i\pi) R(\g_{12}^{+-}-i \pi) R(\g_{12}^{-+}+ i\pi) R(\g_{12}^{-+}-i \pi)} \frac{R^2(\g_{12}^{+-}) R^2(\g_{12}^{-+})}{R^2(\g_{12}^{++})R^2(\g_{12}^{--})} \,,
\eal
and its kernel is given by
\bal
&2K^{n n'}_{\hl}(v,u)=K^{n n'}_{\rm Barnes}(v,u)+\widetilde{K}^{n n'}_{\rm Barnes}(v,u)=\\
&\frac{1}{2 \pi i} \log \biggl( \frac{R(\g_{12}^{++}+ i\pi) R(\g_{12}^{++}-i \pi) R(\g_{12}^{--}+ i\pi) R(\g_{12}^{--}-i \pi)}{R(\g_{12}^{+-}+ i\pi) R(\g_{12}^{+-}-i \pi) R(\g_{12}^{-+}+ i\pi) R(\g_{12}^{-+}-i \pi)} \frac{R^2(\g_{12}^{+-}) R^2(\g_{12}^{-+})}{R^2(\g_{12}^{++})R^2(\g_{12}^{--})} \biggr) \,.
\eal

\subsection{Mixed-mass and massless kernels}

For the mixed mass scattering, we find
\bal
\label{eq:leftmassive_massless_in_Y0_from_TBA}
K^{n0}(v, u)= &-2 \, K_\bes^{n0}(v, u)-\frac{1}{2}  K_{n}(v-u) \\
&+\frac{1}{i \pi}  \left( \g^{-n \circ}_{v u} - \frac{i \pi}{2}\right) K(v- \frac{i}{h}n,u)-\frac{1}{i \pi}  \left( \g^{+n \circ}_{v u} - \frac{i \pi}{2}\right) K(v+ \frac{i}{h}n,u) \,,
\eal
\bal
\label{eq:rightmassive_massless_in_Y0_from_TBA}
\tK^{n0}(v, u)=&-2 \, K_\bes^{n0}(v, u)+\frac{1}{2} K_{n}(v-u)\\
&+\frac{1}{i \pi} \left( \g^{-n \circ}_{v u} + \frac{i \pi}{2} \right) K(v- \frac{i}{h}n,u)-\frac{1}{i \pi} \left( \g^{+n \circ}_{v u} - \frac{3}{2} i \pi\right) K(v+ \frac{i}{h}n,u) \,.
\eal
For the massless-massless scattering, we find
\bal
\label{eq:massless_massless_in_Y0_from_TBA}
K^{00}(v, u)=-2 \, K_\bes^{00}(v, u)+\frac{2 i}{\pi} K(v, u) \g^{\circ \circ}_{vu} \,.
\eal

\section{Discontinuities for auxiliary \texorpdfstring{$Y_{\pm }^{(\alpha)}$}{Ypmalpha}}
\label{appendixYpm}

\subsection{Analytic continuation across the cuts on the real axis}
In this appendix, we prove the relations~\eqref{eq:Yaux_product} and  \eqref{eq:Yaux_ratio}.
We will discuss the analytic continuation of $Y_-^{(\alpha)}$ in detail, since the other relations can be obtained similarly.

\paragraph{Computing $\left(Y_-^{(\alpha)}\right)^{\bar\gamma}$.}

We start with $u \in (-2, +2)$ in the TBA equation~\eqref{Y-particles} and go around the branch point at  $u=+2$ clockwise until we return to the starting point. In this process, we pass under the cut $(+2, +\infty)$ from above.

 As a result of this analytic continuation, we need to pick up a residue from the following integral appearing in the massless term of the TBA equation~\eqref{Y-particles}:
\bal
\label{Ypmcont}
\left(\log\left(1+Y_{0}\right)\cstar K_+^{0y} \right)(u)= \left(\int_{-\infty}^{-2} + \int_{+2}^{+\infty} \right) {\rm d}v \log\left(1+Y_{0}(v)\right) K_+^{0y}(v,u)\,.
\eal
Starting with $u \in (-2, +2)$, we move to the upper edge of the mirror cut  $(+2, +\infty)$. When $u$ is above the cut, we have
\bal
x(v+i0)=x(u)\,,
\eal
and the kernel ${K}_+^{0 y}( v, u)$ develops a pole with residue equal to $-1$.
This pole crosses the integration line from above when $u$ approaches the line $(+2, +\infty)$ from above. Then we have the following continuation for~\eqref{Ypmcont} right before $u$ enters the mirror cut:
\bal
\left(\log\left(1+Y_{0}\right)\cstar K_+^{0y} \right)(u) \to \left(\log\left(1+Y_{0}\right)\cstar K_+^{0y} \right)(u) - \log\left(1+Y_{0}(u)\right)\,.
\eal
After this pole crosses the integration line, $u$ crosses the cut and so $x(u) \to \frac{1}{x(u)}$, while $Y_0(u)\rightarrow Y^{\bar\gamma}_0(u)$. Then we do not encounter any further singularity. Moreover, under the same analytic continuation, we have in the TBA equations \eqref{Y-particles} the changes in the massive-auxiliary kernels $K_+^{ny} \to K_-^{ny}$ and $K_-^{ny} \to K_+^{ny}$; hence:
\bal
\left(\log\left(1+Y_{0}\right)\cstar K_+^{0y} \right)(u) \to \left(\log\left(1+Y_{0}\right)\cstar K_-^{0y} \right)(u) - \log\left(1+Y_{0}^{\bar\gamma}\right)(u)\,.
\eal
The first term, together with the other terms in \eqref{Y-particles} (with $K_\pm^{ny}\rightarrow K_\mp^{ny}$), gives rise to the TBA equation~\eqref{Y+particles} for $Y_+^{(\alpha)}$.
Overall, after exponentiating, we obtain
\bal
\label{eq:app_discYm_non_LR}
\left(Y_-^{(\a)}\right)^{\bar\gamma} = \frac{Y_+^{(\a)}}{ \left(1 + Y^{\bar\gamma}_0\right)}.
\eal
We are keeping the same index $\a$ on both sides of the equation. However, since the TBA equation for $Y^{(\a)}_\pm$ does not depend on $\a$ one can write the continuation as
\bal
\left(Y_-^{(\a)}\right)^{\bar\gamma} = \frac{Y_+^{(\beta)}}{ \left(1 + Y^{\bar\gamma}_0\right)}, \qquad \a , \beta \in \{1, 2\} \,.
\eal
This is because the ground state is left-right symmetric, with $Y^{(1)}_\pm = Y^{(2)}_\pm$, and the discontinuity equations derived from it are restricted to this left-right symmetric sector. The natural generalisation to states that are not left-right symmetric is~\eqref{eq:app_discYm_non_LR}, where the same index $\a$ must appear both on the LHS and RHS of the equation.

\paragraph{Computing $\left(Y_-^{(\alpha)}\right)^{\gamma}$.}

Let us now start with $u \in (-2, +2)$ and continue the same equation in the counterclockwise direction across the cut $(+2, +\infty)$. In this case, we encounter a pole of ${K}_+^{0 y}( v, u)$ with residue equal to $+1$ \emph{after} passing through the cut $(+2, +\infty)$ from below. Then we obtain the following continuation for~\eqref{Ypmcont}:
\bal
\left(\log\left(1+Y_{0}\right)\cstar K_+^{0y} \right)(u) \to \left(\log\left(1+Y_{0}\right)\cstar K_-^{0y} \right)(u) - \log\left(1+Y_{0}(u)\right)\,,
\eal
with the same modifications to the rest of the TBA equation as in the $\bar\gamma$ continuation described in the last paragraph. Therefore, we obtain
\bal
\left(Y_-^{(\a)}\right)^\gamma = \frac{Y_+^{(\a)}}{ \left(1 + Y_0 \right)}.
\eal
The same index considerations just mentioned for $\left(Y_-^{(\alpha)}\right)^{\bar\gamma}$ apply also to this case.

\paragraph{Computing $\left(Y_+^{(\alpha)}\right)^{\bar\gamma}$ and $\left(Y_+^{(\alpha)}\right)^{\gamma}$.}

The same calculations can be repeated for the equations involving $Y^{(\alpha)}_+$, obtaining:
\beq
\left(Y_+^{(\a)}\right)^{\bar\gamma} = {Y_-^{(\a)}}{ \left(1 + Y^{\bar\gamma}_0\right)},\qquad \left(Y_+^{(\a)}\right)^{\gamma} = {Y_-^{(\a)}}{ \left(1 + Y_0\right)}.
\eeq
Relations~\eqref{eq:Yaux_product} and \eqref{eq:Yaux_ratio} come by respectively multiplying and dividing the expressions for $Y_+^{(\alpha)}$ and $Y_-^{(\alpha)}$.

\subsection{Analytic continuation across the cuts at \texorpdfstring{$\Im(u)=\pm 2M/h$}{Imupm2Movh}}
\label{appendixdiscpm}

In this appendix, we derive the discontinuity relations across the cuts at $\Im(u)=\pm 2M/h$ for the auxiliary Y functions.

We start with the TBA equation for $Y_-^{(\alpha)}$ \eqref{Y-particles} and consider $u$ just above the interval $(-2,+2)$.
Then we shift $u$ upward by $i\frac{2M}{h}$; in the TBA equation the pole located at
\bal
v=u-\frac{i}{h}n\,,
\eal
coming from the denominator in~$\log K^{ny}_+(v,u)$ (see definition in~\eqref{eq:SQy_def}) crosses the integration line for all values of $n<2M$. Therefore we need to pick the residue of this pole, obtaining
\bal
\la{eq:disc_LogYpM1}
\log Y_-^{(\a)}(u+\frac{i}{h}2M ) =  &- \left(\log\left(1+Y_{n} \right)\star K^{ny}_- \right)(u+\frac{i}{h} 2M )\\
&+ \left(\log\left(1+ \bY_{n} \right)\star K^{ny}_+ \right)(u+\frac{i}{h} 2M)   \\
&+  \left(\log\left(1+Y_{0}\right)\cstar K_+^{0y} \right)(u+\frac{i}{h}2M )\\
&+\sum^{2M}_{n=1}\log\left(1+\bY_{n}(u+\frac{i}{h}(2M-n) ) \right)\,.
\eal
Performing the same continuation on the TBA equation for $Y_+^{(\alpha)}$ \eqref{Y+particles} we get:
\bal
\la{eq:disc_LogYmM1}
\log Y_+^{(\a)}(u+\frac{i}{h}2M ) =  &- \left(\log\left(1+Y_{n} \right)\star K^{ny}_+ \right)(u+\frac{i}{h} 2M )\\
&+ \left(\log\left(1+ \bY_{n} \right)\star K^{n y}_- \right)(u+\frac{i}{h} 2M )   \\
&-  \left(\log\left(1+Y_{0}\right)\cstar K_+^{0y} \right)(u+\frac{i}{h} 2M )\\
&-\sum^{2M}_{n=1}\log\left(1+Y_{n}(u+\frac{i}{h}(2M-n) ) \right)\,.
\eal
Using the definition in~\eqref{eq:def_discontin_f_N} together with the fact that the shifted kernels involved in the expression above have no cuts for real $u$, and that the functions $Y_{n}$ and $\bY_{n}$ are analytic in the strip $-\frac{n}{h}<u<\frac{n}{h}$ we obtain the following discontinuity relations
\bal
\la{eq:disc_LogYpM12dis}
\left[ \log Y_-^{(\a)} \right]_{2M}=  &+\sum^{M}_{n=1} \left[\log\left(1+ \bY_{n} \right) \right]_{2M-n}\,,\\
\left[ \log Y_+^{(\a)} \right]_{2M}=  &-\sum^{M}_{n=1} \left[\log\left(1+Y_{n} \right) \right]_{2M-n}.
\eal
Performing the opposite shift $u\rightarrow u-i\dfrac{2M}{h}$, we obtain the same result; we just need to replace $2M-n\to -(2M-n)$.
Taking the difference and the sum of the two expressions above, we obtain the relations in~\eqref{eq:disc_Ymp_2M}, respectively.

\section{BES and kernel discontinuities}
\label{appendixBES}

In this appendix, we derive different properties of the BES kernel for massive and massless particles. We also provide useful relations for the discontinuities of the kernels of physical particles.

\subsection{BES kernel as a double convolution}
Let us recall the definition of the improved BES kernel
\bal
\label{eq:app_BES_kernel}
K^{n n'}_{\bes}(v,u) &\equiv \frac{1}{2 \pi i} \frac{d}{d v} \log \Sigma^{n n'}_{\bes}(v, u)\,.
\eal
Using the definition of the improved BES phase in~\eqref{eq:improved_BES_Q_Qp}, we can write
\bal
K^{n n'}_{\bes}(v,u) &=\frac{1}{2\pi i}\int_{\text{cuts}}\frac{ \text{d} v_1}{2\pi i}\int_{\text{cuts}} \frac{\text{d} v_2}{2\pi i} \log \frac{\Gamma \bigl[1+\frac{ih}{2}\big(v_1-v_2\bigl)\bigl]}{\Gamma \bigl[1-\frac{ih}{2}\big(v_1-v_2\bigl)\bigl]}\\
&\frac{d}{d v} \biggl( +\frac{x'(v_1) \, x'(v_2)}{(x(v_1)-x^{+n}(v))(x(v_2)-x^{+n'}(u))}+\frac{x'(v_1) \, x'(v_2)}{(x(v_1)-x^{-n}(v))(x(v_2)-x^{-n'}(u))}\\
&- \frac{x'(v_1) \, x'(v_2)}{(x(v_1)-x^{+n}(v))(x(v_2)-x^{-n'}(u))}-\frac{x'(v_1) \, x'(v_2)}{(x(v_1)-x^{-n}(v))(x(v_2)-x^{+n'}(u))} \biggr)\,.
\eal
Integrating by parts wrt $v_1$ we obtain:
\bal
K^{n n'}_{\bes}(v,u) =\frac{1}{2\pi i}\int_{\text{cuts}}&\frac{ \text{d} v_1}{2\pi i}\int_{\text{cuts}} \frac{\text{d} v_2}{2\pi i} \frac{d}{d v_1}\log \frac{\Gamma \bigl[1+\frac{ih}{2}\big(v_1-v_2\bigl)\bigl]}{\Gamma \bigl[1-\frac{ih}{2}\big(v_1-v_2\bigl)\bigl]}\\
&\frac{d}{d v} \log \left( \frac{x(v_1) - x^{+n}(v)}{x(v_1) - x^{-n}(v)} \right) \ \frac{d}{d v_2} \log \left( \frac{x(v_2) - x^{-n'}(u)}{x(v_2) - x^{+n'}(u)} \right) \,.
\eal
We notice that the expression above is equivalent to
\bal
K^{n n'}_{\bes}(v,u) =&\frac{1}{2\pi i}\int_{\text{cuts}}\frac{ \text{d} v_1}{2\pi i}\int_{\text{cuts}} \frac{\text{d} v_2}{2\pi i} \frac{d}{d v_1}\log \frac{\Gamma \bigl[1+\frac{ih}{2}\big(v_1-v_2\bigl)\bigl]}{\Gamma \bigl[1-\frac{ih}{2}\big(v_1-v_2\bigl)\bigl]}\\
&\frac{d}{d v} \log \left( \sqrt{\frac{x^{+n}(v)}{x^{-n}(v)}} \frac{x(v_1) - x^{-n}(v)}{x(v_1) - x^{+n}(v)} \right) \ \frac{d}{d v_2} \log \left( \sqrt{\frac{x^{-n'}(u)}{x^{+n'}(u)}} \frac{x(v_2) - x^{+n'}(u)}{x(v_2) - x^{-n'}(u)} \right)\,.
\eal
The additional terms $\sqrt{\frac{x^{+n}(v)}{x^{-n}(v)}}$ and $\sqrt{\frac{x^{-n'}(u)}{x^{+n'}(u)}}$ only contribute to the integral with total derivatives and are harmless.
Finally, we obtain
\bal
\la{eq:masmas_BES_Kernel_as_double_conv}
K^{n n'}_{\bes}(v,u) =\frac{1}{2\pi i}\int_{\text{cuts}}& \text{d} v_1\int_{\text{cuts}} \text{d} v_2 \ K_+^{ny}(v, v_1) \ \frac{d}{d v_1}\log \frac{\Gamma \bigl[1+\frac{ih}{2}\big(v_1-v_2\bigl)\bigl]}{\Gamma \bigl[1-\frac{ih}{2}\big(v_1-v_2\bigl)\bigl]} \ K_+^{y n'}(v_2, u)\,,
\eal
where the kernels can be read from~\eqref{eq:SQy_def} and~\eqref{eq:SyQ_def}. The integrals are performed around the long cuts $(-\infty, -2)$ and $(+2, + \infty)$, and we integrate to the right on the lower edge of the cuts (opposite to the contour of figure \ref{im:path_gamma_for_iniv_Y0aux}).

\begin{figure}
\begin{center}
\includegraphics*[width=1\textwidth]{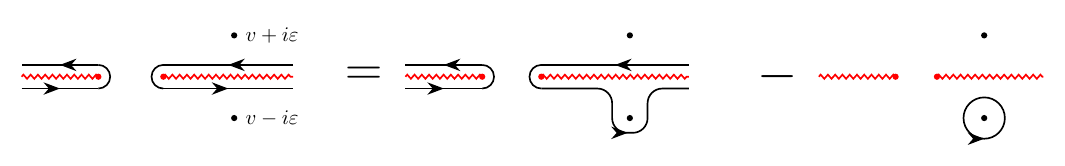} 
\end{center}
\caption{Contour deformation to move $v - i\epsilon$ above the cut in the BES kernel $K^{0P}_{\bes}(v,u)$.} 
\label{im:bes_def_for_massless}
\end{figure}

If the first (and/or second) particle is massless, we just need to take the limit $n \to 0^+$ ($n' \to 0^+$). We recall that the massless Zhukovsky variables $x^{\pm 0}(u)=x(u\pm i \epsilon)$
are evaluated just above and below the integration lines around the long cuts.
To better deal with the BES phase for massless particles, we write both Zhukovsky variables as functions of $u+ i\epsilon$ (a point above the integration contour around the cuts):
\bal
x^{+0}(u)=x(u+i \epsilon)\,, \qquad x^{-0}(u)=x(u-i \epsilon)=\frac{1}{x(u+i \epsilon)}\,.
\eal
In this manner both Zhukovsky variables are evaluated at the same point, which we consider to be slightly above the integration contour around the mirror cuts.
However, when doing this operation we need to pick up the residues of the integrand.
Suppose for example that the first particle is massless; then from~\eqref{eq:SQy_def} in the limit $n \to 0^+$ we have 
\bal
K^{ny}_+(v, v_1) \to \frac{1}{2 \pi i } \frac{d}{d v} \ \log \left(\sqrt{\frac{x(v+ i \eps)}{x(v - i \eps)}} \ \frac{x(v-i \eps) - x(v_1)}{x(v+i \eps) - x(v_1)} \right)\,,
\eal
and this kernel has poles at $ v_1 = v \pm i \epsilon$, above and below the integration contour around the mirror cut (see figure~\ref{im:bes_def_for_massless}).
To write both Zhukovsky variables as functions of $v+ i\eps$, we deform the contour around the point $v- i \eps$, as shown on the r.h.s. of figure~\ref{im:bes_def_for_massless}. Then we can write $x(v-i \epsilon)=\frac{1}{x(v+i \epsilon)}$, since moving $v$ above the cut ensures that no poles intersect the contour. In this manner, we obtain
\bal
\la{eq:BES_Kernel_as_double_conv_0P}
K^{0n'}_{\bes}(v,u) =&-\frac{1}{2\pi i}\int_{\text{cuts}} \text{d} v_1\int_{\text{cuts}} \text{d} v_2 \ K(v, v_1) \ \frac{d}{d v_1}\log \frac{\Gamma \bigl[1+\frac{ih}{2}\big(v_1-v_2\bigl)\bigl]}{\Gamma \bigl[1-\frac{ih}{2}\big(v_1-v_2\bigl)\bigl]} \ K_+^{yn'}(v_2, u)\\
&-\int_{\text{cuts}} \text{d} v_2  \ \frac{d}{d v}\log \frac{\Gamma \bigl[1+\frac{ih}{2}\big(v-v_2\bigl)\bigl]}{\Gamma \bigl[1-\frac{ih}{2}\big(v-v_2\bigl)\bigl]} \ K_+^{y n'}(v_2, u) \,,
\eal
where we used the identity in~\eqref{eq:Kpm_to_universal} and $v$ is a point above the integration contour.
Performing a similar computation for the second variable, we get
\bal
\la{eq:BES_Kernel_as_double_conv_Q0}
K^{n0}_{\bes}(v,u) =&+\frac{1}{2\pi i}\int_{\text{cuts}} \text{d} v_1\int_{\text{cuts}} \text{d} v_2 \ K_+^{ny}(v, v_1) \ \frac{d}{d v_1}\log \frac{\Gamma \bigl[1+\frac{ih}{2}\big(v_1-v_2\bigl)\bigl]}{\Gamma \bigl[1-\frac{ih}{2}\big(v_1-v_2\bigl)\bigl]} \ K(v_2, u)\\
&+\frac{1}{2\pi i}\int_{\text{cuts}} \text{d} v_1 \ K_+^{ny}(v, v_1) \ \frac{d}{d v_1}\log \frac{\Gamma \bigl[1+\frac{ih}{2}\big(v_1-u\bigl)\bigl]}{\Gamma \bigl[1-\frac{ih}{2}\big(v_1-u\bigl)\bigl]}\,,
\eal
with $u$ above the integration contour
and
\bal
\la{eq:BES_Kernel_as_double_conv_00}
K^{00}_{\bes}(v,u) =&-\frac{1}{2\pi i}\int_{\text{cuts}} \text{d} v_1\int_{\text{cuts}} \text{d} v_2 \ K(v, v_1) \ \frac{d}{d v_1}\log \frac{\Gamma \bigl[1+\frac{ih}{2}\big(v_1-v_2\bigl)\bigl]}{\Gamma \bigl[1-\frac{ih}{2}\big(v_1-v_2\bigl)\bigl]} \ K(v_2, u)\\
&-\frac{1}{2 \pi i}\int_{\text{cuts}} \text{d} v_2  \ \frac{d}{d v}\log \frac{\Gamma \bigl[1+\frac{ih}{2}\big(v-v_2\bigl)\bigl]}{\Gamma \bigl[1-\frac{ih}{2}\big(v-v_2\bigl)\bigl]} \ K(v_2, u) \\
&-\frac{1}{2 \pi i}\int_{\text{cuts}} \text{d} v_1 \ K(v, v_1) \ \frac{d}{d v_1}\log \frac{\Gamma \bigl[1+\frac{ih}{2}\big(v_1-u\bigl)\bigl]}{\Gamma \bigl[1-\frac{ih}{2}\big(v_1-u\bigl)\bigl]}-\frac{1}{2 \pi i} \frac{d}{dv}\frac{\Gamma \bigl[1+\frac{ih}{2}\big(v-u\bigl)\bigl]}{\Gamma \bigl[1-\frac{ih}{2}\big(v-u\bigl)\bigl]}\,,
\eal
with both $v$ and $u$ above the contour.

\subsection{Massless and mixed-mass BES as single integrals}

Let us perform the integral wrt $v_2$ in the kernel $K^{n0}_{\bes}(v,u)$. By using Cauchy's theorem, we obtain
\bal
\la{eq:BES_Kernel_as_double_conv_Q02}
K^{n0}_{\bes}(v,u) =&\sum_{j=1}^{+\infty}\int_{\text{cuts}} \text{d} v_1   K_+^{ny}(v, v_1) \left( K(v_1+\frac{2i}{h} j, u) - K(v_1-\frac{2i}{h} j, u) \right)\,,
\eal
where the pole at $v_2=u$ cancels the second line of~\eqref{eq:BES_Kernel_as_double_conv_Q0}. The result is a sum of infinitely many terms, associated with the poles of the derivatives of the $\log \Gamma$ functions. A similar computation in the kernel $K^{00}_{\bes}(v,u)$ leads to
\bal
\la{eq:BES_Kernel_as_double_conv_002}
K^{00}_{\bes}(v,u) =&+\sum^{+\infty}_{j=1}\int_{\text{cuts}} \text{d} v_1 \, K(v, v_1)  \, \left(K(v_1-\frac{2i}{h}j, u) - K(v_1+\frac{2i}{h}j, u) \right)\\
&+ \sum^{+\infty}_{j=1} \left(K(v- \frac{2i}{h}j, u) - K(v+ \frac{2i}{h}j, u) \right) \,.
\eal

\subsection{Discontinuities of mixed-mass BES kernels}

Let us consider the kernel $K^{n0}_{\bes}(v,u)$ in~\eqref{eq:BES_Kernel_as_double_conv_Q02} and shift $u \to u+\frac{2i}{h} N + i 0$, keeping $|\Re(u)|<2$. {Then, we continue $u \rightarrow u^{\gamma}$. In this process, the pole $v_1=u$ of the first term in parentheses in~\eqref{eq:BES_Kernel_as_double_conv_Q02} crosses the integration contour along the lower edge of the cut, yielding the residue $-K^{ny}_{+}(v,u)$.
To close the analytic continuation following $\gamma$, this residue needs to be dragged through the cut and  become $
 -K^{ny}_{-}(v,u)
$
 evaluated above the cut in $u$. Moreover, the same pole of the integration kernel in \eqref{eq:BES_Kernel_as_double_conv_Q02} also passes through the integration contour on the upper edge of the cut, from below. In this case, it picks up a residue $+K_+^{ny}(v, u)$. 
 After this continuation, the kernel comes back to itself, so that the remaining convolution reproduces the one in \eqref{eq:BES_Kernel_as_double_conv_Q02}. 
}
 Overall, we get, for $|\Re(u)|>0$, $\Im(u) = 0$: 
\bal
\label{eq:kerancont1}
&K_\bes^{n0}(v , u + \frac{2i}{h}N -i 0)= K_\bes^{n0}(v , u + \frac{2i}{h}N +i 0)+K^{ny}_{+}(v,u)-K^{ny}_{-}(v,u)\,.
\eal
If we do the same procedure for the cut at $\Im(u)=-\frac{2}{h} N$, we obtain instead
\bal
\label{eq:kerancont2}
&K_\bes^{n0}(v , u - \frac{2i}{h}N -i 0) = K_\bes^{n0}(v , u - \frac{2i}{h}N +i 0)+K^{ny}_{-}(v,u)-K^{ny}_{+}(v,u)\,.
\eal
From the relations above, we obtain therefore, taking the discontinuity with respect to the $u$ variable:
\bal
 [K_\bes^{n0}(v,u)]_{\pm 2N}&=K_\bes^{n0}(v , u\pm \frac{2i}{h}N +i 0) - K_\bes^{n0}(v , u\pm \frac{2i}{h}N -i 0)\\
 &= \mp \left(K^{ny}_+(v, u) - K^{ny}_-(v, u) \right) \,,
\eal
where the second term after the first equality is given by \eqref{eq:kerancont1} if we are at $+2N$ and by \eqref{eq:kerancont2} if we are at $-2N$.
It also holds that, for any function $f(v)$ that makes the convolution convergent: 
\bal
\label{eq:app_conv_KQ0f_discpm2N}
 &[ f \star K_\bes^{n0}(u)]_{\pm 2N}= \mp  f  \star \left(K^{ny}_+  - K^{ny}_-  \right)(u) \,,
\eal
where we used that no residues are generated in the continuation.
On the other hand, it is possible to prove that on the real axis the discontinuity is of square root type:
\bal
K^{n0}_{\bes}(u_1,u_2) - K^{n0}_{\bes}(u_1,u^\gamma_2)=2K^{n0}_{\bes}(u_1,u_2)\,.
\eal
This simply follows from the continuation of
$K(v_1 \pm \frac{2i}{h} j, u)$ across their square-root cut in the $u$ plane (see definition in~\eqref{def:Kuniversal}).
An immediate consequence is that the convolution of any function $f(v)$ with the kernel in \eqref{eq:BES_Kernel_as_double_conv_Q02} has no symmetrised discontinuity on the real axis:
\beq
\label{eq:disc_real_axis_K0nbes}
\{f \cstar K^{n0}_{\bes}\}_0=0\,,
\eeq
where we also used the fact that no residues are generated in the continuation.

\subsection{Discontinuities of massless BES kernels}

The discontinuities of the massless BES kernel are similarly computed starting from~\eqref{eq:BES_Kernel_as_double_conv_002}.
The only difference compared with the mixed mass case is that now the second row of~\eqref{eq:BES_Kernel_as_double_conv_002} (which was absent in the mixed-mass kernel) produces an extra term when it is convoluted with a certain function $f$ on the mirror cuts. Let us consider this term, since the rest of the computation is identical to that just performed for $K^{n0}$. 
Suppose to have
\bal
\label{eq:app_extra_K00_disc1}
\sum^{+\infty}_{j=1} \left(\int^{-2}_{-\infty}+ \int^{+ \infty}_{+2} \right)  dv \, f(v) \left(K(v- \frac{2i}{h}j, u) - K(v+ \frac{2i}{h}j, u) \right)
\eal
with $u$ above the integration contour. We shift $u \to u+\frac{2i}{h}N + i 0$ and then continue $u\rightarrow u^{\gamma}$. 
 While doing this continuation, the pole $v=u$ due to the second term in~\eqref{eq:app_extra_K00_disc1} (for $j=N$) crosses the integration contour from below, and the expression above must be continued as follows
\bal
\label{eq:app_extra_K00_disc2}
\sum^{+\infty}_{j=1} \left(\int^{-2}_{-\infty}+ \int^{+ \infty}_{+2} \right)  dv \, f(v) \left(K(v- \frac{2i}{h}j, u+\frac{2i}{h}N) - K(v+ \frac{2i}{h}j, u+\frac{2i}{h}N) \right)+f(u)\,.
\eal
Similarly, if we shift $u \to u - \frac{2i}{h}N$ and cross the $-2N$ cut from below, we get a residue from the first term in~\eqref{eq:app_extra_K00_disc1} and obtain
\bal
\label{eq:app_extra_K00_disc3}
\sum^{+\infty}_{j=1} \left(\int^{-2}_{-\infty}+ \int^{+ \infty}_{+2} \right)  dv \, f(v) \left(K(v- \frac{2i}{h}j, u-\frac{2i}{h}N) - K(v+ \frac{2i}{h}j, u-\frac{2i}{h}N) \right)-f(u)\,.
\eal
Our final expression for the discontinuity across the long cuts at $\pm 2N$ is
\bal
\label{eq:app_conv_K00f_discpm2N}
 &[f(v) \cstar K_\bes^{00}(v,u)]_{\pm 2N}= \pm 2  f(v) \cstar K(v, u) \mp f(u) \,.
\eal
On the other hand, like in the mixed mass case, the kernels $K(v\pm \frac{2i}{h} j,u)$ are continued to $-K(v\pm \frac{2i}{h} j,u)$ when $u$ crosses the cut $(+2,+\infty)$. Then, using~\eqref{eq:BES_Kernel_as_double_conv_002}, also in this case we have
\bal
K^{00}_{\bes}(u_1,u_2) - K^{00}_{\bes}(u_1,u^\gamma_2)=2K^{00}_{\bes}(u_1,u_2)\,.
\eal
As in the previous section, we obtain
\beq
\label{eq:disc_real_axis_K00bes}
\{f \cstar K^{00}_{\bes}\}_0=0\,.
\eeq

\subsection{Massless BES as a discontinuity of massive BES}
\label{app:masslessBES_as_disc_massiveBES}

We start by defining the discontinuity of the BES kernel $K^{n1}$ along the cuts at $\Im (u)=1/h$. Defining $u^\gamma$ to be the same point as $u$ but after having crossed the cut $(+2+\frac{i}{h}, +\infty+\frac{i}{h})$ from below (or equivalently the cut $(-\infty+\frac{i}{h}, -2+\frac{i}{h})$ from below) we find that
\bal
\la{eq:disc_BES_KQmassive}
&K^{n1}_{\bes}(v,u+\frac{i}{h}+i 0) - K^{n1}_{\bes}(v,u^\gamma+\frac{i}{h}+i 0)=-K_\bes^{n0}(v, u)\,,\\
&K^{01}_{\bes}(v,u+\frac{i}{h}+i 0) - K^{01}_{\bes}(v,u^\gamma+\frac{i}{h}+i 0)=-K_\bes^{00}(v, u) \,.
\eal
These quantities can then be analytically continued to any $u\in \mathbb{C}$. In what follows, we explicitly check the relation in the first line of the expression above. The second line is just the $n \to 0^+$ limit of the first one.

We start with~\eqref{eq:masmas_BES_Kernel_as_double_conv}, from which we can write
\bal
K^{n1}_{\bes}(v,u) =\frac{1}{2\pi i}\int_{\text{cuts}}& \text{d} v_1\int_{\text{cuts}} \text{d} v_2 \ K_+^{ny}(v, v_1) \ \frac{d}{d v_1}\log \frac{\Gamma \bigl[1+\frac{ih}{2}\big(v_1-v_2\bigl)\bigl]}{\Gamma \bigl[1-\frac{ih}{2}\big(v_1-v_2\bigl)\bigl]} \ K_+^{y1}(v_2, u)\,.
\eal
When we cross the BES integration contour on the lower edge of the long cut $(+2+\frac{i}{h}, +\infty+\frac{i}{h})$ with $u+\frac{i}{h}$ we encounter a pole of $K_+^{y1}(v_2, u+\frac{i}{h})$ and the kernel is continued as follows 
\bal
K^{n1}_{\bes}(v,u^\gamma+\frac{i}{h}) &=\frac{1}{2\pi i}\int_{\text{cuts}} \text{d} v_1\int_{\text{cuts}} \text{d} v_2 \ K_+^{ny}(v, v_1) \ \frac{d}{d v_1}\log \frac{\Gamma \bigl[1+\frac{ih}{2} v_{12}\bigl]}{\Gamma \bigl[1-\frac{ih}{2} v_{12} \bigl]} \ K_+^{y1}(v_2, u^\gamma+\frac{i}{h})\\
&+\frac{1}{2\pi i}\int_{\text{cuts}} \text{d} v_1  \ K_+^{ny}(v, v_1) \ \frac{d}{d v_1}\log \frac{\Gamma \bigl[1+\frac{ih}{2}\big(v_1-u\bigl)\bigl]}{\Gamma \bigl[1-\frac{ih}{2}\big(v_1-u\bigl)\bigl]} \,,
\eal
where the second line in the expression above corresponds to the residue of the pole. In the first line $K_+^{y1}(v_2,u^\gamma+\frac{i}{h})$ is the continuation of the kernel under $x^-(u+\frac{i}{h}) \to \frac{1}{x^-(u+\frac{i}{h})}$, which is
\begin{multline}
K_+^{y1}(v_2,u+\frac{i}{h})=\frac{1}{2 \pi i} \frac{d}{d v_2} \log \left( \sqrt{\frac{x^-(u+\frac{i}{h})}{x^+(u+\frac{i}{h})}} \ \frac{x(v_2)  -x^+(u+\frac{i}{h})}{x(v_2) - x^-(u+\frac{i}{h})} \right)  \\
\to K_+^{y1}(v_2,u^\gamma+\frac{i}{h})=\frac{1}{2 \pi i} \frac{d}{d v_2} \log \left( \sqrt{\frac{x^-(u)}{x^+(u)}}  \ \frac{x(v_2) -x^+(u+\frac{i}{h})}{x(v_2)x^-(u+\frac{i}{h}) - 1} \right)\,.
\end{multline}
Noting that
\bal
K_+^{y1}(v_2,u^\gamma+\frac{i}{h})&=K_+^{y1}(v_2,u+\frac{i}{h}) + \frac{1}{2 \pi i} \frac{d}{d v_2} \log \left( \frac{x(v_2) - x(u)}{x(v_2) x(u) -1} \right)\\
&=K_+^{y1}(v_2, u+\frac{i}{h})+ K(v_2, u)\,,
\eal
then we obtain the following discontinuity relation for the BES kernel across the cut $(2+\frac{i}{h}, +\infty+\frac{i}{h})$ from below
\bal
\la{eq:KQbes_masmas_computed}
&K^{n1}_{\bes}(v,u+\frac{i}{h}+ i0) -K^{n1}_{\bes}(v,u+\frac{i}{h}-i0)=\\
&-\frac{1}{2\pi i}\int_{\text{cuts}} \text{d} v_1\int_{\text{cuts}} \text{d} v_2 \ K_+^{ny}(v, v_1) \ \frac{d}{d v_1}\log \frac{\Gamma \bigl[1+\frac{ih}{2} v_{12}\bigl]}{\Gamma \bigl[1-\frac{ih}{2} v_{12}\bigl]} \, K(v_2, u)\\
&-\frac{1}{2\pi i}\int_{\text{cuts}} \text{d} v_1  \ K_+^{ny}(v, v_1) \ \frac{d}{d v_1}\log \frac{\Gamma \bigl[1+\frac{ih}{2}\big(v_1-u\bigl)\bigl]}{\Gamma \bigl[1-\frac{ih}{2}\big(v_1-u\bigl)\bigl]} \,.
\eal
In the expression above, $u$ is just above the integration line on the upper edge of the cut $(+2, +\infty)$. Comparing the expression above with~\eqref{eq:BES_Kernel_as_double_conv_Q0}, we find that the relation in the first line of~\eqref{eq:disc_BES_KQmassive} holds.

The above discussion extends to bound states as well, and we have the following universal property
\bal
\la{eq:disc_BES_KQmassiveP}
&K^{n n'}_{\bes}(v,u \pm \frac{i}{h}n'+i 0) - K^{n n'}_{\bes}(v,u^\gamma \pm \frac{i}{h} n'+i 0)=\mp K_\bes^{n0}(v, u)\,,\\
&K^{0 n'}_{\bes}(v,u \pm \frac{i}{h}n'+i 0) - K^{0 n'}_{\bes}(v,u^\gamma \pm \frac{i}{h} n'+i 0)= \mp K_\bes^{00}(v, u) \,.
\eal

\subsection{Discontinuities of massive kernels}

The massive kernels have the following discontinuity properties
\bal
\label{eq:app_massiveK_disc1}
&K_{sl}^{n n'}( v, u \pm \frac{i}{h} n'+ i0)-K_{sl}^{n n'}( v, u^\gamma \pm \frac{i}{h} n'+ i0)=\mp K^{n0}(v, u)\mp \frac{1}{4} \left( K_n(2-v)+K_n(2+v) \right)\,,\\
&\tK_{su}^{n n'}( v, u \pm \frac{i}{h} n'+ i0)-\tK_{su}^{n n'}( v, u^\gamma \pm \frac{i}{h} n'+ i0)=\mp \tK^{n0}(v, u)\pm \frac{1}{4} \left( K_n(2-v)+K_n(2+v) \right)\,,
\eal
\bal
\label{eq:app_massiveK_disc2}
&\tK_{sl}^{n n'}( v, u \pm \frac{i}{h} n'+ i0)-\tK_{sl}^{n n'}( v, u^\gamma \pm \frac{i}{h} n'+ i0)=\\
&\mp K^{n0}(v,u) +K^{ny}_+(v,u) -  K^{ny}_-(v,u) \mp K_n(v-u) \pm \frac{1}{4} \left(K_n(2-v) + K_n(2+v) \right)\,,\\
&K_{su}^{n n'}( v, u \pm \frac{i}{h} n'+ i0)-K_{su}^{n n'}( v, u^\gamma \pm \frac{i}{h} n'+ i0)=\\
&\mp \tK^{n0}(v,u) +  K^{ny}_+(v,u) -  K^{ny}_-(v,u) \pm K_n(v-u) \mp \frac{1}{4} \left(K_n(2-v) + K_n(2+v) \right)\,.
\eal
Below, we check the relations in~\eqref{eq:app_massiveK_disc1}; the remaining discontinuities can be computed similarly.
We start computing the discontinuities of the `Barnes' kernels $K_R$ and $K^\pm_R$. Let us continue $u$ across the cut $(2, + \infty)$ from below. 
Under this continuation
\bal
\g(u^\gamma)=\g(u) + i \pi \,, \qquad K(v,u^\gamma)=-K(v,u) \,,
\eal
and (using the definitions in~\eqref{subsec:massive_kernels} for the kernels) we get
\bal
&K_{R}(v,u)-K_{R}(v,u^\gamma)=\frac{1}{4}K_{\rm mas}(v,u) - \frac{i}{2\pi} \left( \g_{vu} - i \frac{\pi}{2} \right) \, K(v,u) \,,\\
&K^+_{R}(v,u)-K^+_{R}(v,u^\gamma)=\frac{1}{4}K_{\rm mas}(v,u) + \frac{i}{2\pi} \left( \g_{vu} + i \frac{\pi}{2} \right) \, K(v,u) \,,\\
&K^-_{R}(v,u)-K^-_{R}(v,u^\gamma)=\frac{1}{4}K_{\rm mas}(v,u) + \frac{i}{2\pi} \left( \g_{vu} - i \frac{3 \pi}{2} \right) \, K(v,u) \,.
\eal
From the discontinuities above and those for BES (see equation~\eqref{eq:disc_BES_KQmassiveP}), we can read off the discontinuities of the dressing factors. 
We get
\bal
&\frac{1}{2 \pi i} \frac{d}{d v} \Sigma^{n n'}(v, u \pm \frac{i}{h}n' + i0) - \frac{1}{2 \pi i} \frac{d}{d v}  \Sigma^{n n'}(v, u^\gamma \pm \frac{i}{h}n' + i0)=\pm 2 K_\bes^{n0}(v, u)\\
& \mp \frac{1}{i\pi} \left( \g_{vu}^{- n \circ} - \frac{i \pi}{2} \right) K(v - \frac{i}{h}n, u) \pm \frac{1}{i \pi} \left( \g_{vu}^{ + n \circ} - \frac{i \pi}{2} \right) K(v + \frac{i}{h}n, u)\\
&\pm \frac{1}{2} \left(K_{\rm mas}(v - \frac{i}{h}n,u) - K_{\rm mas}(v + \frac{i}{h}n,u) \right) \,,
\eal
and
\bal
&\frac{1}{2 \pi i} \frac{d}{d v} \widetilde{\Sigma}^{n n'}(v, u \pm \frac{i}{h}n' + i0) - \frac{1}{2 \pi i} \frac{d}{d v}  \widetilde{\Sigma}^{n n'}(v, u^\gamma \pm \frac{i}{h}n' + i0)=\pm 2 K_\bes^{n0}(v, u)\\
&-\frac{1}{i \pi} \left( \g_{vu}^{\mp n \circ} - \frac{i \pi}{2} \right) K(v \mp \frac{i}{h}n, u)+\frac{1}{i\pi} \left( \g_{vu}^{ \pm n \circ} - \frac{i \pi}{2} \right) K(v \pm \frac{i}{h}n, u)\\
&+\frac{1}{2} K_{\rm mas}(v \pm \frac{i}{h}n,u) - \frac{1}{2} K_{\rm mas}(v \mp \frac{i}{h}n,u) \,.
\eal
Using that
\bal
\frac{1}{2} K_{\rm mas}(v+\frac{i}{h}n,u) - \frac{1}{2} K_{\rm mas}(v-\frac{i}{h}n,u)=\frac{1}{4} \left( K_n(2-v)+K_n(2+v) \right)-\frac{1}{2} K_n(v-u)
\eal
and formulas~(\eqref{eq:leftmassive_massless_in_Y0_from_TBA}, \eqref{eq:rightmassive_massless_in_Y0_from_TBA}) for the mixed-mass kernels, then
\bal
\label{eq:app_dressing_mass_mass_disc}
&\frac{1}{2 \pi i} \frac{d}{d v} \Sigma^{n n'}(v, u \pm \frac{i}{h}n' + i0) - \frac{1}{2 \pi i} \frac{d}{d v}  \Sigma^{n n'}(v, u^\gamma \pm \frac{i}{h}n' + i0)=\\
&\mp K^{n0}(v,u) \mp \frac{1}{4} \left(K_n(2-v) + K_n(2+v) \right) \,,\\
&\frac{1}{2 \pi i} \frac{d}{d v} \widetilde{\Sigma}^{n n'}(v, u \pm \frac{i}{h}n' + i0) - \frac{1}{2 \pi i} \frac{d}{d v}  \widetilde{\Sigma}^{n n'}(v, u^\gamma \pm \frac{i}{h}n' + i0)=\\
& \mp \tK^{n0}(v,u) \pm K(v-\frac{i}{h}n, u) \pm K(v+\frac{i}{h}n, u) \pm \frac{1}{4} \left(K_n(2-v) + K_n(2+v) \right)\,.
\eal
We have obtained the discontinuities of the dressing factors.
To obtain the discontinuities of the massive kernels, we still need to add the discontinuity of the remaining normalization.
In the left-left sector, the normalization only contains a ratio of $u$ terms and no further discontinuity appears. Then we have
\bal
&K_{sl}^{n n'}( v, u \pm \frac{i}{h} n'+ i0)-K_{sl}^{n n'}( v, u^\gamma \pm \frac{i}{h} n'+ i0)=\\
&\mp K^{n0} (v,u) \mp \frac{1}{4} \left( K_n(2-v)+K_n(2+v) \right)\,.
\eal
In the remaining kernels, we need to take into account the extra normalization. For example, in the right-left element we have
\bal
{\St}_{su}^{n n'}(  v, u)    &=\frac{x_u^{-}}{x_u^{+}}
  \frac{1-\frac{1}{x^-_v x^-_u}}{1-\frac{1}{x^+_v x^+_u}}
    \frac{1-\frac{1}{x^+_v x^-_u}}{1-\frac{1}{x^-_v x^+_u}} \big(\tSi^{n n'}(v,u)\big)^{-2} \,.
\eal   
Continuing the kernel across $+n'$ cut we get that $x^{-n'}(u+\frac{i}{h}n') \to \frac{1}{x^{-n'}(u+\frac{i}{h}n')}= \frac{1}{x(u)}$
\bal
\left[\frac{1}{2 \pi i} \frac{d}{dv} \log \left( \frac{x_u^{-}}{x_u^{+}}
  \frac{1-\frac{1}{x^-_v x^-_u}}{1-\frac{1}{x^+_v x^+_u}}
    \frac{1-\frac{1}{x^+_v x^-_u}}{1-\frac{1}{x^-_v x^+_u}} \right) \right]_{n'} &= -\frac{1}{2 \pi i} \, \frac{d}{d v} \log \left( \frac{(x(u) - x^-_v) (x(u) - x^+_v)}{(1-x(u) x^-_v) (1-x(u) x^+_v)} \right) \\
    &= -K(v-\frac{i}{h}n, u)-K(v+\frac{i}{h}n, u) \,.
\eal
Continuing the kernel across $-n'$ cut we get that $x^{+n'}(u-\frac{i}{h}n') \to \frac{1}{x^{+n'}(u-\frac{i}{h}n')}= \frac{1}{x(u)}$ and we obtain
\bal
\left[\frac{1}{2 \pi i} \frac{d}{dv} \log \left( \frac{x_u^{-}}{x_u^{+}}
  \frac{1-\frac{1}{x^-_v x^-_u}}{1-\frac{1}{x^+_v x^+_u}}
    \frac{1-\frac{1}{x^+_v x^-_u}}{1-\frac{1}{x^-_v x^+_u}} \right) \right]_{-n'} &= +K(v-\frac{i}{h}n, u)+K(v+\frac{i}{h}n, u) \,.
\eal
Then, including this normalization in the second relation in~\eqref{eq:app_dressing_mass_mass_disc}, we obtain
\bal
&\tK_{su}^{n n'}(v, u, u \pm \frac{i}{h}n' + i0)-\tK_{su}^{n n'}(v, u, u^\gamma \pm \frac{i}{h}n' + i0)=\\
&\mp \tK^{n0}(v, u) \pm \frac{1}{4} \left( K_n(2-v)+K_n(2+v) \right) \,.
\eal
This concludes the check of the relations in~\eqref{eq:app_massiveK_disc1}. The remaining relations can be checked similarly.

\subsection{Discontinuities of mixed-mass kernels}

It is possible to check that the mixed mass kernels have the following discontinuities
\bal
\label{eq:app_mixedmK_disc}
&K^{0 n'}( v, u \pm \frac{i}{h} n'+ i0)-K^{0 n'}( v, u^\gamma \pm \frac{i}{h} n'+ i0)=\mp K^{00}(v, u)\,,\\
&\tK^{0 n'}( v, u \pm \frac{i}{h} n'+ i0)-\tK^{0 P}( v, u^\gamma \pm \frac{i}{h} n'+ i0)=\mp K^{00}(v, u)-2 K(v,u+ i0)\,.
\eal

\section{Discontinuities of massive Y functions}
\label{appendix_discontinuities_Y1}
Using the previously discussed kernel discontinuities, we derive the formulas for $[\log Y_1]_{\pm 1}$ and $[\log \bY_1]_{\pm 1}$ in this appendix. This is a remarkable simplification that, in contrast to what happened in the case of other AdS/CFT Y-systems, allows us to obtain a completely local expression for these discontinuities.

\subsection{Discontinuities of left Y functions}
\label{app_disc_left_Y}

Let us evaluate the discontinuity of the function $\log  Y_{n'}$ on the long cuts at $\pm n'$, from which we can write $[\log Y_1]_{\pm 1}$ as a particular case.
From~\eqref{TbaQ}, the TBA equation of this function is
\bal
\log  Y_{n'} =& -L\, \tE_{n'} + \log\left(1+Y_{n} \right)\star K_{sl}^{n n'} +  \log\left(1+ \bY_{n} \right)\star \tK_{su}^{n n'} 
+  \log  \left(1+Y_0 \right)\cstar K^{0 n'} \\
 &+ \log \prod_{\a=1,2} \left(1-{e^{i \mu_\a}\ov Y_{+}^{(\a)}} \right)\hstar K^{y n'}_+ +  \log \prod_{\a=1,2} \left(1-{e^{i \mu_\a}\ov Y_{-}^{(\a)}} \right)\hstar K^{yn'}_-\,.~~~~~~
\eal
Given the expression for the energy in~\eqref{eq:energy_Q_particle} we find that
\bal
\label{eq:left_energy_disc}
\left[ -L \tE_{n'} \right]_{\pm n'}  = \mp L \log x^2(u) = \pm L \tE_0(u) \,.
\eal
From the relations in~\eqref{eq:app_massiveK_disc1} we get 
\bal
&\left[ \log\left(1+Y_{n} \right) \star K_{sl}^{n n'} \right]_{\pm n'}=\\
&\mp \log\left(1+Y_{n} \right) \star K^{n0} \mp \frac{1}{4} \log\left(1+Y_{n}(v) \right) \star (K_{n}(2-v)+K_{n}(2+v)) + \dots\,,
\eal
and
\bal
&\left[ \log\left(1+\bY_{n} \right) \star \tK_{su}^{n n'} \right]_{\pm n'}=\\
&\mp \log\left(1+\bY_{n} \right) \star \tK^{n0} \pm \frac{1}{4} \log\left(1+\bY_{n}(v) \right) \star (K_{n}(2-v)+K_{n}(2+v)) + \dots \,.
\eal
The ellipses in the expressions above contain eventual additional residues coming from poles intersecting the convolution contours when performing the shifts $u \to u \pm \frac{i}{h} n'$.
Similarly, from~\eqref{eq:app_mixedmK_disc} we obtain
\bal
\left[\log \left(1+Y_0 \right)\cstar K^{0 n'}\right]_{\pm n'}= \mp \log  \left(1+Y_0 \right) \cstar K^{00}(v, u) + \dots \,,
\eal
where, once again, we neglect possible residues contained within the ellipses.
Finally, from~\eqref{eq:app_disc_auxmas_kernels} we can write 
\bal
&\left[\log \prod_{\a=1,2} \left(1-{e^{i \mu_\a}\ov Y_{+}^{(\a)}} \right)\hstar K^{yn'}_+ \right]_{\pm n'}= \mp \log \prod_{\a=1,2} \left(1-{e^{i \mu_\a}\ov Y_{+}^{(\a)}} \right)\hstar  K(v,u) + \dots\,,\\
&\left[\log \prod_{\a=1,2} \left(1-{e^{i \mu_\a}\ov Y_{-}^{(\a)}} \right)\hstar K^{y n'}_- \right]_{\pm n'}= \mp  \log \prod_{\a=1,2} \left(1-{e^{i \mu_\a}\ov Y_{-}^{(\a)}} \right)\hstar  K(v,u) + \dots \,.
\eal
Combining all terms above, one gets
\bal\label{eq:discfromTBA}
&\left[ \log  Y_{n'} \right]_{\pm n'}= \mp \biggl( -L \tE_0(u) + \log\left(1+Y_{n} \right) \star K^{n0}  +\log\left(1+\bY_{n} \right) \star \tK^{n0}\\
&+\log  \left(1+Y_0 \right) \cstar K^{00}(v, u) + \log \prod_{\a=1,2} \left(1-{e^{i \mu_\a}\ov Y_{+}^{(\a)}} \right) \left(1-{e^{i \mu_\a}\ov Y_{-}^{(\a)}} \right) \hstar  K(v,u) \biggr)\\
&\mp \frac{1}{4} \log \frac{1+Y_{n}(v)}{1+\bY_{n}(v)} \star (K_{n}(2-v)+K_{n}(2+v)) + \texttt{residues}_{n',\pm},
\eal
where we omit the explicit computation of the residues.  These generated in a similar way to what already discussed in appendix~\ref{appendixdiscpm}, and we performed their explicit computation only in the case $n'=1$, where we find
\bal\label{eq:residues}
\texttt{residues}_{1,+} &= + \log  (1+Y_0) +\log  \prod_{\a=1,2}\left(1-{e^{i \mu_\a}\ov Y_{+}^{(\a)}} \right) \, ,\\
\texttt{residues}_{1,-} &=  - \log  (1+Y_0) -\log  \prod_{\a=1,2}\left(1-{e^{i \mu_\a}\ov Y_{+}^{(\a)}(u^\gamma ) } \right) \,.
\eal
In the expression \eqref{eq:discfromTBA}, moreover, we recognise the TBA equation of $\log Y_0$ (see~\eqref{TbabQ}), so that we get
\bal
\label{eq:app_disc_Yn_general}
&\left[ \log  Y_{n'} \right]_{\pm n'}= \mp \log Y_0 \mp \frac{1}{4} \log \frac{1+Y_{n}(v)}{1+\bY_{n}(v)} \star (K_{n}(2-v)+K_{n}(2+v)) + \texttt{residues}_{n',\pm} \,.
\eal
In the case $n'=1$, which is the only one we will need explicitly, eventually we get
\bal
\label{eq:app_disc_Y1_general}
\left[ \log  Y_1 \right]_{+ 1}= &- \log Y_0 - \frac{1}{4} \log \frac{1+Y_{n}(v)}{1+\bY_{n}(v)} \star (K_{n}(2-v)+K_{n}(2+v)) \\
&  + \log  (1+Y_0) +\log  \prod_{\a=1,2}\left(1-{e^{i \mu_\a}\ov Y_{+}^{(\a)}} \right) \,,
\eal
and
\bal
\label{eq:app_disc_Y1_m1_general}
\left[ \log  Y_1 \right]_{- 1}= &+ \log Y_0 + \frac{1}{4} \log \frac{1+Y_{n}(v)}{1+\bY_{n}(v)} \star (K_{n}(2-v)+K_{n}(2+v)) \\
&  - \log  (1+Y_0) -\log  \prod_{\a=1,2}\left(1-{e^{i \mu_\a}\ov Y_{+}^{(\a)}(u^\gamma ) } \right) \,.
\eal
Let us prove the expression for the residues in the  case $n'=1$. 
We start with
\bal
\log  Y_1 =& -L\, \tE_{1} + \log\left(1+Y_{n} \right)\star K_{sl}^{n 1} +  \log\left(1+ \bY_{n} \right)\star \tK_{su}^{n 1} 
+  \log  \left(1+Y_0 \right)\cstar K^{01} \\
 &+ \log \prod_{\a=1,2} \left(1-{ \frac{e^{i \mu_\a}}{ Y_{+}^{(\a)}} } \right)\hstar K^{y1}_+ +  \log \prod_{\a=1,2} \left(1-\frac{e^{i \mu_\a}} {Y_{-}^{(\a)}} \right)\hstar K^{y1}_-\,,~~~~~~
\eal
and compute
\bal
\log Y_1(u+\frac{i}{h}) - \log Y_1(u^\gamma+\frac{i}{h})\,, 
\eal
where we consider $u$ to be a point just above the real line, with $\Re(u)>2$. The part of the result coming from the kernel discontinuities was already obtained in~\eqref{eq:app_disc_Yn_general} for arbitrary values of $n'$. Here we just evaluate the additional contributions coming from poles intersecting the integration lines of the convolutions.
First we evaluate $\log Y_1(u+\frac{i}{h})$, which is the function obtained by reaching the point $u+\frac{i}{h}$ while keeping $|\Re(u)|<2$ during the shift. In this case, the auxiliary kernel
\bal
{K}_+^{y 1}( v, u) &= \frac{1}{2 \pi i } \frac{d}{d v} \ \log \left(\sqrt{\frac{x^{-n}(u)}{x^{+n}(u)}} \ \frac{x(v) - x(u+\frac{i}{h})}{x(v) - x(u-\frac{i}{h})} \right)
\eal
has a pole at $v = u - \frac{i}{h}$.
When shifting $u \to u +\frac{i}{h}$ keeping $|\Re(u)|<2$ this pole intersect the integration line of $\log \prod_{\a=1,2} \left(1-{ \frac{e^{i \mu_\a}}{ Y_{+}^{(\a)}} } \right)\hstar K^{y1}_+$ from below. Picking up the associated residue, we have the following continuation for the function above the cut:
\bal
\label{eq_Y1abovecut_app}
\log Y_1(u+\frac{i}{h}) =& \biggl[-L\, \tE_{1} + \log\left(1+Y_{n} \right)\star K_{sl}^{n 1} +  \log\left(1+ \bY_{n} \right)\star \tK_{su}^{n 1} 
+  \log  \left(1+Y_0 \right)\cstar K^{01} \\
 &+ \log \prod_{\a=1,2} \left(1-{ \frac{e^{i \mu_\a}}{ Y_{+}^{(\a)}} } \right)\hstar K^{y1}_+ +  \log \prod_{\a=1,2} \left(1-\frac{e^{i \mu_\a}} {Y_{-}^{(\a)}} \right)\hstar K^{y1}_- \biggr](u+\frac{i}{h})\\
&+ \log  \prod_{\a=1,2}\left(1-{e^{i \mu_\a}\ov Y_{+}^{(\a)}(u) } \right) \,.
\eal
If we instead evaluate $Y_1(u^\gamma+\frac{i}{h})$, this contribution is not present because, in performing the shift, we are now keeping $\Re(u)>2$ and the pole does not intersect the integration line. In $\log \prod_{\a=1,2} \left(1-{ \frac{e^{i \mu_\a}}{ Y_{+}^{(\a)}} } \right)\hstar K^{y1}_+$ we are indeed integrating $v$ on the interval $(-2,+2)$.

The other convolution responsible for poles intersecting the contour is $\log  \left(1+Y_0 \right)\cstar K^{01}$. The S-matrix entering this convolution is
\bal
{S}^{0 1}(  v, u)  =\,\frac{1}{i}\, \sqrt{\frac{x_u^{+}}{x_u^-}}   \frac{x^-_u-x_v}{x^+_u x_v-1} \big(\Sigma^{0 1}(v,u) \big)^{-2} \,,
\eal
where
\bal
\left(\Sigma^{0 1}(v, u) \right)^{-2}=-i \coth \left(\frac{\g^{\circ +}_{v u}}{2} \right) \frac{R^2(\gamma_{v u}^{\circ -})}{R^2(\gamma_{vu}^{\circ+})}  \  \frac{R(\gamma_{vu}^{\circ+}+ i \pi) R(\gamma_{v u}^{\circ+}- i\pi)}{R(\gamma_{vu}^{\circ -} + i \pi) R(\gamma_{vu}^{\circ -}- i \pi)} \big(\Sigma_\bes^{01}(v, u) \big)^{-2}\,.
\eal
In this case, when we perform the shift $u \to u+ \frac{i}{h}$ keeping $\Re(u)<2$ we do not intersect the integration contour with any poles. Note indeed that the integration domain of the massless Y functions is $|v|>2$. When performing the shift keeping $\Re(u)>2$ the following facts happen. First, we enter the mirror cut of the Zhukovsky variable, and we must continue
\bal
{S}^{0 1}(  v, u^{\gamma}+\frac{i}{h})  =\,\frac{1}{i}\, \sqrt{x_u^{+2} x_u}   \frac{\frac{1}{x_u}-x_v}{\frac{1}{x_u} x_v-1} \big(\Sigma^{0 1}(v,u^\gamma + \frac{i}{h}) \big)^{-2} \,,
\eal
where (using that $\gamma^\gamma_u = \gamma_u + i \pi$), we have
\bal
&\left(\Sigma^{0 1}(v, u^\gamma+\frac{i}{h}) \right)^{-2}=-i \coth \left(\frac{\g(v)- \g(u+\frac{2i}{h})}{2} \right) \frac{R^2(\g(v) - \g(u) - i \pi)}{R^2(\g(v) - \g(u+\frac{2i}{h}))}  \\  &\frac{R(\g(v) - \g(u+\frac{2i}{h})+ i \pi) R(\g(v) - \g(u+\frac{2i}{h}) - i\pi)}{R(\g(v) - \g(u)) R(\g(v) - \g(u) -2 i \pi)} \big(\Sigma_\bes^{01}(v, u^\gamma+\frac{i}{h}) \big)^{-2}\,.
\eal
Then, shifting further slightly above, we encounter the singularity of $R(\g(v) - \g(u) -2 i \pi)$, which has a pole at $v=u$ (see~\eqref{eq:R_pole_location}). Picking the residue of this pole into account, we obtain the following continuation
\bal
\label{eq_Y1belowcut_app}
\log Y_1(u^\gamma+\frac{i}{h}) =\biggl[&-L\, \tE_{1} + \log\left(1+Y_{n} \right)\star K_{sl}^{n 1} +  \log\left(1+ \bY_{n} \right)\star \tK_{su}^{n 1} 
+  \log  \left(1+Y_0 \right)\cstar K^{01} \\
 &+ \log \prod_{\a=1,2} \left(1-{ \frac{e^{i \mu_\a}}{ Y_{+}^{(\a)}} } \right)\hstar K^{y1}_+ +  \log \prod_{\a=1,2} \left(1-\frac{e^{i \mu_\a}} {Y_{-}^{(\a)}} \right)\hstar K^{y1}_- \biggr](u^\gamma+\frac{i}{h})\\
 &- \log (1+Y_0(u)) \,.
\eal
Taking the difference of~\eqref{eq_Y1abovecut_app} and~\eqref{eq_Y1belowcut_app}, and substituting $n'=1$ in~\eqref{eq:app_disc_Yn_general} we obtain the expression in~\eqref{eq:app_disc_Y1_general}.
A similar computation leads to~\eqref{eq:app_disc_Y1_m1_general}.

\subsection{Discontinuities of right Y functions}

A similar computation for the right Y functions leads to
\bal
\label{eq:app_disc_dY1_general}
&\left[ \log  \bY_{n'} \right]_{\pm n'}= \mp \log Y_0 \pm 2 \log Y^{(\a)}_{\mp} \pm \frac{1}{4} \log \frac{1+Y_{n}(v)}{1+\bY_{n}(v)} \star (K_{n}(2-v)+K_{n}(2+v)) + \dots \,,
\eal
where, again, we omitted residues arising from poles that intersect the integration contour when shifting $u$. Restoring these poles and setting $n'=1$ we get
\bal
\left[ \log  \bY_1 \right]_{+ 1}=&- \log Y_0 + \frac{1}{4} \log \frac{1+ Y_n}{1+ \bY_n} \star (K_{n}(2-v)+K_{n}(2+v))\\
& + \log (1+Y_0) +  \log  \prod_{\a=1,2}\left(1-{e^{-i \mu_\a} Y_{-}^{(\a)}} \right)\,,
\eal
\bal
\left[ \log  \bY_1 \right]_{- 1}=&+ \log Y_0- \frac{1}{4} \log \frac{1+ Y_n}{1+ \bY_n} \star (K_{n}(2-v)+K_{n}(2+v))\\
& - \log (1+Y_0) -  \log  \prod_{\a=1,2}\left(1-{e^{-i \mu_\a} Y_{-}^{(\a)}(u^\gamma )} \right)\,.
\eal

\section{Calculation of massless discontinuities}
\label{appendixY0disc}
In this appendix, we derive the discontinuities of the massless Y functions. In particular, we compute the discontinuities $[\log Y_0]_{\pm 2N}$ on the cuts in the UHP and LHP, while for the real axis we compute the ``symmetrised discontinuity'' $\{\log Y_0\}_0$.

\subsection{Massless discontinuities on the real axis}

We compute the symmetrised discontinuity on the real axis $\{\log Y_0\}_0\equiv \log Y_0 + \log Y_0^\gamma$, starting from the TBA equation for $\log{Y_0}$~\eqref{Tba0}.
First, we derive the contribution from the physical particles to the TBA equation~\eqref{Tba0}:
\begin{equation}
    - \log\left(1+Y_{0}\right)\cstar K^{00} -  \log\left(1+Y_{n} \right)\star K^{n 0} -  \log\left(1+ \bY_{n} \right)\star \tK^{n 0} \,.
\end{equation}
We split each kernel into a BES, Barnes and rational part  
\bal
K^{A0}=K^{A0}_{\bes}+K^{A0}_{Barnes}+K^{A0}_{Rational} \,, \qquad A=0 \ \text{or} \ A=n\,. 
\eal
Then, we compute the contribution coming from the auxiliary particles.

\paragraph{Massless particles.}
Here we need to analyse the kernels:
\bal
&K^{00}_{\bes}\equiv-2 \frac{1}{2 i \pi} \frac{d}{dv}\log{\Sigma_{\bes}^{00}} \,,\\
&K^{00}_{Barnes}\equiv  \frac{1}{2 i \pi} \frac{d}{dv}\log{\left( \frac{R^2( \g^{\circ \circ}_{12} - i \pi) R^2( \g^{\circ \circ}_{12} + i \pi)}{R^4(\g^{\circ \circ}_{12})}\right)} \,,\\
&
K^{00}_{Rational}\equiv0 \,.
\eal
We  start by continuing $K^{00}_{Barnes}$ through the branch cut on the real axis, obtaining:
\beq
K^{00}_{Barnes}(v,u-i0)=\frac{1}{2 i \pi} \frac{d}{dv}\log\left(\frac{R^2( \g^{\circ \circ}_{12} - 2 i \pi) R^2( \g^{\circ \circ}_{12})}{R^4(\g^{\circ \circ}_{12}- i\pi)} \right)\,.
\eeq
 The function $R^2(\g_{vu}^{\circ\circ}-2 i \pi)$ has a double pole at $u=v$, yielding the only residue term here, with an overall prefactor of $2$. 
Using the identities \eqref{eq:properties_R_functions}, we obtain that:
\begin{align}
    &\log(1+Y_0)\cstar\left(K^{00}_{Barnes}(v,u+i0)+K^{00}_{Barnes}(v,u-i0)\right)=\\
    &-\log(1+Y_0)\cstar\frac{1}{2 i \pi}\frac{d}{dv}\log{\left(\tanh^2{\frac{\g_{vu}
    ^{\circ\circ}}{2}}\right)}
    +2\log{(1+Y_0)(u)}
   \,.
\end{align}
In appendix~\ref{appendixBES} we showed that the BES kernel does not contribute to the symmetrised discontinuity (see equation~\eqref{eq:disc_real_axis_K00bes}).
By an explicit calculation, it can be seen that the kernel here is nothing else than the square-root type Cauchy  kernel in \eqref{def:Kuniversal}, and one obtains 
\begin{align}
&\log(1+Y_0)\cstar\left(K^{00}(v,u+i0)+K^{00}(v,u-i0)\right)\\=&
    2\log(1+Y_0)\cstar K_+^{0y}(v,u)
    +2\log{(1+Y_0)(u)}
    \,.
\end{align}

\paragraph{Left particles.}
Here we need to analyse the kernels:
\bal
&K^{n0}_{\bes}\equiv-2 \frac{1}{2 i \pi} \frac{d}{dv}\log{\Sigma_{\bes}^{n0}} \,,\\
&K^{n0}_{Barnes}\equiv  \frac{1}{2 i \pi} \frac{d}{dv}\log{\tanh \left(\frac{\g^{+ \circ}_{12}}{2} \right) \frac{R^2(\gamma_{12}^{-\circ})}{R^2(\gamma_{12}^{+\circ})}  \  \frac{R(\gamma_{12}^{+\circ}- i \pi) R(\gamma_{12}^{+\circ}+ i\pi)}{R(\gamma_{12}^{-\circ}- i \pi) R(\gamma_{12}^{-\circ}+ i \pi)}} \,,\\
&
K^{n0}_{Rational}\equiv \frac{1}{2 i \pi} \frac{d}{dv}\log{ \sqrt{\frac{x(v)^-}{x(v)^+}} \frac{x(v)^+x(u)-1}{x(v)^--x(u)}} \,.
\eal
For the rational part $K^{n0}_{Rational}$, no poles intersect the integration contour upon analytic continuation. We also have that:
\beq 
K^{n0}_{Rational}(v,u-i0)=i \sqrt{\frac{x(v)^-}{x(v)^+}} \frac{x(v)^+-x(u)}{x(u)x^-(v)-1} \,.
\eeq
Thus it is immediate to see that:
\begin{align}
    &K^{n0}_{Rational}(v,u+i0)+K^{n0}_{Rational}(v,u-i0)=\\&
    \frac{1}{2 i \pi}\frac{d}{dv}\log{ \frac{x(v)^-}{x(v)^+} \frac{(x(v)^+x(u)-1)(x(v)^+-x(u))}{(x(v)^--x(u))(x(v)^-x(u)-1)}} \,.
\end{align}
From the Barnes part, we get no residues when analytically continuing, and using the identities in~\eqref{eq:properties_R_functions} we find that:
\beq
K_{Barnes}^{n0}(v,u+i0)+K_{Barnes}^{n0}(v,u-i0)=\frac{1}{2 \pi i} \frac{d}{dv} \log \tanh \left(\frac{\gamma_{vu}^{-\circ} }{2} \right) \coth \left(\frac{\gamma_{vu}^{+\circ} }{2}\right) \,.
\eeq
As for the massless-massless part, also in this case the BES kernel does not contribute to the symmetrised discontinuity (due to equation~\eqref{eq:disc_real_axis_K0nbes}) and combining the pieces above one gets
\bal
&\log(1+Y_n)\star(K^{n0}(v,u+i0)+K^{n0}(v,u-i0))=\\
&\log(1+Y_n)\star \frac{1}{2i\pi}\frac{d}{dv}\log{\frac{x(v)^-}{x(v)^+} \frac{(x(v)^+x(u)-1)(x(v)^+-x(u))}{(x(v)^--x(u))(x(v)^-x(u)-1)}\tanh \left(\frac{\gamma_{vu}^{-\circ} }{2} \right) \coth \left(\frac{\gamma_{vu}^{+\circ} }{2}\right)} \,.
\eal
The parameterisation of $\g$ in terms of $x$ is:
\bal
\g(x) = \log \frac{1}{i} \frac{1-x}{1+x}
\eal
Using this fact, one gets
\bal
\tanh \left(\frac{\gamma_{vu}^{-\circ} }{2} \right)= \frac{x(v)^- - x(u)}{ x(v)^- x(u) - 1} \,, \qquad \coth \left(\frac{\gamma_{vu}^{+\circ} }{2} \right)= \frac{x(v)^+ x(u) - 1}{x(v)^+ - x(u)} 
\eal
Plugging these equalities into the expression above, we obtain
\bal
&\log(1+Y_n)\star(K^{n0}(v,u+i0)+K^{n0}(v,u-i0))=-2 \log(1+Y_n)\star K_-^{ny}(u) \,.
\eal

\paragraph{Right particles.}
Here we need to analyse the kernels:
\bal
&K^{n0}_{\bes}\equiv-2 \frac{1}{2 i \pi} \frac{d}{dv}\log{\Sigma_{\bes}^{n0}} \,,\\
&K^{n0}_{Barnes}\equiv  \frac{1}{2 i \pi} \frac{d}{dv}\log{\tanh \left(\frac{\g^{+ \circ}_{12}}{2} \right) \frac{R^2(\gamma_{12}^{-\circ})}{R^2(\gamma_{12}^{+\circ})}  \  \frac{R(\gamma_{12}^{+\circ}- i \pi) R(\gamma_{12}^{+\circ}+ i\pi)}{R(\gamma_{12}^{-\circ}- i \pi) R(\gamma_{12}^{-\circ}+ i \pi)}} \,,\\
&
\tilde{K}^{n0}_{Rational}\equiv \frac{1}{2 i \pi} \frac{d}{dv}\log{e^{\frac{i}{2}p(v)}\frac{x(v)^--x(u)}{x(v)^+x(u)-1}} \,.
\eal
This is identical to the $n$-particle case except that the rational part is inverted. So we obtain
\bal
\log(1+\bY_n)\star \frac{1}{2i\pi}\frac{d}{dv}\log{e^{-i p(v)}\frac{(x(v)^--x(u))(x(v)^-x(u)-1)}{(x(v)^+x(u)-1)(x(v)^+-x(u))}\tanh \left(\frac{\gamma_{vu}^{-\circ} }{2} \right) \coth \left(\frac{\gamma_{vu}^{+\circ} }{2}\right)}\,.
\eal
Once again, the BES contribution is zero, and we can simplify everything down to the following expression
\beq
\log(1+\bY_n)\star(\tilde{K}^{n0}(v,u+i0)+\tilde{K}^{n0}(v,u-i0))=2 \log(1+\bY_n)\star K_+^{ny}(u) \,.
\eeq

\paragraph{Auxiliary particles.}
We next analyse the auxiliary term
\beq
-\sum_{\alpha=1,2} \log \left(1-\frac{e^{i \mu_\alpha}}{Y^{(\alpha)}_+} \right)\left(1-\frac{e^{i \mu_\alpha}}{Y^{(\alpha)}_-} \right) \hstar K_+^{y0}\,.
\eeq
Unlike physical particles, the kernel here contains a single term.
Continuing it around the cut, there will be a pole of $K_+^{y0}(v,u)$ at $u=v$ that hits the integration contour from above, with residue $+1$. This gives an overall $+1$ sign, and the resulting term will then cross the branch cut on the real axis. Furthermore, we have that:
\bal
 {K}_+^{y 0}( v, u-i0) &= \frac{1}{2 \pi i } \frac{d}{d v} \ \log \left(\sqrt{(x(u+i0))^2} \ \frac{x(v) - x(u-i0)}{x(v) - x(u+i0)} \right)\,,\\
\eal
and so ${K}_+^{y 0}( v, u+i0)+{K}_+^{y 0}( v, u-i0)=0$. Therefore the only term we get is a residue:
\beq
\sum_{\alpha=1,2} \log \left(1-\frac{e^{i \mu_\alpha}}{Y^{(\alpha)}_+}\right)\left(1-\frac{e^{i \mu_\alpha}}{Y^{(\alpha)}_-} \right)(u^\gamma) \,.
\eeq

\paragraph{Vanishing energy discontinuity on the real axis.}

The energy term in the TBA equation is given by (see equation~\eqref{eq:mom_en_0_particle})
\bal
\tE_0(u) = -\log x^2(u+i0) =
\begin{cases}
& -2 \log x(u+i0) \,, \hspace{18mm} \text{if} \ u>+2 \,,\\
& -2 (\log x(u+i0)+ i \pi)  \,, \qquad \text{if} \ u<-2 \,.
\end{cases}
\eal
For $u>2$, then $x(u \pm i0)$ are just below the positive side of the real axis, and we obtain
\bal
\tE_0(u+i0) + \tE_0(u-i0) = &-2 \log x(u+i0) - 2 \log x(u-i0)\\
=&-2 \log x(u+i0) - 2 \log \frac{1}{x(u+i0)}=0\,.
\eal
For $u<-2$, then $x(u \pm i0)$ are just below the negative side of the real axis, and we obtain
\bal
\tE_0(u+i0) + \tE_0(u-i0) = &-2 \log x(u+i0) -2 i \pi - 2 \log x(u-i0) -2 i \pi \\
=&-2 \log x(u+i0) -2 i \pi - 2 \left( \log \frac{1}{x(u+i0)} - 2 i \pi \right) - 2 i \pi=0 \,.
\eal
From the results above, we see that for $u \in (-\infty, -2) \cup (+2, +\infty)$ 
\bal
\label{eq:app_disc_Y0_real_axis_en}
-L \tE_0(u+i0)-L \tE_0(u-i0) = 0 \,.
\eal

\paragraph{Final result.}
Putting everything together, we have:
\bal
&\{\log{Y_0}\}_0=\\
&+2\log{(1+Y_0)}\cstar K_+^{0y}(v,u) -2 \log{(1+Y_n)}\star K_-^{ny}(v,u +2 \log{(1+\bY_n)}\star K_+^{ny}(v,u)\\&
+\sum_{\alpha=1,2} \log \left(1-\frac{e^{i \mu_\alpha}}{Y^{(\alpha)}_+}\right)\left(1-\frac{e^{i \mu_\alpha}}{Y^{(\alpha)}_-} \right)(u^\gamma)+2\log{(1+Y_0)(u)}\,.
\eal
The second line is equal to the TBA equation for $2\log{Y^{(\a)}_-}$. Therefore we obtain
\bal
\label{eq:disc_sym_disc_Y0_appendix}
\{\log{Y_0}\}_0=&+ 2 \log{Y^{(\a)}_-}(u)
+\sum_{\alpha=1,2} \log \left(1-\frac{e^{i \mu_\alpha}}{Y^{(\alpha)}_+}\right)\left(1-\frac{e^{i \mu_\alpha}}{Y^{(\alpha)}_-} \right)(u_{\gamma})\\&
    +2\log{(1+Y_0)(u)} \,,
\eal
which corresponds to the result reported in section~\ref{sec:disc_massless}.

\subsection{Discontinuities on UHP and LHP of 
\texorpdfstring{$\log Y_0$}{lnY0}}

We now compute the discontinuities across the cuts at $\Im(u)=\pm 2N/h$ in the TBA equation~\eqref{Tba0}.
We will first compute the contribution from the BES part of the kernels, then address the remaining terms in the TBA equation.

\paragraph{BES discontinuities of $\log Y_0$.}

From equation~\eqref{Tba0} and the relations in~\eqref{eq:app_conv_KQ0f_discpm2N} and~\eqref{eq:app_conv_K00f_discpm2N} we see that
\bal
[\log Y_0]^\bes_{\pm 2N}(u)=&\pm 4 \log \left(1+Y_{0}(v)\right) \cstar K^{0y}_+(v, u)   \pm 2 \log \left(1+Y_{0}(u)\right) \\
&\mp 2\log\left(1+Y_{n} \right) \left(1+\bY_{n} \right)\star \left(K^{ny}_-(v, u) - K^{ny}_+(v, u)\right)\\
=& \pm 2 \log \frac{Y^{(\a)}_-}{Y^{(\a)}_+}\pm 2 \log \left(1+Y_{0} \right)\,,
\eal
where in the second equality we have used the TBA equations for the combination $\log Y^{(\a)}_-/Y_+^{(\a)}$.

\paragraph{Auxiliary particles.}

Consider the auxiliary term in the TBA equations \eqref{Tba0}
\bal
\log \prod_{\alpha=1,2} \left(1-\frac{e^{i \mu_\alpha}}{Y^{(\alpha)}_+} \right) \left(1-\frac{e^{i \mu_\alpha}}{Y^{(\alpha)}_-} \right) \hstar K_+^{y0}
\eal
with 
\bal
K_+^{y0}(v,u)= \frac{1}{2 \pi i} \frac{d}{d v} \log \left(\frac{x(v) - x(u)}{1 - x(v) x(u)} \right) \,.
\eal
If we shift $u$ above, we do not encounter any poles; therefore, we obtain
\bal
\biggl[\log \prod_{\alpha=1,2} \left(1-\frac{e^{i \mu_\alpha}}{Y^{(\alpha)}_+} \right) \left(1-\frac{e^{i \mu_\alpha}}{Y^{(\alpha)}_-} \right) \hstar K_+^{y0} \biggr]_{2N} =0 \,.
\eal
If instead we start with $u$ just above the mirror cuts and shift $u$ below, passing through the interval $(-2, +2)$, then the pole at $v=u$ crosses the integration contour from above and we must continue
\bal
\log \prod_{\alpha=1,2} \left(1-\frac{e^{i \mu_\alpha}}{Y^{(\alpha)}_+} \right) \left(1-\frac{e^{i \mu_\alpha}}{Y^{(\alpha)}_-} \right) \hstar K_+^{y0} &\to \log \prod_{\alpha=1,2} \left(1-\frac{e^{i \mu_\alpha}}{Y^{(\alpha)}_+} \right) \left(1-\frac{e^{i \mu_\alpha}}{Y^{(\alpha)}_-} \right) \hstar K_+^{y0}\\
&+\log \prod_{\alpha=1,2} \left(1-\frac{e^{i \mu_\alpha}}{Y^{(\alpha)}_+(u)} \right) \left(1-\frac{e^{i \mu_\alpha}}{Y^{(\alpha)}_-(u)} \right)\,.
\eal
Therefore we obtain
\bal
&\biggl[\log \prod_{\alpha=1,2} \left(1-\frac{e^{i \mu_\alpha}}{Y^{(\alpha)}_+} \right) \left(1-\frac{e^{i \mu_\alpha}}{Y^{(\alpha)}_-} \right) \hstar K_+^{y0} \biggr]_{-2N} \\
&=\biggl[\log \prod_{\alpha=1,2} \left(1-\frac{e^{i \mu_\alpha}}{Y^{(\alpha)}_+(u)} \right) \left(1-\frac{e^{i \mu_\alpha}}{Y^{(\alpha)}_-(u)} \right)\biggr]_{-2N} \,.
\eal

\paragraph{Left particles.}
Consider the term
\bal
\log (1+Y_n(v)) \star K^{n 0}(v,u)
\eal
where $K^{n0}$ denotes the massless-massive kernel with its BES part removed.
The argument of this kernel has a pole at $v=u+ \frac{i}{h}n$ due to the rational factor and a zero at $v=u-\frac{i}{h} n$, due to the $\tanh$ term in~\eqref{eq:massive_massless_full_dressing}.

If we shift $u \to u'$, with $\frac{2N}{h}< \Im(u') < \frac{2N+1}{h}$, then we must pick up the residues associated with the zero at $v=u-\frac{i}{h}n$, which crosses the contour from below. There is one such residue for each $n=1, 2\dots, 2N$.
Then we have the following continuation
\bal
\log (1+Y_n(v)) \star K^{n 0}(v,u) \to \log (1+Y_n(v)) \star K^{n 0}(v,u) - \sum^{2N}_{n=1}  \log \left(1+Y_n(u' - \frac{i}{h}n) \right) \,.
\eal
Using that $Y_n$ is smooth in the strip $-\frac{n}{h}<\Im(u)<\frac{n}{h}$, the discontinuity at the $2N$ cut is given by
\bal
\left[\log (1+Y_n(v)) \star K^{n 0}(v,u)\right]_{2N}=- \left[\sum^{N}_{n=1}  \log \left(1+Y_n \right) \right]_{2N-n} \,.
\eal

If we instead shift $u \to u'$, with $-\frac{2(N+1)}{h}< \Im(u') < -\frac{2N}{h}$, then we must pick up the residues associated with the pole at $v=u+\frac{i}{h}n$, which crosses the contour from above. There is one such residue for each $n=1, 2\dots, 2N$.
Then we must continue
\bal
\log (1+Y_n(v)) \star K^{n 0}(v,u) \to \log (1+Y_n(v)) \star K^{n 0}(v,u) - \sum^{2N}_{n=1}  \log \left(1+Y_n(u' + \frac{i}{h}n) \right) \,.
\eal
Using that $Y_n$ is smooth in the strip $-\frac{n}{h}<\Im(u)<\frac{n}{h}$, the discontinuity at the $-2N$ cut is given by
\bal
\left[\log (1+Y_n(v)) \star K^{n 0}(v,u)\right]_{-2N}=- \left[\sum^{N}_{n=1}  \log \left(1+Y_n \right) \right]_{-2N+n} \,.
\eal

\paragraph{Right particles.}
Consider the term
\bal
\log (1+\bY_n(v)) \star \tK^{n 0}(v,u)
\eal
where $\tK^{n0}$ denotes the full kernel excluding its BES part.
The argument of this kernel has a zero at $v=u+ \frac{i}{h}n$ due to the rational part and a zero at $v=u-\frac{i}{h} n$, due to the $\tanh$ in~\eqref{eq:massive_massless_full_dressing}.
Then the computation of the discontinuities is identical to the one just performed for $n$-particles. We just need to change a sign for the discontinuities at $-2N$, because the S-matrix now has a zero at $v=u+ \frac{i}{h}n$ instead of a pole. Then we get
\bal
\left[\log (1+\bY_n(v)) \star K^{n 0}(v,u)\right]_{\pm 2N}= \mp \left[\sum^{N}_{n=1}  \log \left(1+\bY_n \right) \right]_{ \pm (2N-n)} \,.
\eal

\paragraph{Massless particles.}
Finally, we consider the term
\bal
\log\left(1+Y_{0}(v)\right)\cstar K^{00}(v, u) \,.
\eal
When we shift $u$ in the UHP or in the LHP no pole is encountered, and therefore we have
\bal
\left[ \log \left(1+Y_{0}(v)\right)\cstar K^{00}(v, u) \right]_{\pm 2N}=0\,.
\eal

\subsubsection*{Combining the different terms}

Combining the different terms, we obtain the following discontinuities for the cuts in the UHP
\bal
{[\log Y_0]}_{+2N}= &- \sum^{N}_{n=1} \left[ \log(1+Y_n) \right]_{2N-n} - \sum^{N}_{n=1} \left[ \log(1+\bY_n) \right]_{2N-n}\\
&+ 2 \log \frac{Y^{(\a)}_-}{Y^{(\a)}_+}+ 2 \log \left(1+Y_{0} \right)\,.
\eal
For those in the LHP we get instead
\bal
[\log{Y_0}]_{-2N}= &- \sum^{N}_{n=1} \left[ \log(1+Y_n) \right]_{-2N+n} + \sum^{N}_{n=1} \left[ \log(1+\bY_n) \right]_{-2N+n}\\
&- 2 \log \frac{Y^{(\a)}_-}{Y^{(\a)}_+}- 2 \log \left(1+Y_{0} \right)\\
&+\left[ \log \prod_{\a=1,2} \left(1-\frac{e^{i \mu_\a}}{Y^{(\a)}_+} \right) \left( 1-\frac{e^{i \mu_\a}}{Y^{(\a)}_-} \right) \right]_{-2N}\,.
\eal

\section{Useful relations}
\label{app:useful_rel}

In this appendix, we collect useful results for inverting the discontinuity relations of the different Y functions and deriving the TBA equations from them.

\subsection{Formulas to invert discontinuities}
\label{app:formula_disc_inversion}

These results are known from~\cite{Cavaglia:2010nm}. However, we provide a derivation here for completeness.
\begin{prop}
\label{disc_prop1}

Let $I(u)$ be
\bal
\label{eq:app_def_IPNzK}
I(u) \equiv \sum_{N=1}^{\infty}\sum_{Q=1}^N \left( \int_{-\infty}^{-2} + \int_{+2}^{+\infty} \right) dz \,  [V_Q ]_{2 N-Q} (z)   K(z+ \frac{2i}{h} N,u) \,,
\eal
and assume the following conditions
\begin{enumerate}
    \item $-1/h<\Im(u)<1/h$;
    \item $V_Q$ is analytic in the strip $\frac{Q}{h}+\frac{2n}{h}<\Im(u)<\frac{Q}{h}+\frac{2(n+1)}{h}$, with $n=1, 2, \dots$;
    \item $V_Q$ is analytic for $-2< \Re(u)<2$;
    \item $V_Q$ is analytic for $-Q/h<\Im(u)<Q/h$.
\end{enumerate}
Then it holds that
\begin{equation}
I(u)= -\sum_{Q=1}^{\infty} \int^{+\infty}_{-\infty} dz \, V_Q(z)   K(z + i Q/h, u) 
\end{equation}
\end{prop}
\begin{proof}
{The original integral defining $I(u)$ integrates discontinuities of a function along the mirror cuts. Since the function is free of singularities for  $-2<\Re(z)<2$, we rewrite this in the following way: 
\bal
I(u)=\sum_{N=1}^{\infty} \sum^N_{Q=1} \int_{-\infty}^{+\infty} dz \, & \big(  V_Q(z+ i0 + i (2 N - Q)/h)\\&- V_Q(z- i0 + i (2 N - Q)/h) \big)\,K(z+\frac{2i}{h}N,u) \,.
\eal
}

We define the functions
\bal
f_Q(z) \equiv V_Q(z) \, K(z+ \frac{i}{h} Q,u) \,,
\eal
so that we can write $I(u)$ as
\bal
I(u)=&+ \int_{-\infty}^{+\infty} dz \,  \left(-f_1(z+\frac{i}{h}-i0){\color{purple}+f_1(z+\frac{i}{h}+i0)} \right)\\
&+  \int_{-\infty}^{+\infty} dz \,  \left({\color{purple}-f_1(z+\frac{3i}{h}-i0)}-f_2(z+\frac{2i}{h}-i0){\color{orange}+f_2(z+\frac{2i}{h}+i0)} {\color{brown}+f_1(z+\frac{3i}{h}+i0)} \right)\\
&+  \int_{-\infty}^{+\infty} dz \,  \left({\color{brown}-f_1(z+\frac{5i}{h} -i0)} {\color{orange}-f_2(z+\frac{4i}{h}-i0)}-f_3(z+\frac{3i}{h}-i0)+ \dots \right)+ \dots \,.
\eal
Terms of the same colour cancel since the function $f_Q(u)$ is analytic in the intervals $Q/h<\Im(u)<(Q+2)/h$ by our starting hypothesis, and we can shift the integration lines by $\frac{2i}{h}$. Then we obtain
\bal
I(u)=&-\sum_{Q=1}^{\infty}   \int_{-\infty}^{+\infty} dz f_Q(z+\frac{i}{h}Q-i0)=-\sum_{Q=1}^{\infty}   \int_{-\infty}^{+\infty} dz V_Q(z+\frac{i}{h} Q-i0) \, K(z+ \frac{2i}{h} Q,u)
\eal
Since both $V_Q(z+\frac{i}{h} Q)$ and $ K(z+ \frac{2i}{h} Q,u)$ are analytic for $-Q/h<\Im(z)<0$ by hypothesis, then we can shift the integration lines by $-\frac{i}{h} Q$ and write
\bal
I(u)=-\sum_{Q=1}^{\infty}   \int_{-\infty}^{+\infty} dz V_Q(z) \, K(z+ \frac{i}{h} Q,u) \,.
\eal
\end{proof}

\begin{prop}
\label{disc_prop2}

Let $I(u)$ be
\bal
\label{eq:app_def_IPNzK2}
I(u) \equiv \sum_{N=1}^{\infty}\sum_{Q=1}^N \left( \int_{-\infty}^{-2} + \int_{+2}^{+\infty} \right) dz \,  [V_Q ]_{-(2 N-Q)} (z)   K(z- \frac{2i}{h} N,u) \,,
\eal
and assume the following conditions
\begin{enumerate}
    \item $-1/h<\Im(u)<1/h$;
    \item $V_Q$ is analytic in the strip $-\frac{Q}{h}-\frac{2(n+1)}{h}<\Im(u)<-\frac{Q}{h}-\frac{2 n}{h}$, with $n=1, 2, \dots$;
    \item $V_Q$ is analytic for $-2< \Re(u)<2$;
    \item $V_Q$ is analytic for $-Q/h<\Im(u)<Q/h$.
\end{enumerate}
Then it holds that
\begin{equation}
I(u)= \sum_{Q=1}^{\infty} \int^{+\infty}_{-\infty} dz \, V_Q(z)   K(z - i Q/h, u) 
\end{equation}
\end{prop}
\begin{proof}
The proof is analogous to the one of property~\ref{disc_prop1}, and we leave it to the reader.
\end{proof}

\subsection{Kernel convolutions on short cuts}
\label{sec:convolutions_shortcuts}

Let $t$, $u \in \mathbb{R}$. Then it is possible to prove the following results for the convolutions on the short cuts:
\bal
&K^{Qy}_{-}(t,z) \hstar K^{yP}_{-}(z, u) =\\
&-\frac{1}{2 \pi i} \left( \g_t^{+Q} + i \pi \right) K^{yP}_+(t+ \frac{i}{h} Q,u) +\frac{1}{2 \pi i} \left( \g_t^{-Q} + i \pi \right) K^{yP}_+(t- \frac{i}{h} Q,u)\\
&-\frac{1}{2 \pi i} \left( \g_u^{+P} + i \pi \right) K^{Qy}_+(t,u+\frac{i}{h} P) +\frac{1}{2 \pi i} \left( \g_u^{-P} + i \pi \right) K^{Qy}_+(t,u-\frac{i}{h}P) \,,
\eal
\bal
&K^{Qy}_{+}(t,z) \hstar K^{yP}_{+}(z, u) = +\frac{1}{2 \pi i} \left( \g_t^{+Q} + 2 i \pi \right) K^{yP}_+(t+ \frac{i}{h} Q,u) -\frac{1}{2 \pi i} \g_t^{-Q}  K^{yP}_+(t- \frac{i}{h} Q,u)\\
&+\frac{1}{2 \pi i} \left( \g_u^{+P} + 2 i \pi \right) K^{Qy}_+(t,u+\frac{i}{h} P) -\frac{1}{2 \pi i}  \g_u^{-P}  K^{Qy}_+(t,u-\frac{i}{h}P) \,,
\eal
\bal
&K^{Qy}_{+}(t,z) \hstar K^{yP}_{-}(z, u) = +\frac{1}{2 \pi i} \left( \g_t^{+Q} + 2 i \pi \right) K^{yP}_-(t+ \frac{i}{h} Q,u) -\frac{1}{2 \pi i} \g_t^{-Q}  K^{yP}_-(t- \frac{i}{h} Q,u)\\
&-\frac{1}{2 \pi i} \left( \g_u^{+P} + i \pi \right) K^{Qy}_-(t,u+\frac{i}{h} P) +\frac{1}{2 \pi i}  \left(\g_u^{-P} + i \pi \right)  K^{Qy}_-(t,u-\frac{i}{h}P) \,,
\eal
\bal
&K^{Qy}_{-}(t,z) \hstar K^{yP}_{+}(z, u) =\\
&-\frac{1}{2 \pi i} \left( \g_t^{+Q} + i \pi \right) K^{yP}_-(t+ \frac{i}{h} Q,u) +\frac{1}{2 \pi i} \left( \g_t^{-Q} + i \pi \right)  K^{yP}_-(t- \frac{i}{h} Q,u)\\
&+\frac{1}{2 \pi i} \left( \g_u^{+P} + 2 i \pi \right) K^{Qy}_-(t,u+\frac{i}{h} P) -\frac{1}{2 \pi i}  \g_u^{-P}  K^{Qy}_-(t,u-\frac{i}{h}P) \,.
\eal

\subsection{Kernel convolutions on long cuts}
\label{sec:app_kern_conv_on_long_cuts}

Let $t$, $u \in \mathbb{R}$.  Then it is possible to prove the following results for the convolutions on the long cuts:
\bal
&K^{Qy}_{-}(t,z+ i 0) \cstar K^{yP}_{-}(z+ i0, u) =\\
& +\frac{1}{2 \pi i} \left( \g_t^{+Q} + i \pi \right) K^{yP}_+(t+ \frac{i}{h} Q,u) -\frac{1}{2 \pi i} \left( \g_t^{-Q} + i \pi \right) K^{yP}_+(t- \frac{i}{h} Q,u)\\
&+\frac{1}{2 \pi i} \left( \g_u^{+P} + i \pi \right) K^{Qy}_+(t,u+\frac{i}{h} P) -\frac{1}{2 \pi i} \left( \g_u^{-P} + i \pi \right) K^{Qy}_+(t,u-\frac{i}{h}P) \,,
\eal
\bal
K^{Qy}_{+}(t,z+i0) \cstar K^{yP}_{+}(z+i0, u) =& -\frac{1}{2 \pi i} \g_t^{+Q}  K^{yP}_+(t+ \frac{i}{h} Q,u) +\frac{1}{2 \pi i} \g_t^{-Q}  K^{yP}_+(t- \frac{i}{h} Q,u)\\
&-\frac{1}{2 \pi i}  \g_u^{+P}  K^{Qy}_+(t,u+\frac{i}{h} P) +\frac{1}{2 \pi i}  \g_u^{-P}  K^{Qy}_+(t,u-\frac{i}{h}P) \,,
\eal
\bal
&K^{Qy}_{+}(t,z+i0) \cstar K^{yP}_{-}(z+i0, u) =\\
& -\frac{1}{2 \pi i}  \g_t^{+Q}  K^{yP}_-(t+ \frac{i}{h} Q,u) +\frac{1}{2 \pi i} \g_t^{-Q}  K^{yP}_-(t- \frac{i}{h} Q,u)\\
&+\frac{1}{2 \pi i} \left( \g_u^{+P} + i \pi \right) K^{Qy}_-(t,u+\frac{i}{h} P) -\frac{1}{2 \pi i}  \left(\g_u^{-P} + i \pi \right)  K^{Qy}_-(t,u-\frac{i}{h}P) \,,
\eal
\bal
&K^{Qy}_{-}(t,z+i0) \cstar K^{yP}_{+}(z+i0, u) =\\
&+\frac{1}{2 \pi i} \left( \g_t^{+Q} + i \pi \right) K^{yP}_-(t+ \frac{i}{h} Q,u) -\frac{1}{2 \pi i} \left( \g_t^{-Q} + i \pi \right)  K^{yP}_-(t- \frac{i}{h} Q,u)\\
&-\frac{1}{2 \pi i}  \g_u^{+P}  K^{Qy}_-(t,u+\frac{i}{h} P) +\frac{1}{2 \pi i}  \g_u^{-P}  K^{Qy}_-(t,u-\frac{i}{h}P) \,.
\eal

\subsection{Useful kernel convolutions for the inversion of \texorpdfstring{$\log Y_0$}{lnY0}}
\label{app:useful_convolutions}

Below, we list the necessary convolutions to derive the TBA equations from the Y system's discontinuities. These relations can be obtained from the relations above by taking, in a suitable way, the mass $Q$ or $P$ to zero.
Let $u \in (-2, +2)$, with $\Im(u)>0$, and $t \in \mathbb{R}$. Then it holds that
\bal
&\left(\int^{-2}_{-\infty} + \int_{+2}^{+\infty} \right) dz K^{Qy}_{-} (t, z+ i 0) K (z+ i 0, u)=
\frac{1}{4} K_{Q}(t-u)\\
&-\frac{1}{2 i \pi} \left( \g^{-Q \circ}_{t u} + \frac{i \pi}{2}\right) K(t- \frac{i}{h}Q,u)+\frac{1}{2 i \pi} \left( \g^{+Q \circ}_{t u} + \frac{i \pi}{2}\right) K(t+ \frac{i}{h}Q,u)\,,
\eal
\bal
\label{eq:final_result_KpQyconvK2}
&\left(\int^{-2}_{-\infty} + \int_{+2}^{+\infty} \right) dz K^{Qy}_{+} (t, z+ i 0) K (z+ i 0, u)=\frac{1}{4} K_{Q}(t-u)\\
&+\frac{1}{2 i \pi} \left( \g^{-Q \circ}_{t u} - \frac{i \pi}{2}\right) K(t- \frac{i}{h}Q,u)-\frac{1}{2 i \pi} \left( \g^{+Q \circ}_{t u} - \frac{i \pi}{2}\right) K(t+ \frac{i}{h}Q,u) \,.
\eal
The analytic derivation of these convolutions involves several steps; however, it is quite simple to verify these relations numerically. For $v$ just above the mirror cut and integrating in the $z$ plane at infinitely small distance $\eps$ from the cut (so that $0<\Im(v)< \eps$) we obtain
\bal
\label{eq:app_massless_convolutions}
&\left(\int_{-\infty}^{-2} + \int_{+2}^{+\infty} \right) dz \, K(v,z+ i \epsilon)\, K(z+i \epsilon ,u)  = K(v+ i 0, u) \left(1 -\frac{i}{\pi} \g^{\circ \circ}_{vu} \right) \,.
\eal
These convolutions enter the inversion of the discontinuities of the Y functions and appear, for example, in the computations of section~\ref{sec:inversionY0}.

The relations above can be obtained from the results in appendix~\ref{sec:app_kern_conv_on_long_cuts} in the limit as the mass approaches zero. They can also be obtained by explicitly integrating. Below, we show how one of these integrals can be computed analytically. The remaining integrals can be performed similarly, and we do not provide details of their derivation.

\subsubsection*{Convolution of \texorpdfstring{$K^{ny}_- K$}{KmK}}

Let us compute
\bal
I&=\left(\int^{-2}_{-\infty} + \int_{+2}^{+\infty} \right) dz K^{ny}_{-} (t, z+ i 0) K (z+ i 0, u)\,.
\eal
To perform the integration, we close the contour by adding a large arc in the upper half-plane at infinity. The contribution of the arc is suppressed since $K(z,u) \sim \frac{1}{z^2}$ at infinity. Then we deform the contour to the interval $(-2, +2)$ and get
\bal
I&=-\int^{+2}_{-2}  dz K^{ny}_{-} (t, z) K (z, u) + K^{ny}_{-} (t, u)\\
&=-\int^{+2}_{-2}  dz \frac{1}{2 \pi i } \frac{d}{d t} \ \log \left(\sqrt{\frac{x^{+n}(t)}{x^{-n}(t)}} \ \frac{x^{-n}(t) - \frac{1}{x(z)}}{x^{+n}(t) - \frac{1}{x(z)}} \right) \frac{1}{2 \pi i} \frac{\sqrt{4-u^2}}{\sqrt{4 - z^2}} \, \frac{1}{z-u} + K^{ny}_{-} (t, u)\,.
\eal
The term $K^{ny}_{-} (t, u)$ comes from  the pole of $K (z, u)$ at $z=u$. 
Using the parameterisation~\eqref{eq:gamma_u}, which in this case is
\begin{equation}
\g=\frac{1}{2} \log \left( \frac{2-z}{2+z} \right) \iff z= -2 \tanh \g
\end{equation}
and recalling also~\eqref{eq:x_funct_of_g}
\bal
x(z)=  -\coth \left( \frac{\g}{2} - i\frac{\pi}{4} \right)\,,
\eal
we can write
\bal
\label{eq:I_sum_I0_Km}
I=I_0 + K^{ny}_{-} (t, u)\,,
\eal
where
\bal
I_0 \equiv \int^{+\infty}_{- \infty}  \frac{d \g}{\cosh \g}  \frac{1}{2 \pi i } \frac{d}{d t} \ \log \left(\sqrt{\frac{x^{+n}(t)}{x^{-n}(t)}} \ \frac{x^{-n}(t) +\tanh \left( \frac{\g}{2} - i\frac{\pi}{4} \right)}{x^{+n}(t) +\tanh \left( \frac{\g}{2} - i\frac{\pi}{4} \right)} \right) \frac{1}{2 \pi i} \frac{\sqrt{4-u^2}}{2 \tanh \g+u}\,.
\eal
Now we want to compute $I_0$.

Note that the integrand of $I_0$ is periodic under the shift $\g \to \g + 2 i \pi$ and is exponentially suppressed as $\g \to \pm \infty$.
We define
\bal
I_\beta \equiv \int^{+\infty}_{- \infty}  \frac{d \g}{\cosh \g} e^{ i \beta \g } \frac{1}{2 \pi i } \frac{d}{d t} \ \log \left(\sqrt{\frac{x^{+n}(t)}{x^{-n}(t)}} \ \frac{x^{-n}(t) +\tanh \left( \frac{\g}{2} - i\frac{\pi}{4} \right)}{x^{+n}(t) +\tanh \left( \frac{\g}{2} - i\frac{\pi}{4} \right)} \right) \frac{1}{2 \pi i} \frac{\sqrt{4-u^2}}{2 \tanh \g+u} \,,
\eal
which in the limit $\beta \to 0$ reduces to the integral we want to compute.
Let $\Gamma$ be a closed counterclockwise rectangle having the lower side on the real line and upper side on the line of constant imaginary part $2 i \pi$.
Then it holds that
\bal
&\oint_{\Gamma}  \frac{d \g}{\cosh \g} e^{ i \beta \g } \frac{1}{2 \pi i } \frac{d}{d t} \ \log \left(\sqrt{\frac{x^{+n}(t)}{x^{-n}(t)}} \ \frac{x^{-n}(t) +\tanh \left( \frac{\g}{2} - i\frac{\pi}{4} \right)}{x^{+n}(t) +\tanh \left( \frac{\g}{2} - i\frac{\pi}{4} \right)} \right) \frac{1}{2 \pi i} \frac{\sqrt{4-u^2}}{2 \tanh \g+u}\\
&=I_\beta - e^{ -2 \pi \beta } I_\beta
\eal
from which we can write
\begin{small}
\bal
I_\beta=\frac{1}{1-e^{ -2 \pi \beta }}\oint_{\Gamma}  \frac{d \g}{\cosh \g}  \frac{e^{ i \beta \g }}{(2 \pi i )^2} \frac{d}{d t} \ \log \left(\sqrt{\frac{x^{+n}(t)}{x^{-n}(t)}} \ \frac{x^{-n}(t) +\tanh \left( \frac{\g}{2} - i\frac{\pi}{4} \right)}{x^{+n}(t) +\tanh \left( \frac{\g}{2} - i\frac{\pi}{4} \right)} \right)  \frac{\sqrt{4-u^2}}{2 \tanh \g+u}
\eal
\end{small}
Taylor expanding around $\beta=0$ and using that (due to the periodicity of the integrand)
\bal
\oint_{\Gamma}  \frac{d \g}{\cosh \g}  \frac{1}{2 \pi i } \frac{d}{d t} \ \log \left(\sqrt{\frac{x^{+n}(t)}{x^{-n}(t)}} \ \frac{x^{-n}(t) +\tanh \left( \frac{\g}{2} - i\frac{\pi}{4} \right)}{x^{+n}(t) +\tanh \left( \frac{\g}{2} - i\frac{\pi}{4} \right)} \right) \frac{1}{2 \pi i} \frac{\sqrt{4-u^2}}{2 \tanh \g+u}=0
\eal
we obtain
\begin{small}
\bal
I_0=\frac{i}{2 \pi  }\oint_{\Gamma}  \frac{d \g}{\cosh \g}     \frac{\g}{2 \pi i } \frac{d}{d t} \ \log \left(\sqrt{\frac{x^{+n}(t)}{x^{-n}(t)}} \ \frac{x^{-n}(t) +\tanh \left( \frac{\g}{2} - i\frac{\pi}{4} \right)}{x^{+n}(t) +\tanh \left( \frac{\g}{2} - i\frac{\pi}{4} \right)} \right) \frac{1}{2 \pi i} \frac{\sqrt{4-u^2}}{2 \tanh \g+u}
\eal
\end{small}
The integral is now performed on a closed contour and can be computed using Cauchy's theorem. 

Let us analyse the poles in the contour.
There are potential poles when
\bal
&\cosh \g=0 \rightarrow \g=i \frac{\pi}{2}, \ i \frac{3\pi}{2}\,.
\eal
However, these are not actual poles since at the values where $\cosh \g=0$ we also have $\tanh \g= \infty$ and the poles are cancelled by simultaneous zeros of $\frac{1}{2 \tanh \g+u}$. Then we have poles at
\bal
&-2\tanh \g=u \rightarrow \g=\frac{1}{2} \log \left( \frac{2-u}{2+u} \right) + i\pi, \  \frac{1}{2} \log \left( \frac{2-u}{2+u} \right) + 2 i\pi\,.
\eal
The poles in the contour are obtained by recalling that we are assuming $\Im(u)>0$ and therefore 
\bal
-\frac{\pi}{2}<\Im(  \frac{1}{2} \log \left( \frac{2-u}{2+u} \right) )<0\,.
\eal
Finally, we have the poles
\bal
&x^{-n}(t) +\tanh \left( \frac{\g}{2} - i\frac{\pi}{4} \right)=0 \rightarrow -\coth \left( \frac{\g^{-n}(t)}{2} - i\frac{\pi}{4} \right)+\tanh \left( \frac{\g}{2} - i\frac{\pi}{4} \right)=0\,,\\
&x^{+n}(t) +\tanh \left( \frac{\g}{2} - i\frac{\pi}{4} \right)=0 \rightarrow -\coth \left( \frac{\g^{+n}(t)}{2} - i\frac{\pi}{4} \right)+\tanh \left( \frac{\g}{2} - i\frac{\pi}{4} \right)=0\,.
\eal
Since $\Im(t-\frac{i}{h}n)<0$ then $0< \Im(\g^{-n}(t))< \frac{\pi}{2}$. Since $\Im(t+\frac{i}{h}n)>0$ then $ -\frac{\pi}{2}< \Im(\g^{-n}(t))<0$. The poles associated to the two equations above that are contained in the rectangle are
\bal
&\g=\g^{-n}(t)+i \pi\,,\\
&\g=\g^{+n}(t)+i \pi\, .
\eal

Summing over the residues we obtain
\bal
I_0&=\frac{1}{2 i \pi} \left( \g(u) + i\pi \right) K^{ny}_{+}(t, u)- \frac{1}{2 i \pi} \left( \g(u) +2 i\pi \right) K^{ny}_{-}(t, u)\\
&+\frac{1}{2 i \pi}  \left( \g^{+n}(t) + i \pi \right) K(t+\frac{i}{h}n, u)- \frac{1}{2 i \pi}  \left( \g^{-n}(t) + i \pi \right) K(t-\frac{i}{h}n, u) \,.
\eal
In the first line above are contained the contributions from the poles at $\g=\frac{1}{2} \log \left( \frac{2-u}{2+u} \right) + i\pi$ and $\g=\frac{1}{2} \log \left( \frac{2-u}{2+u} \right) + 2i\pi$. The second line contains the contributions from the poles at $\g=\g^{+n}(t)+i \pi$ and $\g=\g^{-n}(t)+i \pi$. Using the relations in \eqref{Krationalrels}, the expression above can also be written as
\bal
\label{eq:I0KmK}
I_0&=\frac{\g(u)}{2 i \pi} \left( K^{ny}_{+}(t, u) - K^{ny}_{-}(t, u) \right) - \frac{1}{2} K^{ny}_- (t,u)\\
&+\frac{\g^{+n}(t)}{2 i \pi}  K(t+\frac{i}{h}n, u) -\frac{\g^{-n}(t)}{2 i \pi}  K(t-\frac{i}{h}n, u)
\eal
Plugging this expression into~\eqref{eq:I_sum_I0_Km} we obtain
\bal
\label{eq:final_result_KmQyconvK}
&\left(\int^{-2}_{-\infty} + \int_{+2}^{+\infty} \right) dz K^{ny}_{-} (t, z+ i 0) K (z+ i 0, u)= \frac{1}{2} K^{ny}_- (t,u)\\
&+\frac{\g(u)}{2 i \pi} \left( K^{ny}_{+}(t, u) - K^{ny}_{-}(t, u) \right)+\frac{\g^{+n}(t)}{2 i \pi}  K(t+\frac{i}{h}n, u) -\frac{\g^{-n}(t)}{2 i \pi}  K(t-\frac{i}{h}n, u) \,.
\eal
Using the relations~\eqref{Krationalrels} we can express $K^{ny}_{-}$ and $K^{ny}_{+}$ in terms of the universal kernels as follows
\bal
\label{eq:Kp_Km_as_universal_K}
&K^{ny}_{-}(t,u)=\frac{1}{2}K_{n}(t-u)-\frac{1}{2} \left( K(t- \frac{i}{h}n,u) - K(t+ \frac{i}{h}n,u) \right)\,,\\
&K^{ny}_{+}(t,u)=\frac{1}{2}K_{n}(t-u)+\frac{1}{2} \left( K(t- \frac{i}{h}n,u) - K(t+ \frac{i}{h}n,u) \right)
\eal
and obtain
\bal
\label{eq:final_result_KmQyconvK2}
&\left(\int^{-2}_{-\infty} + \int_{+2}^{+\infty} \right) dz K^{ny}_{-} (t, z+ i 0) K (z+ i 0, u)= \frac{1}{4} K_{n}(t-u)\\
&-\frac{1}{2 i \pi} \left( \g^{-n \circ}_{t u} + \frac{i \pi}{2}\right) K(t- \frac{i}{h}n,u)+\frac{1}{2 i \pi} \left( \g^{+n \circ}_{t u} + \frac{i \pi}{2}\right) K(t+ \frac{i}{h}n,u) \,.
\eal

\bibliography{bibliography}
\end{document}